\documentclass[twocolumn]{aastex631}
\usepackage{CJK}

\usepackage{amsmath}
\usepackage{amssymb}
\usepackage{epsfig}
\usepackage{apjfonts}
\usepackage{natbib}
\usepackage{epstopdf}
\usepackage{verbatim}
\usepackage{enumitem}
\usepackage{color}
\setlist[enumerate]{itemsep=-1mm}
\usepackage{soul, xcolor}
\setstcolor{blue}

\usepackage{textcase}

\usepackage{mathtools}

\usepackage{bigints}

\newcommand\new[1]{{\color{black}#1}}          

\newcommand{\dotdeg}{\rlap{.}^\circ}
\newcommand{\dotarcsec}{\rlap{.}''}

\submitjournal{AJ (Received: August 1, 2024; Revised: October 6, 2024; Accepted: October 17, 2024)}

\begin{document}

\begin{CJK*}{UTF8}{gbsn}

\title{Disequilibrium Chemistry, Diabatic Thermal Structure, and Clouds in the Atmosphere of COCONUTS-2b}

\author[0000-0002-3726-4881]{Zhoujian Zhang (张周健)} \thanks{NASA Sagan Fellow}
\affiliation{Department of Astronomy \& Astrophysics, University of California, Santa Cruz, CA 95064, USA}

\author[0000-0003-1622-1302]{Sagnick Mukherjee}
\affiliation{Department of Astronomy \& Astrophysics, University of California, Santa Cruz, CA 95064, USA}

\author[0000-0003-2232-7664]{Michael C. Liu}
\affiliation{Institute for Astronomy, University of Hawai`i at M\={a}noa. 2680 Woodlawn Drive, Honolulu, HI 96822, USA}

\author[0000-0002-9843-4354]{Jonathan J. Fortney}
\affiliation{Department of Astronomy \& Astrophysics, University of California, Santa Cruz, CA 95064, USA}

\author{Emily Mader}
\affiliation{Department of Astronomy \& Astrophysics, University of California, Santa Cruz, CA 95064, USA}

\author[0000-0003-0562-1511]{William M. J. Best}
\affiliation{The University of Texas at Austin, Department of Astronomy, 2515 Speedway, C1400, Austin, TX 78712, USA}

\author[0000-0001-9823-1445]{Trent J. Dupuy}
\affiliation{Institute for Astronomy, University of Edinburgh, Royal Observatory, Blackford Hill, Edinburgh, EH9 3HJ, UK}

\author[0000-0002-3681-2989]{Sandy K. Leggett}
\affiliation{Gemini Observatory/NSFʼs NOIRLab, 670 N. A'ohoku Place, Hilo, HI 96720, USA}

\author[0000-0001-7356-6652]{Theodora Karalidi}
\affiliation{Department of Physics, University of Central Florida, 4111 Libra Dr, Orlando, FL 32816, USA}

\author[0000-0002-2338-476X]{Michael R. Line}
\affiliation{School of Earth and Space Exploration, Arizona State University, Tempe, AZ 85281, USA}

\author[0000-0002-5251-2943]{Mark S. Marley}
\affiliation{Lunar \& Planetary Laboratory, University of Arizona, Tucson, AZ 85721, USA}

\author[0000-0002-4404-0456]{Caroline V. Morley}
\affiliation{The University of Texas at Austin, Department of Astronomy, 2515 Speedway, C1400, Austin, TX 78712, USA}

\author[0000-0001-6041-7092]{Mark W. Phillips}
\affiliation{Institute for Astronomy, University of Edinburgh, Royal Observatory, Blackford Hill, Edinburgh, EH9 3HJ, UK}

\author[0000-0001-5016-3359]{Robert J. Siverd}
\affiliation{Institute for Astronomy, University of Hawai`i at M\={a}noa. 2680 Woodlawn Drive, Honolulu, HI 96822, USA}

\author[0000-0001-9828-1848]{Joseph A. Zalesky}
\affiliation{School of Earth and Space Exploration, Arizona State University, Tempe, AZ 85281, USA}

\begin{abstract}
Located 10.888~pc from Earth, COCONUTS-2b is a planetary-mass companion to a young {(150--800~Myr)} M3 star, with a wide orbital separation (6471~au) and a low companion-to-host mass ratio ($0.021\pm0.005$). We have studied the atmospheric properties of COCONUTS-2b using newly acquired 1.0--2.5~$\mu$m spectroscopy from Gemini/Flamingos-2. The spectral type of COCONUTS-2b is refined to T$9.5 \pm 0.5$ based on comparisons with T/Y dwarf spectral templates. We have conducted an extensive forward-modeling analysis, comparing the near-infrared spectrum and mid-infrared broadband photometry of COCONUTS-2b with sixteen state-of-the-art atmospheric model grids {developed for brown dwarfs and self-luminous exoplanets near the T/Y transition}. The {\texttt{PH$_{3}$-free ATMO2020++},} \texttt{ATMO2020++}, and \texttt{Exo-REM} models best match the specific observations of COCONUTS-2b, regardless of variations in the input spectrophotometry. This analysis suggests the presence of disequilibrium chemistry, along with a diabatic thermal structure and/or clouds, in the atmosphere of COCONUTS-2b. All models predict fainter $Y$-band fluxes than observed, highlighting uncertainties in the alkali chemistry models {and opacities}. We determine a bolometric luminosity of $\log{(L_{\rm bol}/L_{\odot})}=-6.18$~dex, with a 0.5~dex-wide range of $[-6.43,-5.93]$~dex that accounts for various assumptions of atmospheric models. Using several thermal evolution models, we derive an effective temperature of $T_{\rm eff}=483^{+44}_{-53}$~K, a surface gravity of $\log{(g)}=4.19^{+0.18}_{-0.13}$~dex, a radius of $R=1.11^{+0.03}_{-0.04}$~R$_{\rm Jup}$, and a mass of $M=8 \pm 2$~M$_{\rm Jup}$. Various atmospheric model grids consistently indicate that COCONUTS-2b's atmosphere likely has sub- or near-solar metallicity and C/O. These findings provide valuable insights into COCONUTS-2b's formation history and the potential outward migration to its current wide orbit.  
\end{abstract}

\section{Introduction} 
\label{sec:introduction}

There has been a remarkable surge in spectroscopic observations of exoplanets over the past few years, powered by new-generation ground-based facilities and the James Webb Space Telescope \citep[JWST; e.g.,][]{2020A&A...633A.110G, 2020A&A...640A.131M, 2021AJ....162..148W, 2021Natur.595..370Z, 2023Natur.614..664A, 2023ApJ...953L..24A, 2023Natur.618...43B, 2023ApJ...951L..48B, 2023Natur.620..292C, 2023Natur.614..670F, 2023Natur.614..649J, 2023NatAs...7.1317L, 2023ApJ...946L...6M, 2023ApJ...956L..13M, 2023ApJ...948L..11M, 2023Natur.614..659R, 2023Natur.617..483T, 2023Natur.620..746Z, 2024arXiv240708518B, 2024AJ....168....4H, 2024arXiv240606493K, 2024ApJ...966L..11P}. These spectra offer quantitative insights into the atmospheres, formation, and evolution of the diverse population of exoplanets, uncovering rich information about the physical and chemical processes in their atmospheres \citep[for recent reviews, see][]{2015ARA&A..53..279M, 2018arXiv180408149F, 2020RAA....20...99Z, 2022ARA&A..60..159W, 2023ASPC..534..799C, 2024RvMG...90..411K}. These processes include vertical mixing, horizontal transport, cloud condensation, formation of photochemical hazes, hot spots, mass loss, planetesimal accretion, and auroral emission \citep[e.g.,][]{2005MNRAS.364..649F, 2009ApJ...699..564S, 2016Natur.529...59S, 2017Sci...357..683A, 2018AJ....156...10M, 2021NatAs...5.1224M, 2020A&A...640A.131M, 2021JGRE..12606655G, 2021Natur.595..370Z, 2022ApJ...938..107M, 2023AJ....165..238M, 2023AJ....166..198Z, 2023SciA....9F8736Z, 2024AJ....167..142G, 2023AJ....166...85H, 2024AJ....167..218I, 2024ApJ...962...10X, 2024Natur.628..511F}. Moreover, the atmospheric compositions of exoplanets and brown dwarfs, as measured from spectra, exhibit trends that can be contextualized with those observed in our solar system, providing crucial insights into their formation pathways \citep[e.g.,][]{2019ApJ...887L..20W, 2021Natur.595..370Z, 2023Natur.624..263B, 2023AJ....166...85H, 2023AJ....166..198Z, 2024AJ....167..237L, 2024arXiv240403776N, 2024arXiv240513128X}. 

\begin{figure*}[t]
\begin{center}
\includegraphics[width=6.5in]{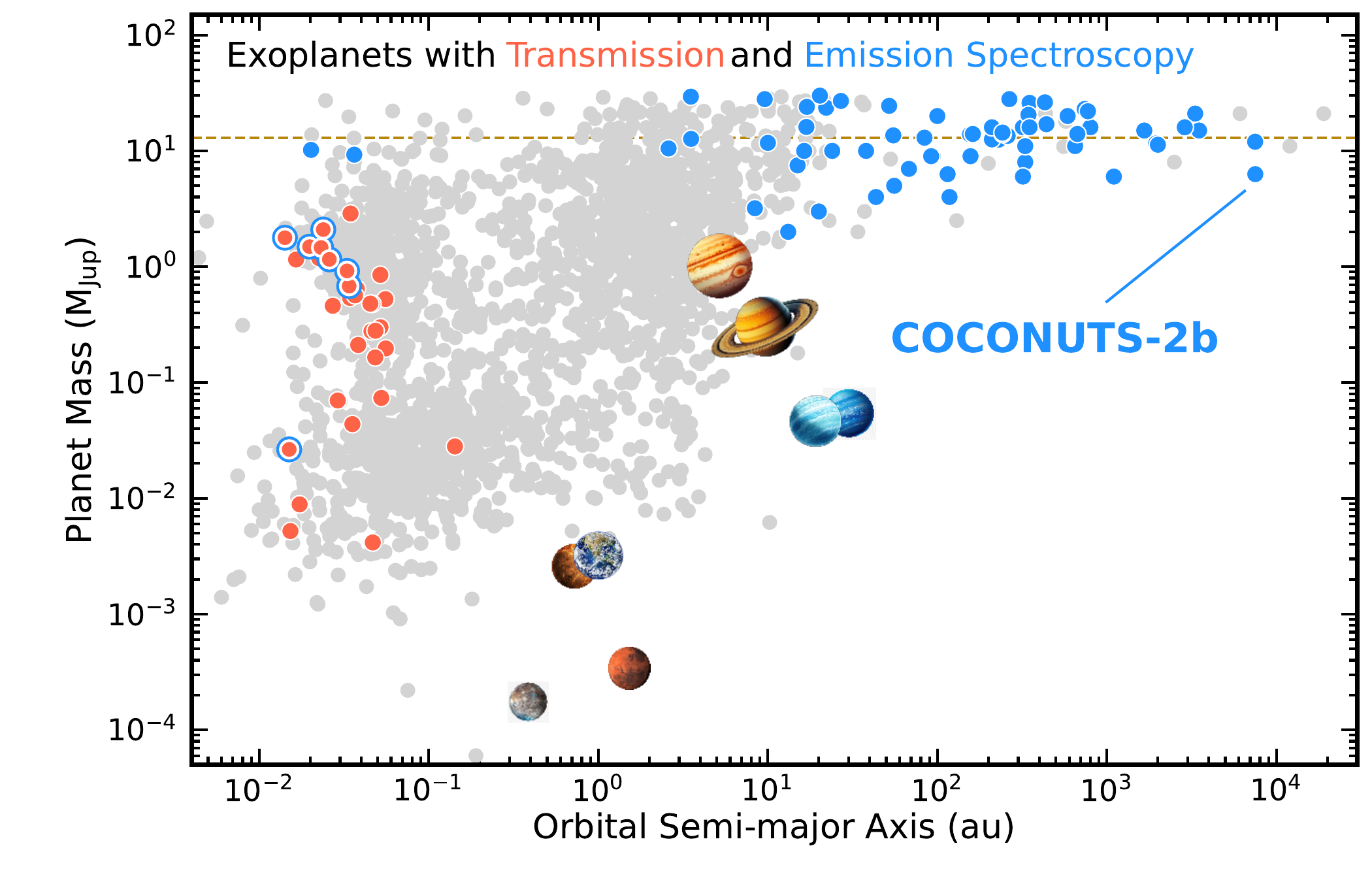}
\includegraphics[width=6.5in]{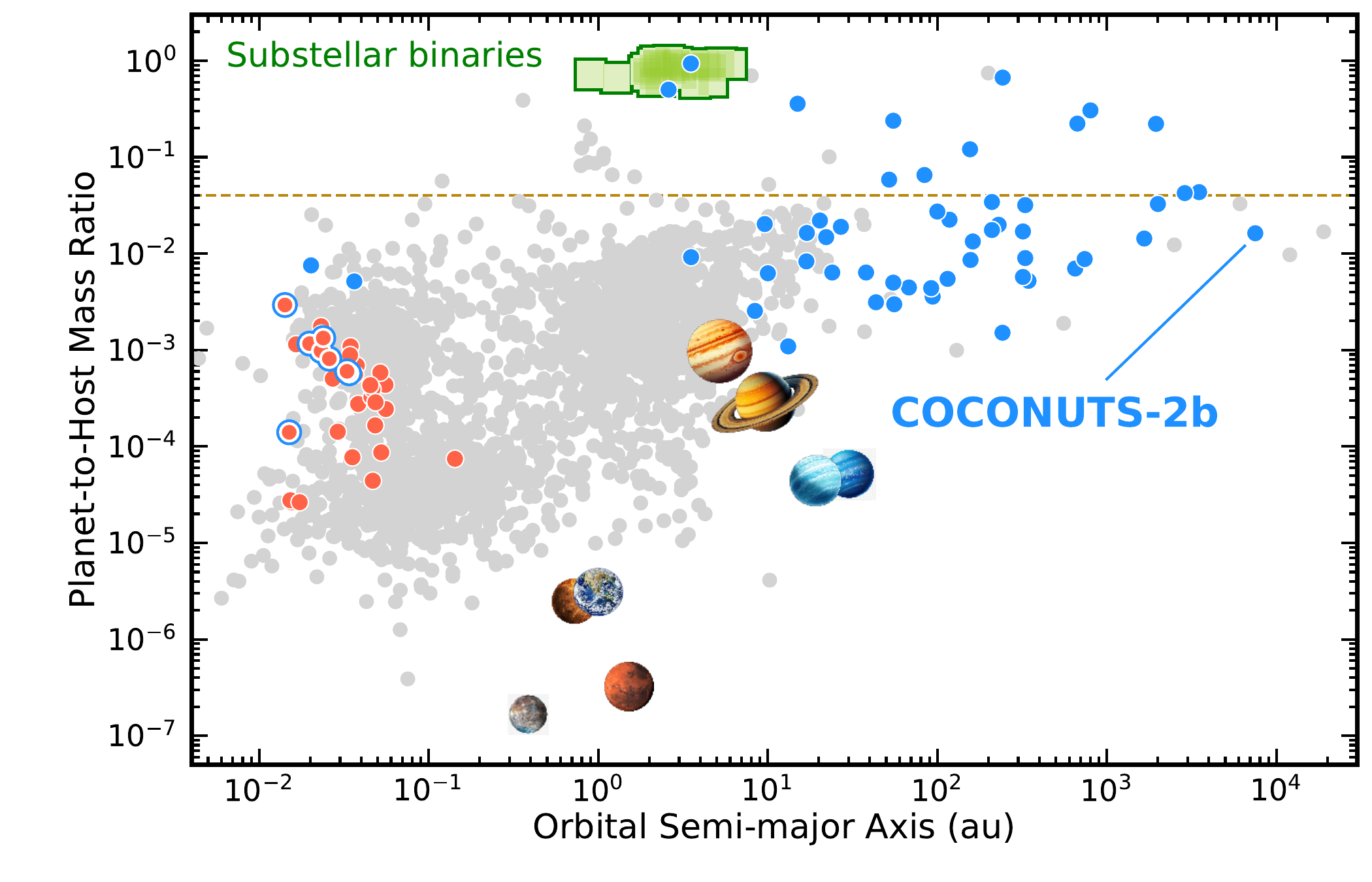}
\caption{Masses (top) or mass ratios (bottom) of confirmed exoplanets {(collected from the NASA Exoplanet Archive on June 21, 2024)} as functions of their orbital separations (grey), with those having observed transmission and emission spectra highlighted in orange and blue, respectively. Substellar binaries with dynamical masses are shown in green \citep[][]{2016ApJ...827...23D, 2019AJ....158..174D, 2023MNRAS.519.1688D, 2017ApJS..231...15D, 2018A&A...618A.111L, 2023A&A...671A.119C}. Solar system planets are plotted with sizes unscaled. Brown dashed lines highlight the planet mass of 13~$M_{\rm Jup}$ and the planet-to-host mass ratio of 0.04, as adopted by the IAU working definition of exoplanets. }
\label{fig:context}
\end{center}
\end{figure*}

However, only a relatively small portion of known exoplanets has had their transmission or emission spectra observed so far (Figure~\ref{fig:context}). \new{Among the 5548 confirmed exoplanets as of June 21, 2024 \citep{10.26133/nea12}, transmission spectra have been obtained for 44 objects, emission spectra for 69 objects, and both types of spectra for 10 objects, with each spectrum containing more than ten data points.} Observing transmission spectra is challenging due to the planets' relatively small transit depths, the activity and starspots of some planet-host stars, and the potential absence of detectable exoplanetary atmospheres. For emission spectroscopy, most data have been collected for directly imaged, self-luminous planets that reside on relatively wide orbits ($\gtrsim 10$~au) with high masses ($\gtrsim 2$~M$_{\rm Jup}$; {for reviews, see \citealt{2010exop.book..111T, 2016PASP..128j2001B, 2023ASPC..534..799C}}). The primary challenge of such observations arises from the contaminating light of these planets' much brighter host stars (over $10^{4}$ times brighter in flux) at small angular separations ($<1\arcsec$).

\begin{figure*}[t]
\begin{center}
\includegraphics[width=6.5in]{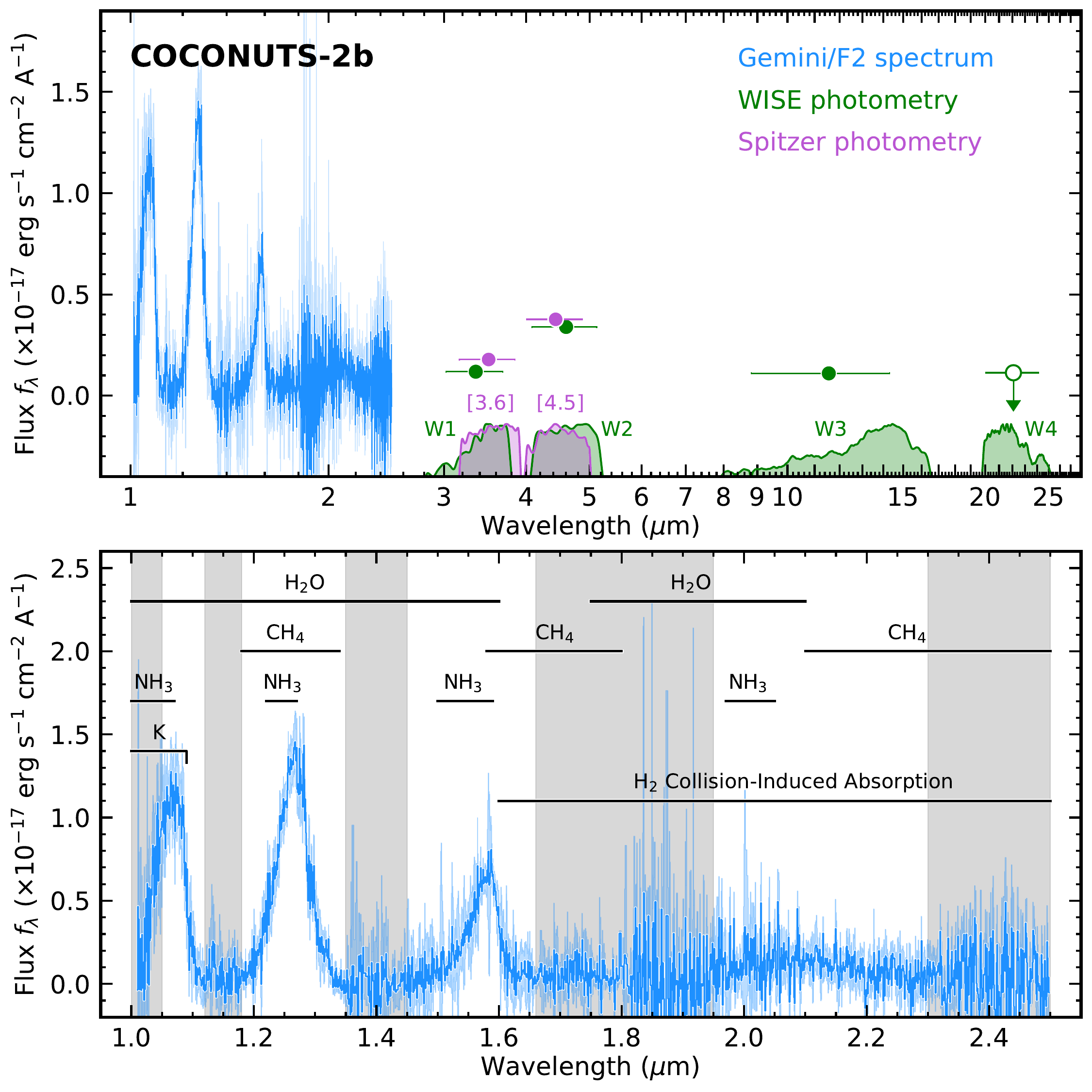}
\caption{{\it Top}: The 1--25~$\mu$m SED of COCONUTS-2b, including our new Gemini/F2 spectrum (blue) and the WISE (green) and Spitzer (purple) photometry. This object's $W1$ and $W2$ photometry are from CatWISE, with proper motion accounted for; the $W4$ photometry is an upper limit. The spectral flux uncertainties are shown as light shades. {\it Bottom}: The Gemini/F2 spectrum with key spectral features labeled. Fluxes in grey shades are masked for the subsequent forward-modeling analysis in Section~\ref{sec:fm}. }
\label{fig:c2b}
\end{center}
\end{figure*}

COCONUTS-2b \citep[WISEPA~J075108.79$-$763449.6;][]{2021ApJ...916L..11Z} is a remarkable self-luminous planetary-mass object suitable for high-quality spectroscopy. This object's thermal emission can be distinctly detected from that of its parent star due to the wide projected separation ($6471$~au or $594''$; Figure~\ref{fig:context}). COCONUTS-2b was initially considered a free-floating T9 brown dwarf as found by \cite{2011ApJS..197...19K}. \cite{2021ApJ...916L..11Z} established its physical association with a well-known young M3 star, COCONUTS-2A (L~34$-$26), based on their consistent parallaxes and proper motions. \cite{2021ApJ...916L..11Z} assigned an age of 150--800~Myr to the COCONUTS-2A+b system, using indicators such as the host star's alkali lines, absence of lithium absorption, strong H$\alpha$ emission, UV and X-ray luminosities, rotation, kinematics, and position in the HR diagram (see their Section~3.2). {According to the BANYAN $\Sigma$ tool\footnote{\url{https://www.exoplanetes.umontreal.ca/banyan/banyansigma.php}} developed by \cite{2018ApJ...856...23G}, COCONUTS-2A is not associated with any known young moving groups. However, an updated version of BANYAN $\Sigma$ suggests that COCONUTS-2A may be a candidate member of the $\sim 400$~Myr-old Ursa Major association \citep{2024ApJ...967..147M}, as proposed by \cite{2020RNAAS...4...92G}. This potential membership supports the age estimates of 150--800~Myr established by \cite{2021ApJ...916L..11Z}. Furthermore, by} combining the stellar radius, rotation period, and the projected rotation velocity, \cite{2021ApJ...916L..11Z} suggested that the spin axis of COCONUTS-2A has a nearly edge-on inclination. \cite{2023AJ....165..164B} quantified this stellar inclination to $80\dotdeg2^{+9\dotdeg7}_{-4\dotdeg3}$, with the probability distribution peaking at $90^{\circ}$. 

With a planetary mass ($8 \pm 2$~M$_{\rm Jup}$) and a low mass ratio relative to its parent star ($0.021 \pm 0.005$),\footnote{The mass and companion-to-host mass ratio of COCONUTS-2b are determined by this work (Section~\ref{subsec:rec_params}), updated from the values reported by \cite{2021ApJ...916L..11Z}.} COCONUTS-2b satisfies the IAU working definition of an exoplanet\footnote{\url{https://www.iau.org/science/scientific_bodies/commissions/F2/info/ documents/}} at the time of writing (Figure~\ref{fig:context}). However, the formation pathway of COCONUTS-2b remains elusive, especially given the wide orbital separation from its parent star \citep[see discussions in][]{2021ApJ...916L..11Z, 2024ApJ...967..147M}. {One possibility is that COCONUTS-2b formed within the circumstellar disk of COCONUTS-2A and subsequently migrated outward through dynamical interactions \citep[e.g.,][]{2013A&A...552A.129V}. Alternatively, COCONUTS-2b may have formed via gravitational instability in a collapsing molecular cloud, similar to the secondary components in stellar binary systems, as its mass is close to the estimated lower mass limit for star formation products \citep[e.g.,][]{2009A&A...493.1149Z, 2021ApJS..253....7K}. Additionally, \cite{2024ApJ...967..147M} discussed that COCONUTS-2b might be a captured object from a flyby; they suggested that while this scenario is possible, it is less likely due to the high relative velocities typically expected between the two objects. }

COCONUTS-2b has extreme properties. Located at $10.888 \pm 0.002$~pc\footnote{Here we use the trigonometric distance of the parent star as the distance of this system, given the consistency in the parallax between COCONUTS-2b \citep[$97.9 \pm 6.7$~mas;][]{2019ApJS..240...19K} and COCONUTS-2A \citep[$91.826 \pm 0.019$~mas;][]{2016A&A...595A...2G, 2021A&A...649A...1G}.} \citep{2021AJ....161..147B}, it is the second-nearest directly imaged planet to the solar system \citep[after Eps~Ind~Ab;][]{matthews2024}. COCONUTS-2b has the second-coldest effective temperature \citep[after WD0806-661~B;][]{2011ApJ...730L...9L} and the second-widest orbit \citep[after TYC~9486-927-1~b;][]{2016MNRAS.457.3191D} among all imaged exoplanets to date. Additionally, the bolometric luminosity and age of COCONUTS-2b are consistent with both hot-start and cold-start thermal evolution models, a characteristic shared by only three other objects among all the known directly imaged exoplanets and brown dwarfs: WD~0806-661~B \citep{2011ApJ...730L...9L}, 51~Eri~b \citep{2015Sci...350...64M}, and AF~Lep~b \citep{2023A&A...672A..94D, 2023ApJ...950L..19F, 2023A&A...672A..93M}. Among these four objects, the post-formation entropies and/or formation epochs of 51~Eri~b and AF~Lep~b can be constrained by warm-start thermal evolution models based on these objects' directly measured luminosities and dynamical masses \citep[][]{2022MNRAS.509.4411D, 2024RNAAS...8..114Z}. However, such analysis cannot be applied to COCONUTS-2b as yet, due to the difficulties in measuring its dynamical mass given its slow orbital motion at such a wide separation from its parent star. 

The atmospheric properties of COCONUTS-2b have been studied through its photometry in the literature. Comparing the [3.6]$-$[4.5] and $J-$[4.5] colors with the atmospheric models of \cite{2015ApJ...804L..17T}, \cite{2017ApJ...842..118L} suggested that this object is likely a subdwarf with a low metallicity. \cite{2021ApJS..253....7K} found that the absolute magnitudes of COCONUTS-2b in the $H$, $W1$, and [3.6] bands are fainter than those of brown dwarfs with similar [3.6]$-$[4.5] colors. Additionally, \cite{2021ApJ...916L..11Z} compared the observed broadband photometry with several atmospheric models, finding evidence of both disequilibrium chemistry and clouds in the atmosphere of COCONUTS-2b. 

We present new near-infrared spectra of COCONUTS-2b collected from the Gemini-South telescope (Section~\ref{sec:observation}). Comparing these data with T and Y brown dwarf spectral templates results in a slightly updated spectral type of T9.5$\pm 0.5$ (Section~\ref{sec:spt}). To estimate the atmospheric properties of COCONUTS-2b, we compare its Gemini spectrum with sixteen state-of-the-art atmospheric model grids of T/Y dwarfs (Section~\ref{sec:fm}). These atmospheric model grids have various assumptions, allowing us to examine what model assumption can best approximate the specific atmosphere of our target. Our extensive forward-modeling analysis suggests the presence of disequilibrium chemistry, a diabatic thermal structure, clouds, a sub-solar metallicity, and a sub- or near-solar carbon-to-oxygen ratio (C/O) in the atmosphere of COCONUTS-2b (Section~\ref{sec:discussion}).

\section{Near-Infrared Spectroscopy with Gemini/FLAMINGOS-2} 
\label{sec:observation}

We acquired near-infrared spectra of COCONUTS-2b using the FLAMINGOS-2 (F2) spectrograph \citep{2004SPIE.5492.1196E, 2008SPIE.7014E..0VE} on the 8.1-m Gemini-South telescope (GS-2021B-FT-111, GS-2021B-FT-204). The adaptive optics system was not needed because this planet is widely separated from its host star ($594\arcsec$) and neighboring detectable sources (mostly $>20\arcsec$ except for one object at $8''$ with comparable $J$-band fluxes). Thirty 200-second frames were taken with the JH grism (0.98--1.80~$\mu$m) on 2021 December 16 UT, and thirty-six 180-second frames were obtained with the HK grism (1.29--2.51~$\mu$m) on 2021 December 19 UT and 23 UT. Data were all collected under photometric conditions using the 2-pixel-wide slit ($0\dotarcsec36 \times 263\arcsec$) and an ABBA nodding pattern, resulting in an average spectral resolution of $R \sim 900$ (with a maximum of $R \sim 1200$). The A0V telluric standard stars HIP~43762 (on December 16 and 23, {2021}) and HIP~28680 (on December 19, {2021}) were observed contemporaneously with COCONUTS-2b with airmass differences of $0.01-0.07$.

Data reduction was performed using the Gemini IRAF package following the standard procedures, including dark subtraction, flat fielding, bad-pixel masking, wavelength calibration, one-dimensional spectral extraction, telluric correction, and correction of the detector's sensitivity.\footnote{\url{https://gemini-iraf-flamingos-2-cookbook.readthedocs.io/en/latest/Tutorial_Longslit.html}} In particular, spectra of COCONUTS-2b were extracted using the traces of the corresponding telluric standard stars as references. This approach allows for more reliable spectral extraction, especially for T/Y dwarfs, whose spectra appear as dashes on the detector (due to the strong H$_{2}$O and CH$_{4}$ absorption in these objects' atmospheres) and cannot be traced along the full wavelengths \citep[e.g.,][]{2016ApJ...824....2L}. Our telluric standard stars were also used to derive the sensitivity function of the detector. We scaled and combined the $JH$ and $HK$ spectra by aligning their weighted average fluxes near the $H$-band peak (1.567--1.602~$\mu$m). {The required scaling factor for the HK spectrum observed on December 19 and 23 is about $0.87$ and $1.73$, respectively.} The resulting 1.0--2.5~$\mu$m spectrum (in vacuum wavelengths) was flux-calibrated using $J_{\rm MKO} = 19.342 \pm 0.048$~mag \citep{2011ApJS..197...19K} and the corresponding filter response curve and zero-point flux from \cite{2002PASP..114..180T}. Figure~\ref{fig:c2b} presents the 1--25~$\mu$m spectral energy distribution (SED) of COCONUTS-2b, including our Gemini/F2 spectrum and the WISE, CatWISE, and Spitzer/IRAC photometry \citep{2011ApJS..197...19K, 2014yCat.2328....0C, 2020ApJS..247...69E, 2021ApJS..253....8M}. In Appendix~\ref{app:comp_archival}, we compare our Gemini/F2 spectrum with an archival spectrum \citep{2011ApJS..197...19K} and the $Y/J/H/K$ broadband photometry.

\begin{figure}[t]
\begin{center}
\includegraphics[width=3.4in]{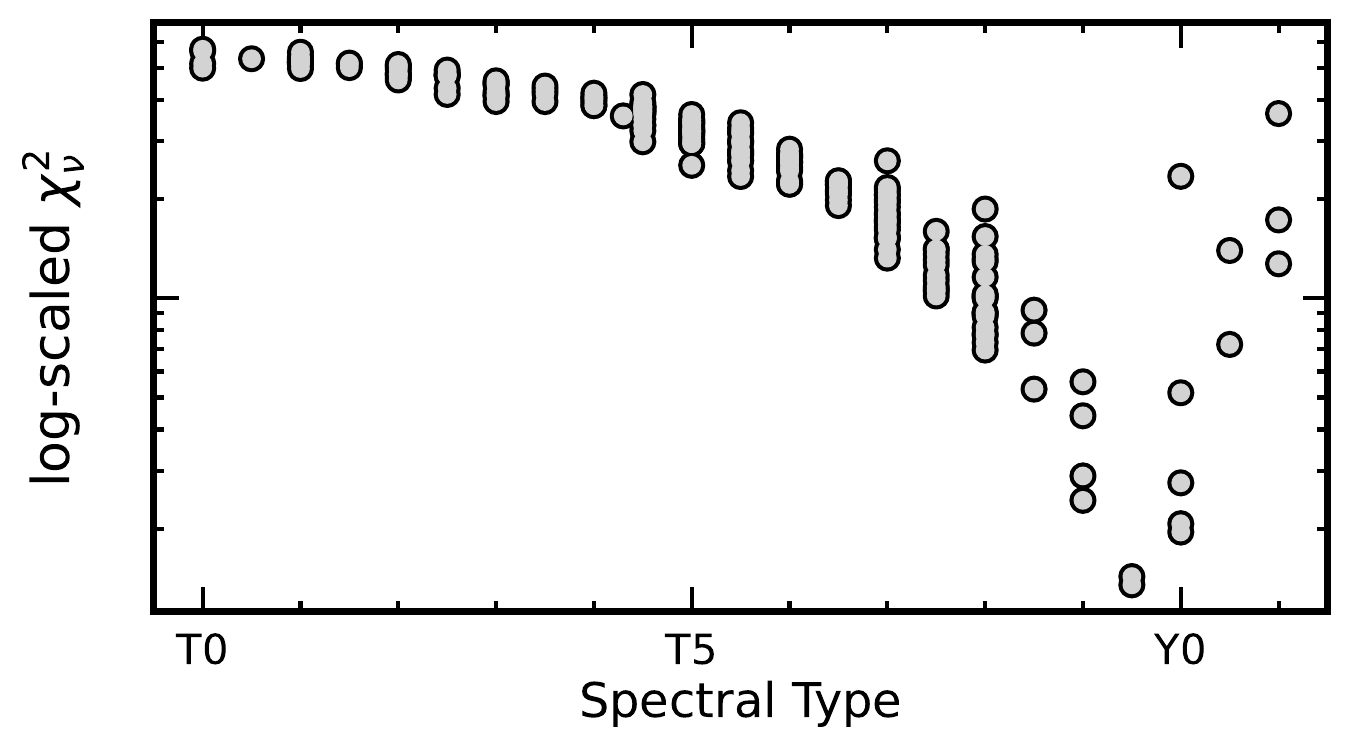}
\includegraphics[width=3.4in]{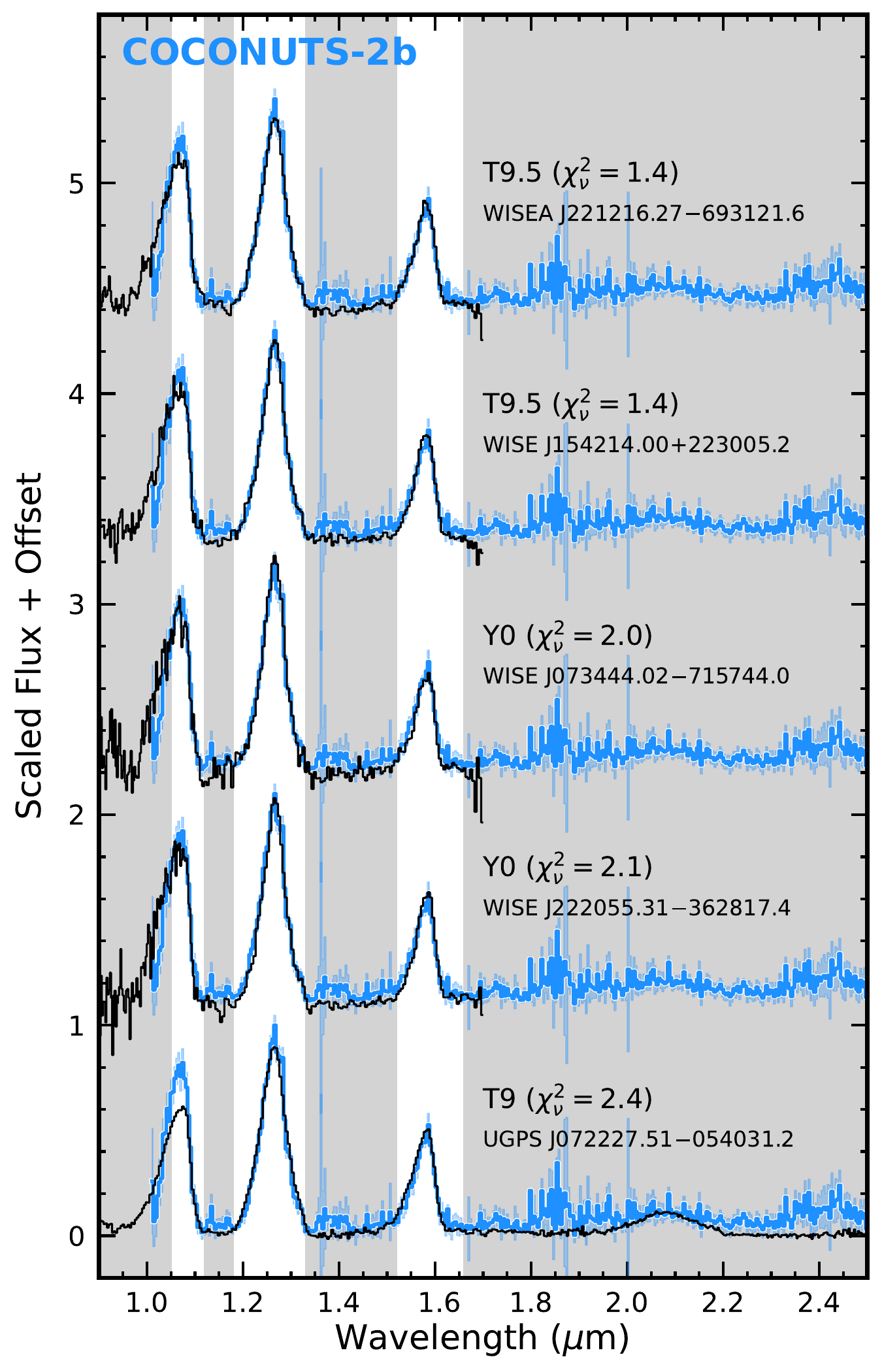}
\caption{{\it Top}: Reduced $\chi^{2}$ values of the fitted T0--Y1 spectral templates. The best-matched template with the lowest $\chi_{\nu}^{2}$ has a T9.5 spectral type. {\it Bottom}: The Gemini/F2 spectrum of COCONUTS-2b (blue; with downgraded spectral resolution) overlaid with five best-fitted spectral templates (black). The masked wavelength ranges in this comparison are highlighted in grey shades, which are different from the grey shades shown in Figure~\ref{fig:c2b}. }
\label{fig:spt_single}
\end{center}
\end{figure}

\section{A Refined Spectral Type} 
\label{sec:spt}

We updated the spectral type of COCONUTS-2b by comparing its Gemini/F2 spectrum with 209 T0--Y1 ultracool dwarf templates. These ultracool dwarfs were first selected from the UltracoolSheet \citep{ultracoolsheet} based on spectral types.  Binaries or candidate binaries (as suggested by their peculiar spectral morphology) with unresolved spectrophotometry were excluded. We then collected the published spectra of these objects from the literature. Most spectral templates near the T/Y transition were observed by HST/WFC3 \citep[$R \sim 130-210$;][]{2015ApJ...804...92S}; the remaining templates are from IRTF/SpeX \citep[$R \sim 80-200$; e.g.,][]{2014ASInC..11....7B, 2015ApJ...814..118B}. 

{To perform the comparison,} we downgraded the resolution of COCONUTS-2b's spectrum, scaled the fluxes of each spectral template following the Equation~2 of \cite{2008ApJ...678.1372C}, and computed the reduced $\chi^{2}$ statistics. This comparison was performed over wavelengths of 1.05--1.12~$\mu$m, 1.18--1.33~$\mu$m, and 1.52--1.66~$\mu$m. Fluxes from wavelengths shorter than $1.05$~$\mu$m were masked to avoid the low sensitivity area near the edge of Gemini/F2's detector; the flux valleys among the $Y/J/H$ peaks were masked due to the residuals from telluric correction and the relatively large flux uncertainties of the Gemini/F2 spectrum; fluxes beyond 1.65~$\mu$m were masked considering the wavelength coverage of HST/WFC3 spectral templates. 

As shown in Figure~\ref{fig:spt_single}, {the five best-matched templates to COCONUTS-2b's spectrum are WISEA~J221216.27$-$693121.6 \citep[T9.5;][]{2015ApJ...804...92S}, WISE~J154214.00$+$223005.2 \citep[T9.5;][]{2013ApJS..205....6M}, WISE~J073444.02$-$715744.0 \citep[Y0;][]{2012ApJ...753..156K}, WISE J222055.31$-$362817.4 \citep[Y0;][]{2012ApJ...753..156K}, and UGPS J072227.51$-$054031.2 \citep[T9;][]{2010MNRAS.408L..56L}. The near-infrared spectra of the first four objects and the last one were obtained by \cite{2015ApJ...804...92S} and \cite{2011ApJ...743...50C}, respectively. Consequently, we assign a new spectral type of T9.5$\pm$0.5 to COCONUTS-2b.}

\section{Forward-Modeling Analysis of COCONUTS-2\lowercase{b}} 
\label{sec:fm}
We compare the spectrophotometry of COCONUTS-2b with sixteen state-of-the-art atmospheric model grids, pre-computed with various assumptions about the atmospheres of T/Y dwarfs. Each model grid is utilized over its entire regular parameter space when possible, with an upper limit of 800~K in the effective temperature ($T_{\rm eff}$).

\subsection{Atmospheric Model Grids} 
\label{subsec:modelgrids}

\subsubsection{ATMO2020}
\label{subsec:atmo2020p20}

\new{The \texttt{ATMO2020} atmospheric models\footnote{Available via the ERC-ATMO Opendata website maintained by P. Tremblin: \url{https://noctis.erc-atmo.eu/fsdownload/zyU96xA6o/phillips2020}} were developed by \cite{2020A&A...637A..38P}}, assuming radiative-convective equilibrium, solar metallicity, and cloud-free atmospheres. These models incorporate recent line profiles for the potassium resonance doublet \citep[][]{2016A&A...589A..21A} that shape the spectra of T and Y dwarfs in the red optical and $Y$ band (e.g., Figure~\ref{fig:c2b}). The grid parameters include $T_{\rm eff}$ and logarithmic surface gravity ($\log{(g)}$).\footnote{Throughout this manuscript, we use ``$\log{()}$'' and ``$\ln{()}$'' for 10-based and natural logarithm, respectively.} ATMO2020 includes a set of rainout chemical equilibrium models (\texttt{ATMO2020 CEQ}), and two sets of chemical disequilibrium models with weak (\texttt{ATMO2020 NEQ weak}) and strong (\texttt{ATMO2020 NEQ strong}) vertical mixing. The mixing is quantified by the eddy diffusion coefficient $K_{\rm zz}$, assumed to vary from $10^{10}$~cm$^{2}/$s or $10^{8}$~cm$^{2}/$s at $\log{(g)} = 2.5$~dex to $10^{4}$~cm$^{2}/$s or $10^{2}$~cm$^{2}/$s at $\log{(g)} = 5.5$~dex for the strong and weak cases, respectively; the $K_{\rm zz}$ is assumed to be constant with pressures. Our forward-modeling analysis uses all three versions of the grids with $T_{\rm eff} \in [200, 800]$~K and $\log{(g)} \in [2.5, 5.5]$~dex.

\subsubsection{ATMO2020++ and PH$_{3}$-free ATMO2020++} 
\label{subsec:atmo2020plusplus}
By expanding the \texttt{ATMO2020} framework \citep{2020A&A...637A..38P}, \cite{2021ApJ...918...11L} and \cite{2023AJ....166...57M} developed the \texttt{ATMO2020++} \new{models,\footnote{Available via the ERC-ATMO Opendata website maintained by P. Tremblin: \url{https://noctis.erc-atmo.eu/fsdownload/Q7MUSoCLR/meisner2023}}} which additionally incorporate a non-adiabatic thermal structure in the atmospheres. Such a diabatic process has been suggested by studies of solar-system gas giant planets and brown dwarfs \citep[e.g.,][]{1994Icar..112..354G, 2015ApJ...804L..17T, 2019ApJ...876..144T, 2017ApJ...842..118L, 2023ApJ...959...86L, 2023A&A...670L...9P}. The \texttt{ATMO2020++} models are obtained by tuning the adiabatic index $\gamma$ ($\approx 1.4$ for pure molecular hydrogen gas) of the \texttt{ATMO2020 NEQ strong} models until the predicted emission spectra qualitatively match the observed data of a handful of T/Y brown dwarfs. The modified adiabatic index can reduce the temperature gradient of the atmospheric temperature-pressure profile (Section~\ref{subsec:properties}). The publicly available grid assumes the adiabatic temperature-pressure profile in the deep atmosphere, with a tuned $\gamma = 1.25$ for pressures below 15~bar. The grid parameters of \texttt{ATMO2020++} include $T_{\rm eff}$, $\log{(g)}$, and metallicity ([M/H]) at solar and sub-solar values. We use the models with $T_{\rm eff} \in [250, 800]$~K, $\log{(g)} \in [2.5, 5.5]$~dex, and [M/H] values of $-1.0$~dex, $-0.5$~dex, $0$~dex, and $+0.3$~dex.

Building upon the \texttt{ATMO2020++} grid, \cite{2024RNAAS...8...13L} published the \texttt{PH$_{3}$-free ATMO2020++} \new{grid.\footnote{Available via the ERC-ATMO Opendata website maintained by P. Tremblin: \url{https://noctis.erc-atmo.eu/fsdownload/9puhIZma2/leggett2024}}} The assumptions of these two model grids are kept the same, except that the latter does not include phosphine, as inspired by the JWST observations of late-T and Y brown dwarfs \citep[e.g.,][]{2023ApJ...951L..48B, 2023ApJ...959...86L, 2024AJ....167....5L}. We use the models with $T_{\rm eff} \in [250, 800]$~K, $\log{(g)} \in [4.0, 5.0]$~dex, and [M/H] values of $-0.5$~dex, $0$~dex, and $+0.3$~dex.

\subsubsection{Sonora Bobcat and Sonora Elf Owl} 
\label{subsec:bobcat}

The \texttt{Sonora Bobcat} \new{models\footnote{On Zenodo --- \citet[][]{marley_2021_5063476}}} were developed by \cite{2021ApJ...920...85M}, assuming radiative-convective equilibrium, rainout equilibrium chemistry, and cloud-free atmospheres. A dedicated comparative analysis between these models and a large spectroscopic dataset was performed by \cite{2021ApJ...916...53Z} and \cite{2021ApJ...921...95Z}. They verified that the Sonora Bobcat models could match the spectrophotometry of late-T dwarfs and discussed the systematic offsets between data and models. We use this grid with $T_{\rm eff} \in [200, 800]$~K, $\log{(g)} \in [3.25, 5.5]$~dex, and [M/H] values of $-0.5$~dex, $0$~dex, and $+0.5$~dex, with the C/O fixed at solar value \citep[C/O$_{\odot} = 0.458$;][]{2010ASSP...16..379L}. 

Within the same model series, \new{the \texttt{Sonora Elf Owl} grid\footnote{
On Zenodo --- Y-type models: \citet[][]{mukherjee_2023_10381250};
T-type models: \citet[][]{mukherjee_2023_10385821};
L-type models: \citet[][]{mukherjee_2023_10385987}} was developed by \cite{2024ApJ...963...73M}.} These models have similar assumptions as \texttt{Sonora Bobcat} but additionally incorporate disequilibrium chemistry and updated alkali opacities  \citep{2016A&A...589A..21A, 2019A&A...628A.120A} with broader ranges of [M/H] and C/O. Our analysis focuses on the models with $T_{\rm eff} \in [200, 800]$~K, $\log{(g)} \in [3.25, 5.5]$~dex, [M/H] $\in [-0.5, +0.5]$~dex, C/O $\in[0.22, 1.14]$, and (the pressure-indenendepnt) $K_{\rm zz} \in [10^{2}, 10^{9}]$~cm$^{2}/$s.

\renewcommand{\arraystretch}{1.25} 
{ 
\begin{deluxetable*}{lcccccccccc}
\setlength{\tabcolsep}{1.3pt} 
\tablecaption{Atmospheric properties of COCONUTS-2b (Gemini/F2 spectrum and $W1$/$W2$/[3.6]/[4.5] photometry)} \label{tab:fm_params_key_runs} 
\tablehead{ \multicolumn{1}{l}{Atmospheric Model Grid}   &   \multicolumn{1}{c}{$\Delta$BIC}\tablenotemark{\scriptsize a}   &   \multicolumn{1}{c}{$T_{\rm eff}$}   &   \multicolumn{1}{c}{$\log{(g)}$}   &   \multicolumn{1}{c}{[M/H]\tablenotemark{\scriptsize b}}   &   \multicolumn{1}{c}{C/O\tablenotemark{\scriptsize b}}   &   \multicolumn{1}{c}{$f_{\rm sed}$\tablenotemark{\scriptsize b}}   &   \multicolumn{1}{c}{$\log{(K_{\rm zz})}$\tablenotemark{\scriptsize b}}   &   \multicolumn{1}{c}{$R$}   &   \multicolumn{1}{c}{$\log{(b)}$}   &   \multicolumn{1}{c}{$\log{(L_{\rm bol}/L_{\odot})}$\tablenotemark{\scriptsize c}}    \\ 
\multicolumn{1}{l}{}   &   \multicolumn{1}{c}{}   &   \multicolumn{1}{c}{(K)}   &   \multicolumn{1}{c}{(dex)}   &   \multicolumn{1}{c}{(dex)}   &   \multicolumn{1}{c}{}   &   \multicolumn{1}{c}{}   &   \multicolumn{1}{c}{(cm$^{2}/$s)}   &   \multicolumn{1}{c}{(R$_{\rm Jup}$)}   &   \multicolumn{1}{c}{(erg s$^{-1}$ cm$^{-2}$ \AA$^{-1}$)}   &   \multicolumn{1}{c}{(dex)}  } 
\startdata 
\multicolumn{11}{c}{disequilibrium chemistry + diabatic thermal structure + no clouds}   \\ 
\hline   
\texttt{PH$_{3}$-free ATMO2020++} & $0$ & $545.5^{+3.3}_{-3.2}$ & $4.174^{+0.024}_{-0.015}$ & $-0.152^{+0.012}_{-0.010}$ & ($0.55$) & $\dots$ & ($6.65^{+0.03}_{-0.05}$) & $0.878^{+0.014}_{-0.015}$ & $-18.52^{+0.06}_{-0.07}$ & $-6.179^{+0.005}_{-0.005}$  \\ 
\texttt{ATMO2020++} & $6$ & $534.5^{+3.9}_{-3.8}$ & $4.220^{+0.036}_{-0.039}$ & $-0.209^{+0.025}_{-0.021}$ & ($0.55$) & $\dots$ & ($6.56^{+0.08}_{-0.07}$) & $0.933^{+0.020}_{-0.020}$ & $-18.54^{+0.06}_{-0.08}$ & $-6.161^{+0.006}_{-0.006}$  \\ 
\hline   
\multicolumn{11}{c}{disequilibrium chemistry + clouds}   \\ 
\hline   
\texttt{Exo-REM} & $111$ & $449.9^{+0.4}_{-0.7}$ & $3.348^{+0.004}_{-0.006}$ & $-0.001^{+0.008}_{-0.009}$ & $0.400^{+0.001}_{-0.000}$ & (microphys.) & (profile) & $1.902^{+0.012}_{-0.011}$ & $-18.34^{+0.04}_{-0.04}$ & $-5.849^{+0.004}_{-0.004}$  \\ 
\hline   
\multicolumn{11}{c}{disequilibrium chemistry + no clouds}   \\ 
\hline   
\texttt{LB23 Clear NEQ} & $557$ & $463.2^{+1.0}_{-1.0}$ & $3.758^{+0.012}_{-0.006}$ & $-0.309^{+0.021}_{-0.021}$ & (0.537) & $\dots$ & (profile) & $1.149^{+0.009}_{-0.009}$ & $-18.28^{+0.04}_{-0.04}$ & $-6.225^{+0.004}_{-0.004}$  \\ 
\texttt{Sonora Elf Owl} & $638$ & $472.7^{+4.6}_{-3.4}$ & $3.235^{+0.007}_{-0.003}$ & $-0.265^{+0.034}_{-0.040}$ & $0.505^{+0.007}_{-0.004}$ & $\dots$ & $8.96^{+0.03}_{-0.06}$ & $1.106^{+0.025}_{-0.031}$ & $-18.33^{+0.04}_{-0.05}$ & $-6.224^{+0.007}_{-0.008}$  \\ 
\texttt{ATMO2020 NEQ strong} & $832$ & $445.6^{+2.6}_{-2.6}$ & $4.810^{+0.035}_{-0.036}$ & (0) & ($0.55$) & $\dots$ & ($5.38^{+0.07}_{-0.07}$) & $1.145^{+0.022}_{-0.022}$ & $-18.38^{+0.04}_{-0.05}$ & $-6.293^{+0.007}_{-0.007}$  \\ 
\texttt{ATMO2020 NEQ weak} & $886$ & $461.7^{+2.4}_{-2.5}$ & $4.808^{+0.038}_{-0.038}$ & (0) & ($0.55$) & $\dots$ & ($3.38^{+0.08}_{-0.08}$) & $1.023^{+0.016}_{-0.016}$ & $-18.40^{+0.05}_{-0.05}$ & $-6.327^{+0.005}_{-0.005}$  \\ 
\hline   
\multicolumn{11}{c}{equilibrium chemistry + clouds}   \\ 
\hline   
\texttt{Linder19 Cloudy} & $402$ & $655.2^{+4.6}_{-5.6}$ & $3.294^{+0.088}_{-0.083}$ & $-0.398^{+0.004}_{-0.002}$ & ($0.55$) & $0.93^{+0.12}_{-0.07}$ & $\dots$ & $0.586^{+0.006}_{-0.006}$ & $-18.11^{+0.03}_{-0.04}$ & $-6.226^{+0.011}_{-0.015}$  \\ 
\texttt{Morley12} & $993$ & $570.3^{+4.9}_{-4.8}$ & $4.720^{+0.054}_{-0.053}$ & (0) & ($0.497$) & $4.97^{+0.02}_{-0.03}$ & $\dots$ & $0.621^{+0.010}_{-0.010}$ & $-18.02^{+0.02}_{-0.02}$ & $-6.406^{+0.005}_{-0.005}$  \\ 
\texttt{Morley14}\tablenotemark{\scriptsize d} & $1522$ & $449.9^{+0.1}_{-0.1}$ & $4.205^{+0.036}_{-0.035}$ & (0) & ($0.497$) & (5) & $\dots$ & $1.024^{+0.005}_{-0.004}$ & $-17.99^{+0.02}_{-0.02}$ & $-6.360^{+0.004}_{-0.004}$  \\ 
\hline   
\multicolumn{11}{c}{equilibrium chemistry + no clouds}   \\ 
\hline   
\texttt{LB23 Clear CEQ} (M1)\tablenotemark{\scriptsize e} & $881$ & $561.7^{+0.7}_{-0.9}$ & $4.630^{+0.013}_{-0.015}$ & $-0.261^{+0.012}_{-0.013}$ & (0.537) & $\dots$ & $\dots$ & $0.630^{+0.004}_{-0.004}$ & $-18.10^{+0.02}_{-0.02}$ & $-6.408^{+0.004}_{-0.004}$  \\ 
\texttt{LB23 Clear CEQ} (M2)\tablenotemark{\scriptsize e} & $886$ & $559.7^{+0.7}_{-0.7}$ & $4.415^{+0.018}_{-0.012}$ & $-0.311^{+0.012}_{-0.012}$ & (0.537) & $\dots$ & $\dots$ & $0.633^{+0.004}_{-0.004}$ & $-18.12^{+0.02}_{-0.02}$ & $-6.411^{+0.004}_{-0.004}$  \\ 
\texttt{Linder19 Clear} & $915$ & $487.6^{+1.6}_{-1.6}$ & $3.223^{+0.051}_{-0.051}$ & $-0.394^{+0.010}_{-0.005}$ & ($0.55$) & $\dots$ & $\dots$ & $0.863^{+0.008}_{-0.008}$ & $-18.37^{+0.04}_{-0.05}$ & $-6.379^{+0.004}_{-0.004}$  \\ 
\texttt{ATMO2020 CEQ} & $972$ & $493.0^{+2.0}_{-1.9}$ & $5.178^{+0.042}_{-0.041}$ & (0) & ($0.55$) & $\dots$ & $\dots$ & $0.834^{+0.008}_{-0.008}$ & $-18.18^{+0.03}_{-0.03}$ & $-6.387^{+0.004}_{-0.004}$  \\ 
\texttt{Sonora Bobcat} & $1189$ & $530.4^{+2.7}_{-2.7}$ & $5.494^{+0.005}_{-0.010}$ & $0.102^{+0.020}_{-0.021}$ & ($0.458$) & $\dots$ & $\dots$ & $0.700^{+0.008}_{-0.008}$ & $-18.19^{+0.03}_{-0.03}$ & $-6.414^{+0.004}_{-0.004}$  \\ 
\enddata 
\tablenotetext{a}{This column lists the BIC value of each model grid relative to that of the \texttt{PH$_{3}$-free ATMO2020++} models.}  
\tablenotetext{b}{Values inside parentheses are fixed by the model grids. In particular, the $\log{(K_{\rm zz})}$ values of the \texttt{ATMO2020}, \texttt{ATMO2020++}, and \texttt{PH$_{3}$-free ATMO2020++} modeling results are linearly interpolated from the inferred $\log{(g)}$ based on Figure~1 of Phillips et al. (2021). }  
\tablenotetext{c}{The bolometric luminosity is computed based on Equation~\ref{eq:lbol} that accounts for the offsets between observed fluxes and the fitted models}  
\tablenotetext{d}{The inferred $T_{\rm eff}$ based on \texttt{Morley14} models is pinned at the maximum grid value at 450~K.}  
\tablenotetext{e}{The parameter posteriors inferred by the \texttt{LB23 Clear CEQ} grid exhibit two modes, with the corresponding results labeled as ``M1'' and ``M2''.   }  
\end{deluxetable*} 
}

\subsubsection{\cite{2012ApJ...756..172M} and \cite{2014ApJ...787...78M} } 
\label{subsec:m12}

While the aforementioned model grids assume cloud-free atmospheres, \cite{2012ApJ...756..172M} proposed that optically thin clouds (e.g., KCl, Na$_{2}$S, ZnS, MnS, and Cr) form in T and Y dwarf atmospheres, impacting their thermal emission spectra \citep[also see][]{1999ApJ...519..793L, 2000ASPC..212..152M, 2001RvMP...73..719B, 2005MNRAS.364..649F}. Compared to cloud-free models, the predictions of their cloudy models are more consistent with the observed near-infrared photometry of mid-to-late-T dwarfs. The \cite{2012ApJ...756..172M} \new{models (\texttt{Morley12})\footnote{Available via the website \url{www.carolinemorley.com/models} maintained by C. Morley: \url{https://www.dropbox.com/s/3lnt7pyitueor2r/sulfideclouds.tar.gz}}} assume radiative-convective equilibrium, rainout equilibrium chemistry, and solar metallicity, parameterized by $T_{\rm eff}$, $\log{(g)}$, and the condensate sedimentation efficiency $f_{\rm sed}$ \citep{2001ApJ...556..872A}. We use the models with $T_{\rm eff} \in [400, 800]$~K, $\log{(g) \in [4.0, 5.5]}$~dex, $f_{\rm sed} \in [2, 5]$.

\new{\cite{2014ApJ...787...78M}\footnote{Available via the website \url{www.carolinemorley.com/models} maintained by C. Morley: \url{https://www.dropbox.com/s/djdm8k8rr1dvvmt/spectra_Morley2014.zip?dl=0}} expanded these \cite{2012ApJ...756..172M} models} toward colder effective temperatures and incorporated the formation of water ice clouds with an inhomogeneous cloud cover \citep[\texttt{Morley14}; also see][]{2010ApJ...723L.117M}. Our analysis uses the models with $T_{\rm eff} \in [200, 450]$~K, $\log{(g) \in [3.0, 5.0]}$~dex, a fixed $f_{\rm sed} = 5$, and a fixed cloud patchiness of $50\%$.

\subsubsection{\cite{2019A&A...623A..85L}}
\label{subsec:L19}

\cite{2019A&A...623A..85L} developed atmospheric models with radiative-convective equilibrium, equilibrium chemistry, and assumptions of both \new{cloud-free (\texttt{Linder19 Clear}) and cloudy (\texttt{Linder19 Cloudy}) atmospheres.\footnote{Available via the website of P. Molli\`{e}re: \url{http://mpia.de/~molliere/online_data/linder_molliere_grid.zip}}} This grid is an extension of the models developed by \cite{2017A&A...603A..57S} when studying the directly imaged exoplanet 51~Eri~b using petitCODE \citep{2017A&A...600A..10M}. The cloudy models incorporate the formation of Na$_{2}$S and KCl condensates; as noted by \cite{2019A&A...623A..85L}, these models ignore water ice clouds that are important for $T_{\rm eff}$ below 400~K \citep[e.g.,][]{2014ApJ...787...78M, 2023ApJ...950....8L}. We use both cloud-free and cloudy models with $T_{\rm eff} \in [300, 800]$~K, $\log{(g)} \in [1.5, 5.0]$~dex, [M/H] $\in [-0.4, +1.4]$~dex, and $f_{\rm sed} \in [0.5, 3.0]$ (for cloudy models).

\subsubsection{\cite{2023ApJ...950....8L}} 
\label{subsec:lb23}

\cite{2023ApJ...950....8L} developed new atmospheric models for Y dwarfs assuming radiative-convective equilibrium, along with the updated opacities of alkali lines \citep{2016A&A...589A..21A, 2019A&A...628A.120A} among others. These models include assumptions of both cloud-free and cloudy atmospheres, and both equilibrium and disequilibrium chemistry, leading to four versions in total \new{(\texttt{LB23 Clear/Cloudy CEQ/NEQ}).\footnote{On Zenodo --- \citet[][]{brianna_lacy_2023_7779180}}} For NEQ models, the $K_{\rm zz}$ is calculated as a function of pressures based on the mixing length theory in the convective zone and is fixed at $10^{6}$~cm$^{2}/$s in the radiative zone. We use the cloud-free CEQ and NEQ models with $T_{\rm eff} \in [200, 600]$~K, $\log{(g)} \in [3.5, 5.0]$~dex, and [M/H] values of $-0.5$~dex, $0$~dex, and $+0.5$~dex. 

The cloudy models include a homogeneous water ice cloud cover, with opacities modulated by the vertical extent and characteristic cloud particle size. These models include a regular grid with $T_{\rm eff} \in [250, 400]$~K, $\log{(g)} \in [4.0, 5.0]$~dex, and [M/H] values of $-0.5$~dex and $0$~dex, along with a few irregularly-spaced models at $T_{\rm eff} = 450$~K. Given that the maximum $T_{\rm eff}$ of the regular grid is too cold for COCONUTS-2b, we do not include these cloudy models in the forward-modeling analysis presented in Section~\ref{subsec:results}. Instead, we compare these models with the Gemini/F2 spectrum of COCONUTS-2b based on $\chi^{2}$, as detailed in Appendix~\ref{app:lb23_cloudy}.

\subsubsection{Exo-REM}
\label{subsec:exorem}

The \texttt{Exo-REM} atmospheric models \citep{2015A&A...582A..83B, 2018ApJ...854..172C, 2021A&A...646A..15B}\footnote{\url{https://lesia.obspm.fr/exorem/YGP_grids/}} assume radiative-convective equilibrium and incorporate disequilibrium chemistry using the self-consistently calculated $K_{\rm zz}$-pressure profiles. The publicly available grid includes iron and silicate clouds with simple microphysics. The Na$_{2}$S, KCl, and water condensates, which are important for the atmospheres of T/Y dwarfs \citep[e.g.,][]{2012ApJ...756..172M, 2014ApJ...787...78M}, are not implemented by these models. Nevertheless, within the \texttt{Exo-REM} framework, the effects of iron and silicate clouds are notable near the T/Y transition at low surface gravities (Section~\ref{subsec:properties}). Our analysis uses the models with $T_{\rm eff} \in [400, 800]$~K, $\log{(g)} \in [3.0, 5.0]$~dex, C/O $\in [0.1, 0.8]$, and [M/H] values of $-0.5$~dex, $0$~dex, $+0.5$~dex, and $+1.0$~dex.

\begin{figure*}[t]
\begin{center}
\includegraphics[width=6.1in]{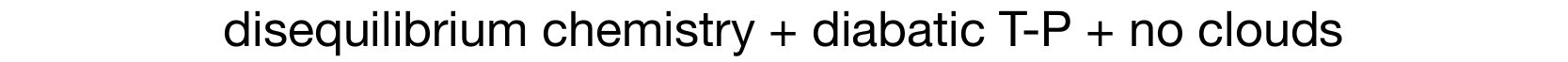}
\includegraphics[width=6.1in]{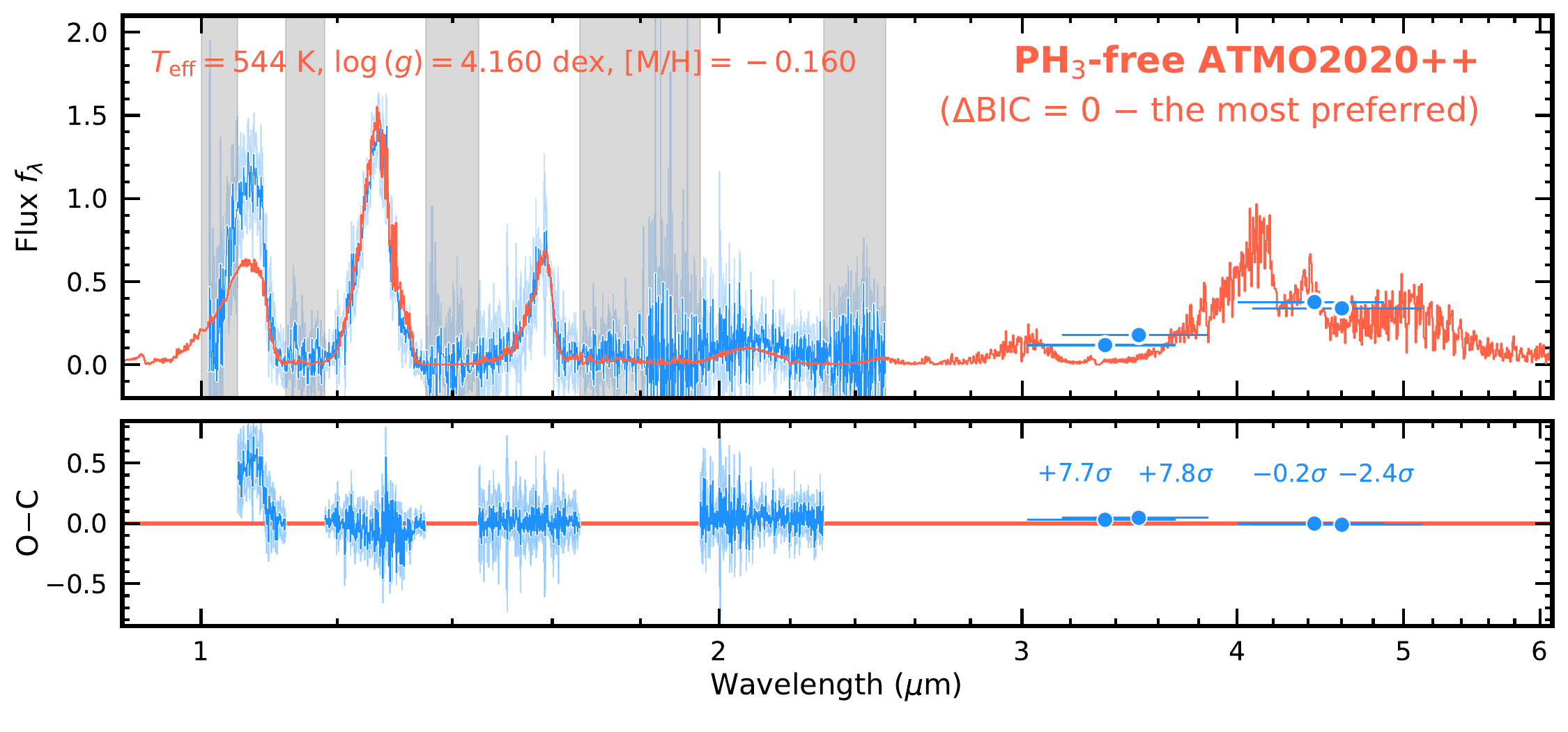}
\includegraphics[width=6.1in]{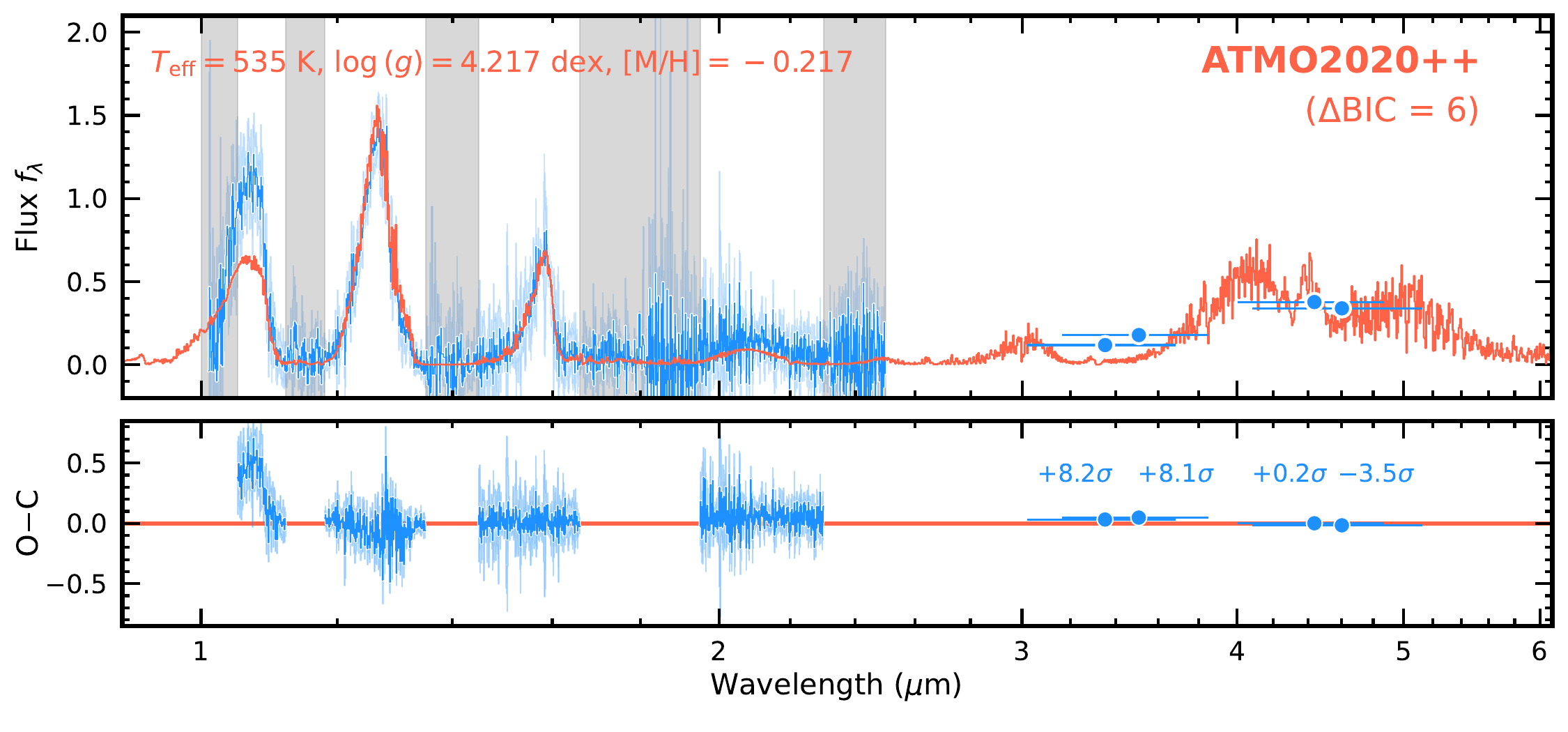}
\caption{Forward-modeling results based on models with disequilibrium chemistry, diabatic temperature-pressure (T-P) profiles, and no clouds. For each grid, the observed spectrophotometry of COCONUTS-2b (blue) and the maximum-likelihood model spectrum (orange) are plotted in the top panel; {the best-fit $T_{\rm eff}$, $\log{(g)}$, and [M/H] values are labeled}. The grey-shaded regions indicate the masked wavelengths in the forward-modeling analysis. The bottom panel shows the data$-$model residuals, with the significance values labeled for the broadband photometry. The unit of plotted fluxes is $10^{-17}$ erg s$^{-1}$ cm$^{-2}$ \AA$^{-1}$. The $\Delta$BIC of a given grid is computed relative to the BIC of the \texttt{PH$_{3}$-free ATMO2020++} results.}
\label{fig:best_neq}
\end{center}
\end{figure*}

\begin{figure*}[t]
\begin{center}
\includegraphics[width=6.1in]{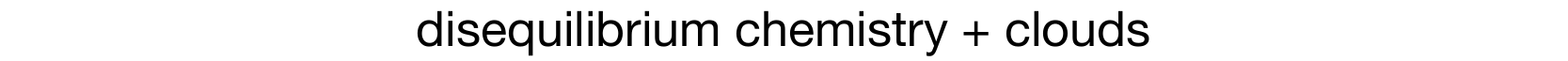}
\includegraphics[width=6.1in]{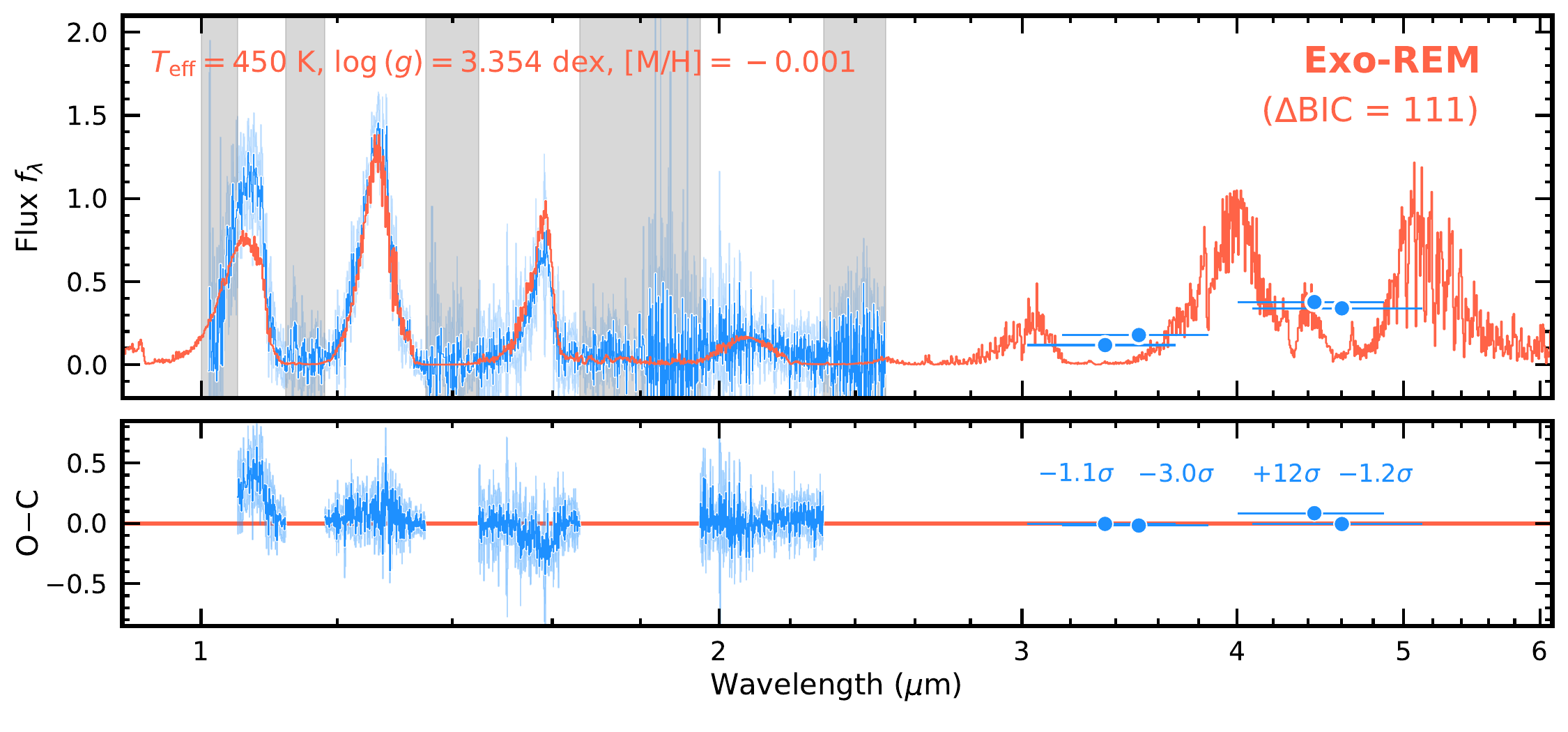}
\caption{Forward-modeling results based on the \texttt{Exo-REM} models that include disequilibrium chemistry and clouds, with the same format as Figure~\ref{fig:best_neq}.  }
\label{fig:clouds_neq}
\end{center}
\end{figure*}

\begin{figure*}[t]
\begin{center}
\includegraphics[width=6.1in]{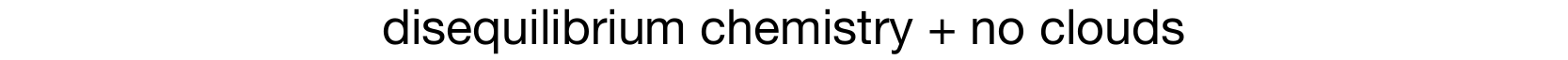}
\includegraphics[width=6.1in]{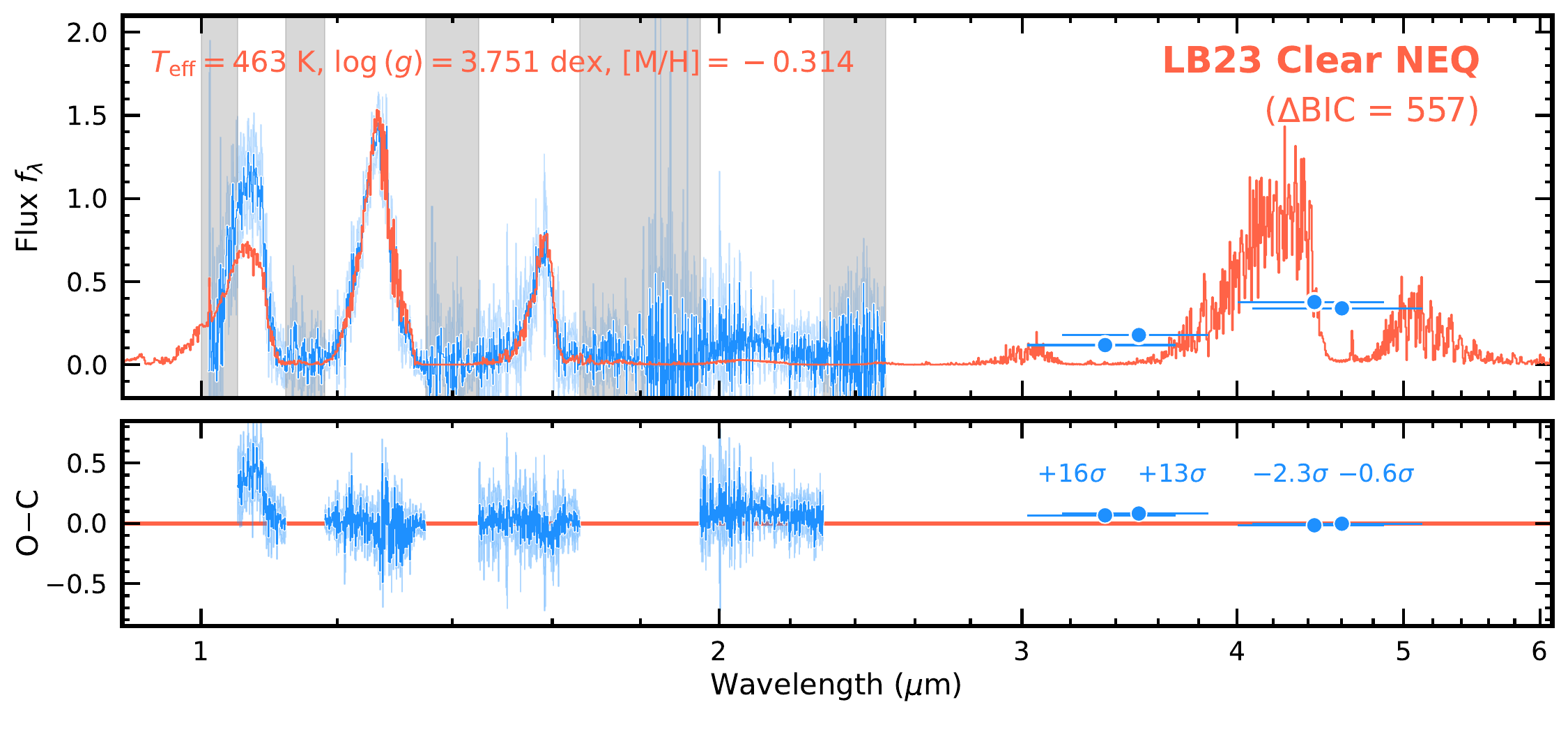}
\includegraphics[width=6.1in]{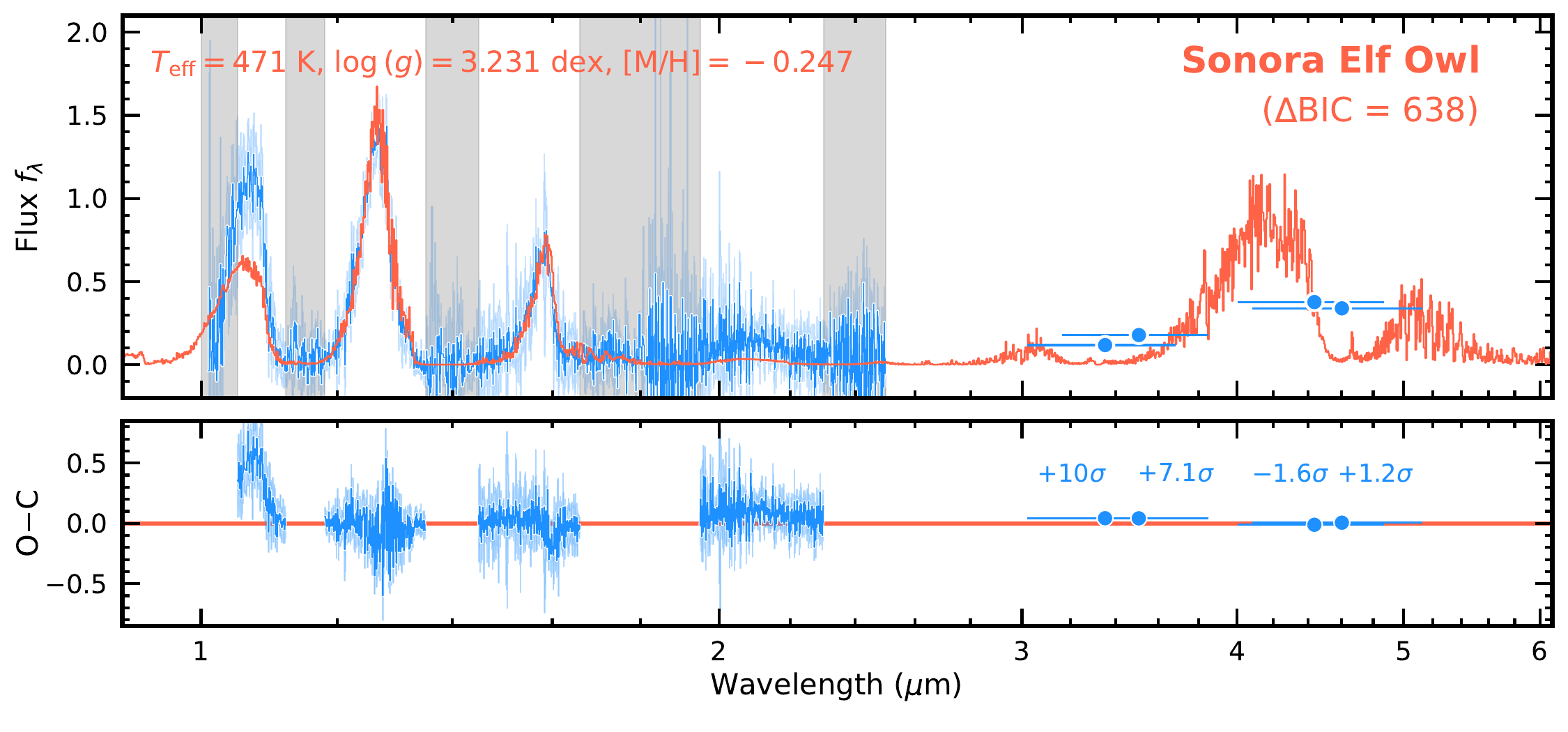}
\caption{Forward-modeling results based on models that include disequilibrium chemistry and no clouds, with the same format as Figure~\ref{fig:best_neq}.  }
\label{fig:clear_neq}
\end{center}
\end{figure*}

\renewcommand{\thefigure}{\arabic{figure}}
\addtocounter{figure}{-1}
\begin{figure*}[t]
\begin{center}
\includegraphics[width=6.1in]{title_Clear_NEQ.pdf}
\includegraphics[width=6.1in]{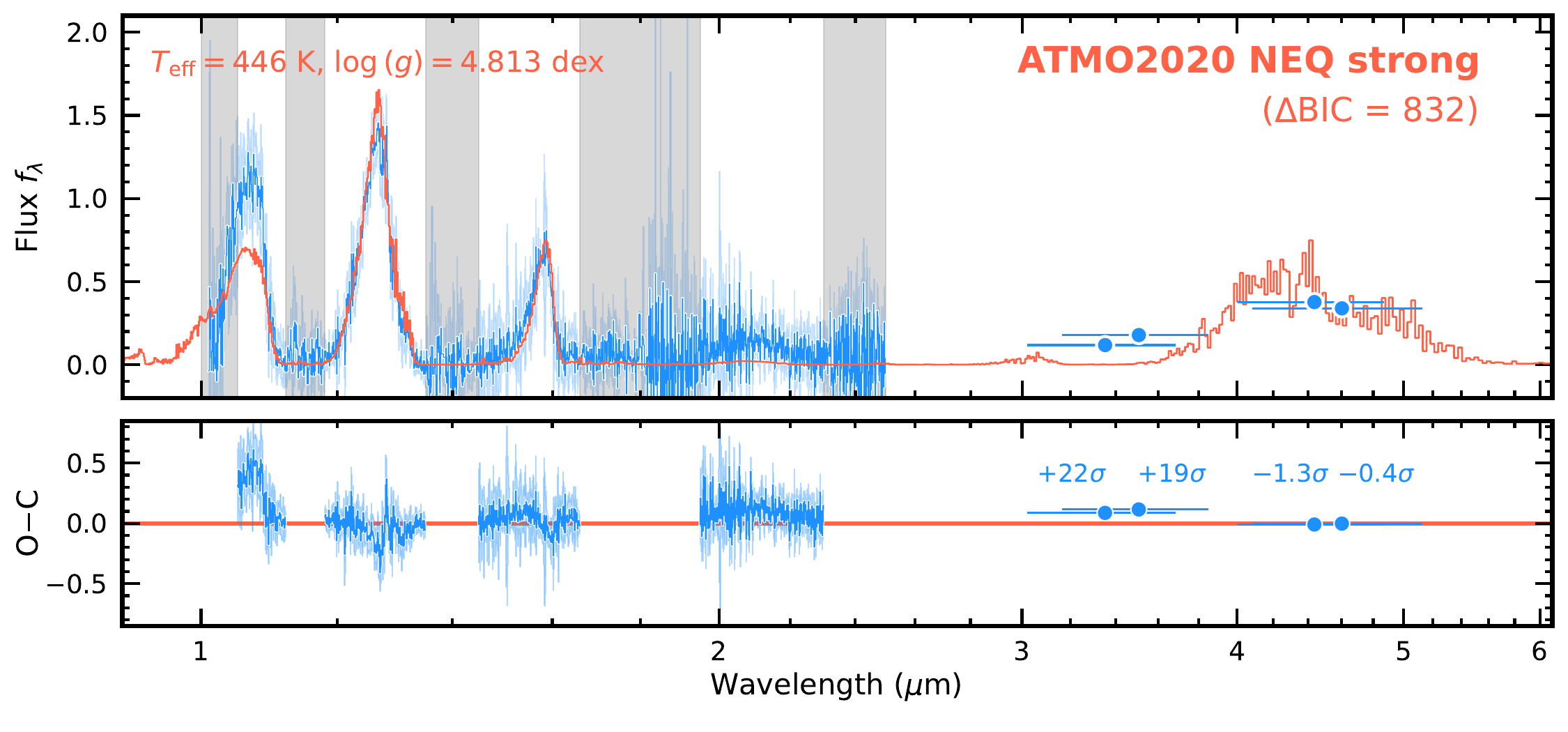}
\includegraphics[width=6.1in]{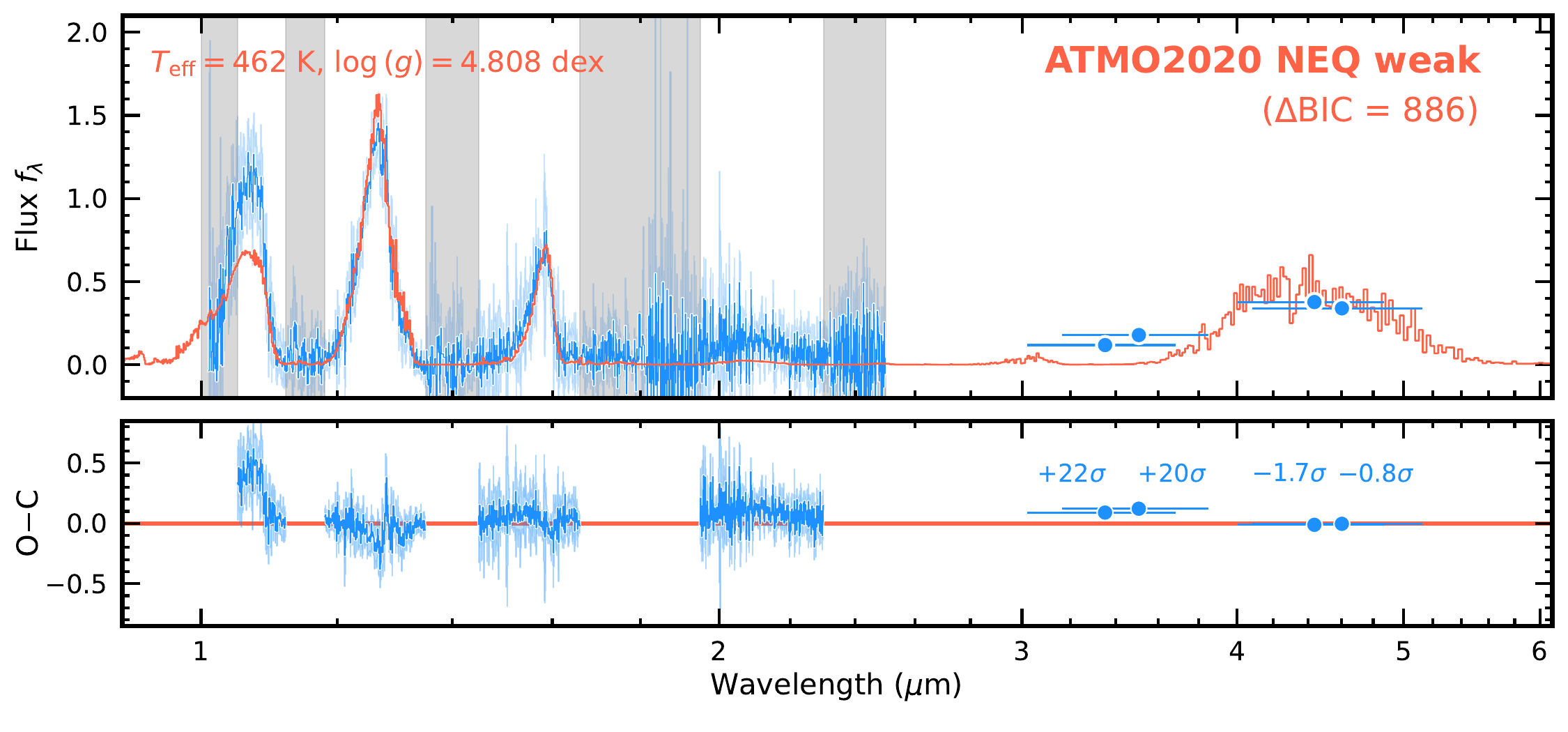}
\caption{Continued. }
\label{fig:clear_neq}
\end{center}
\end{figure*}

\begin{figure*}[t]
\begin{center}
\includegraphics[width=6.1in]{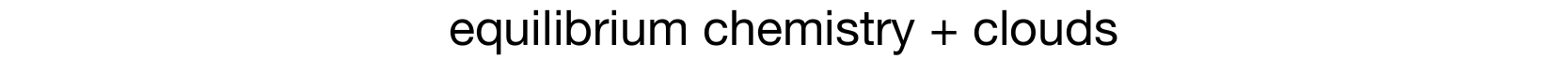}
\includegraphics[width=6.1in]{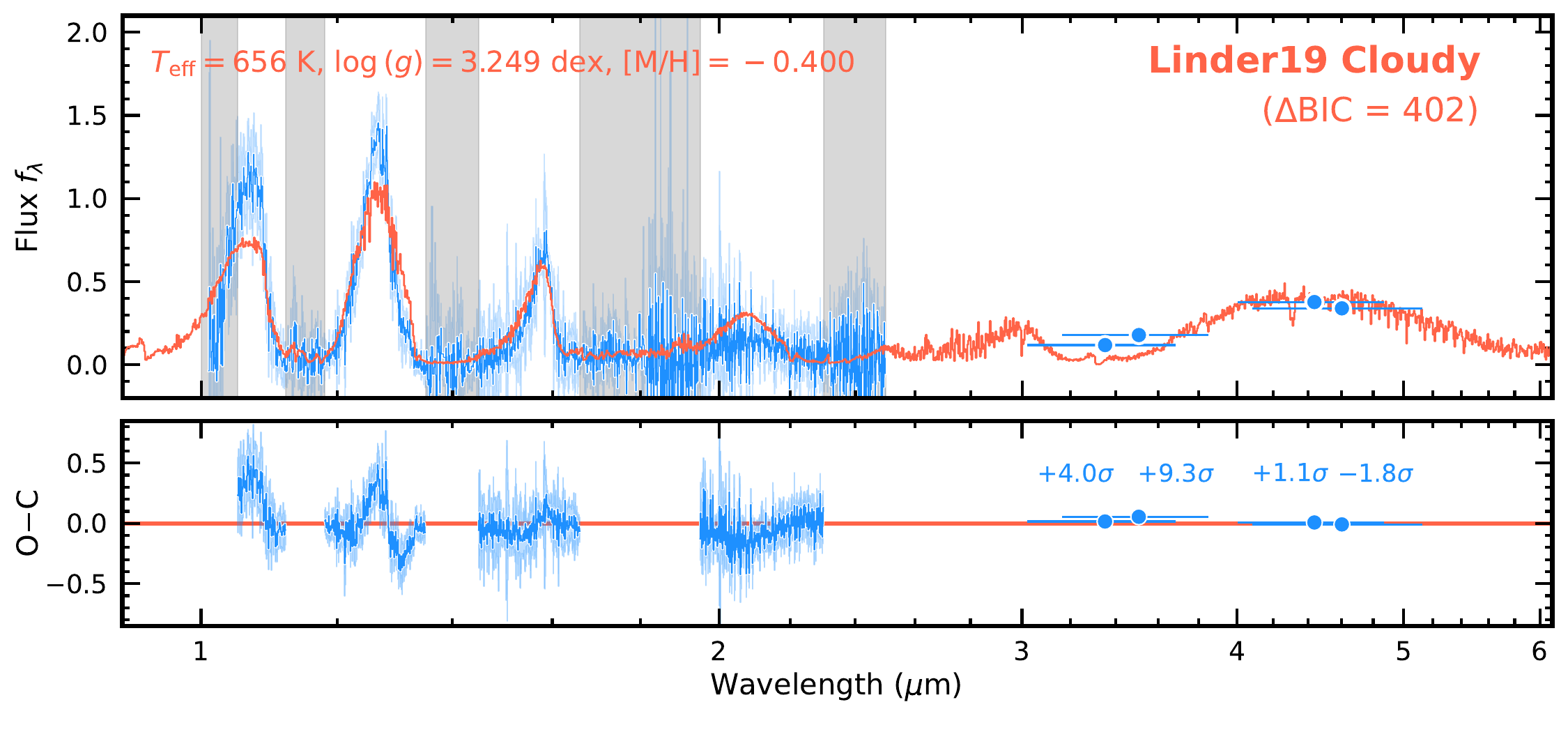}
\includegraphics[width=6.1in]{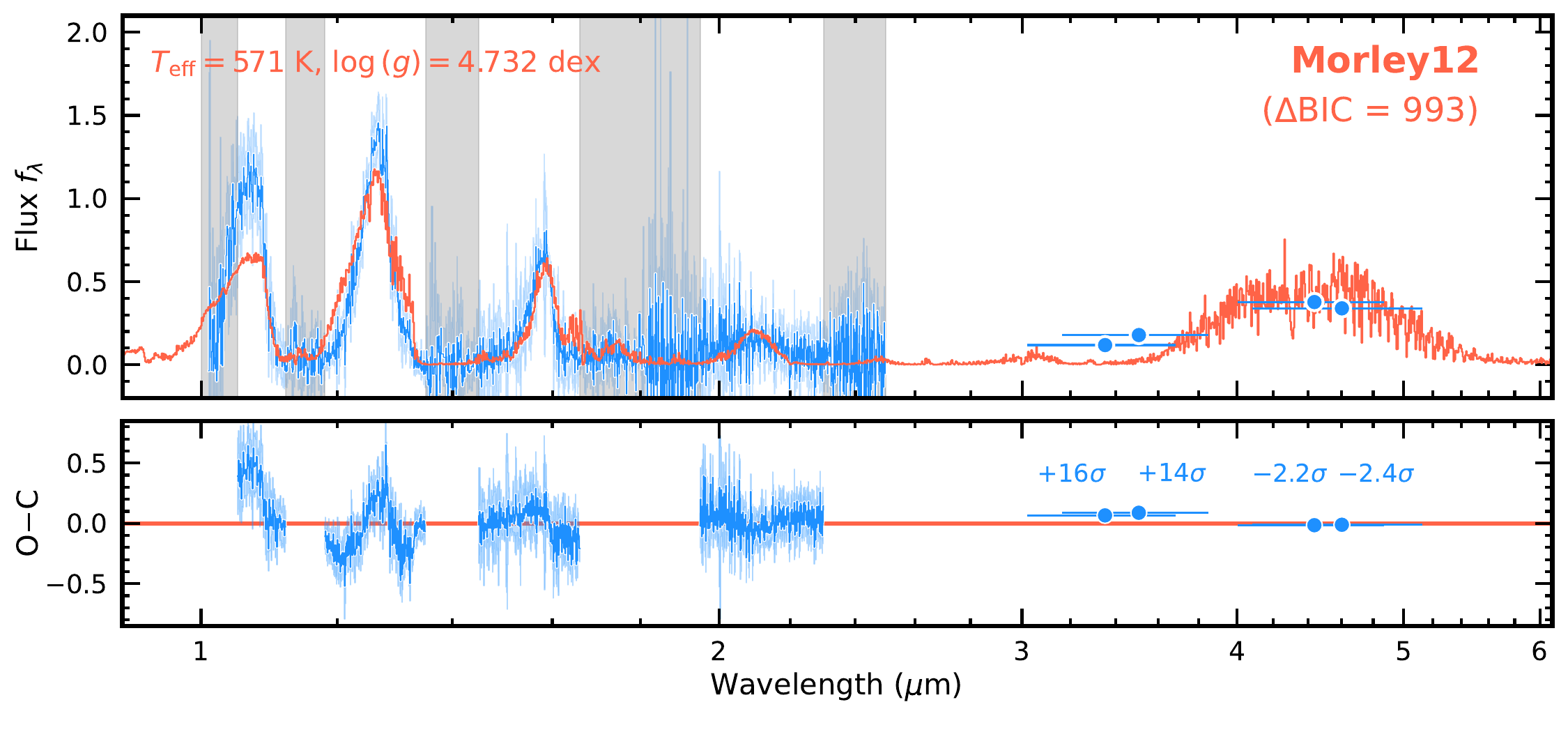}
\includegraphics[width=6.1in]{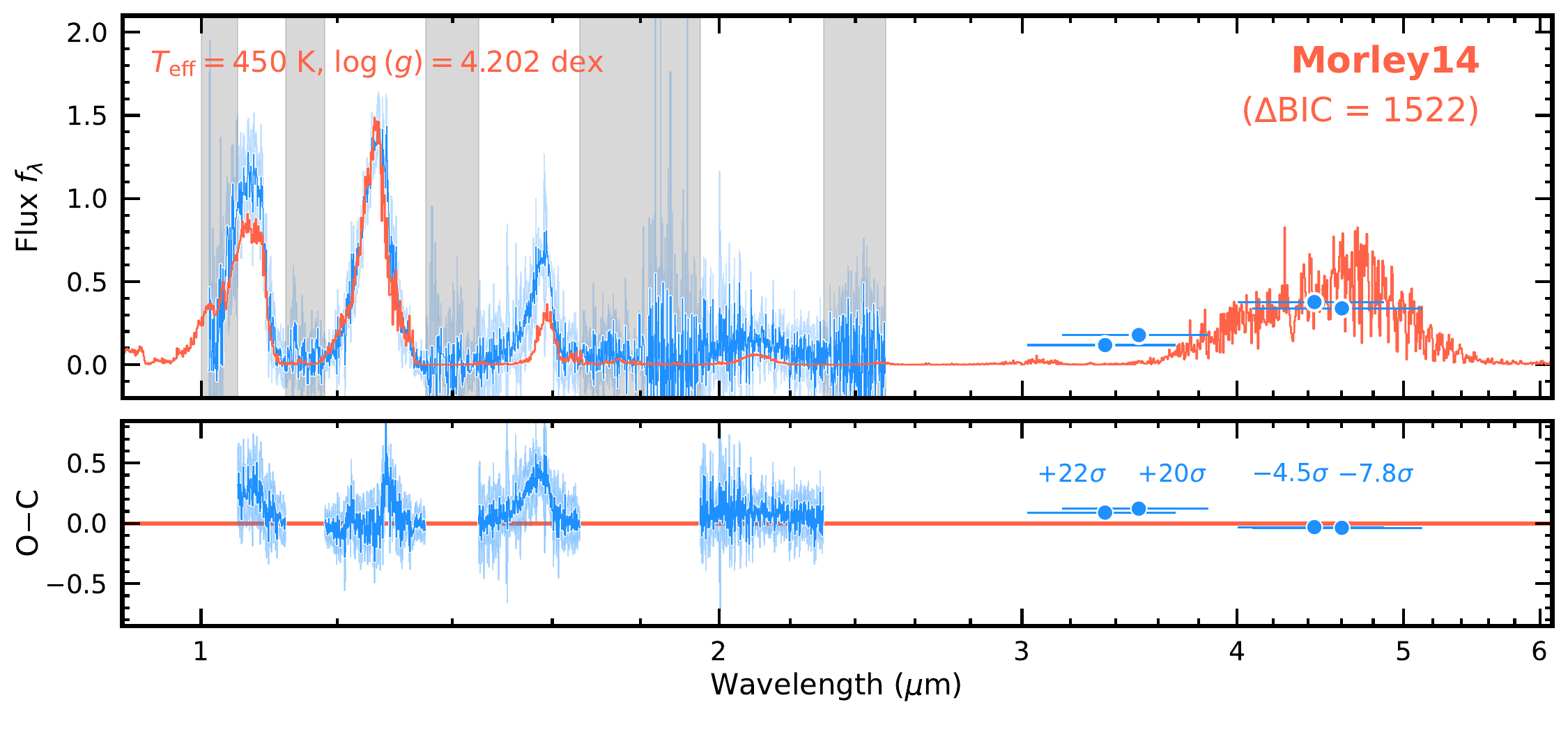}
\caption{Forward-modeling results based on models that include equilibrium chemistry and clouds, with the same format as Figure~\ref{fig:best_neq}.  }
\label{fig:cloudy_ceq}
\end{center}
\end{figure*}

\begin{figure*}[t]
\begin{center}
\includegraphics[width=6.1in]{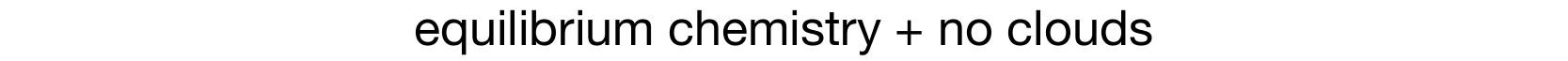}
\includegraphics[width=6.1in]{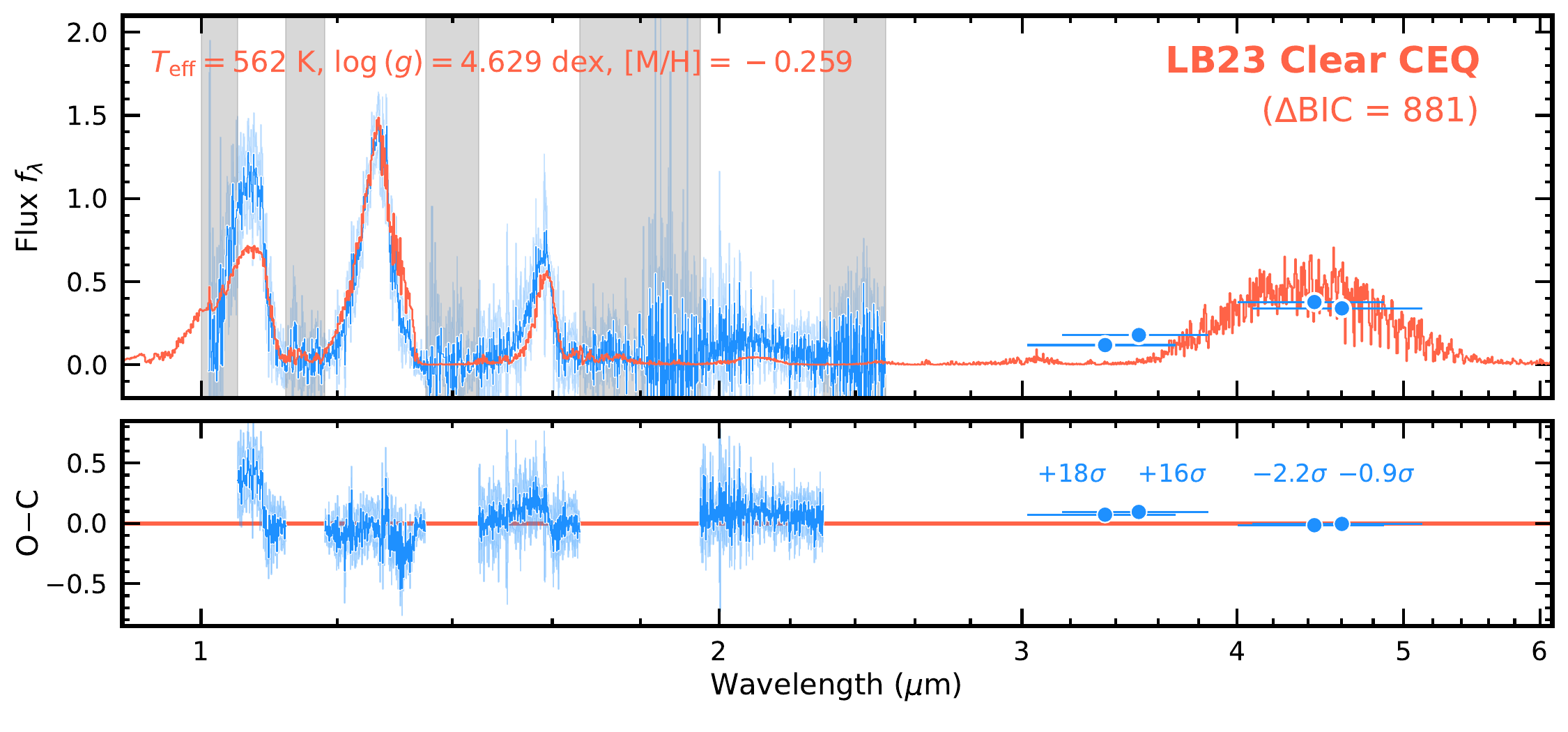}
\includegraphics[width=6.1in]{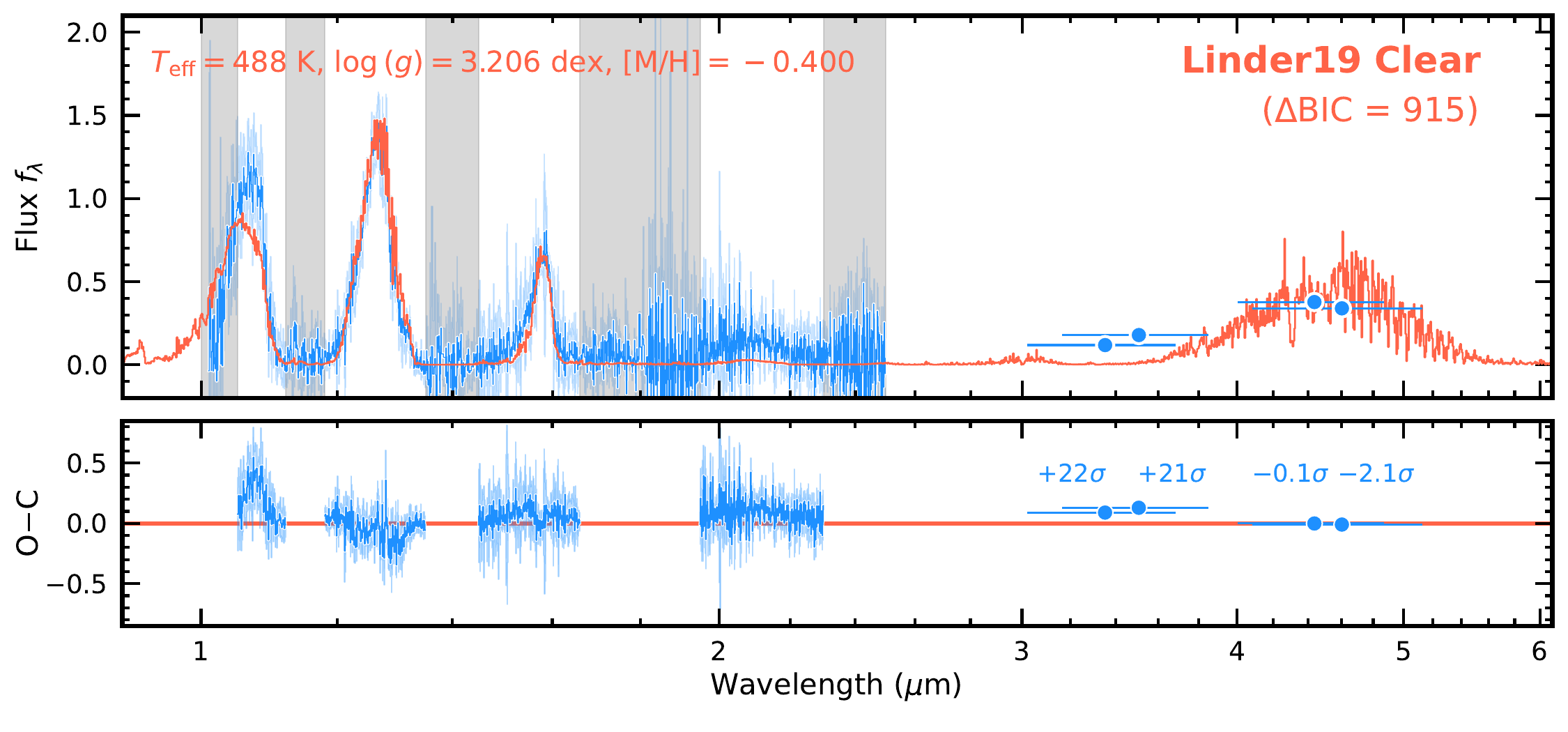}
\caption{Forward-modeling results based on models that include equilibrium chemistry and no clouds, with the same format as Figure~\ref{fig:best_neq}.  }
\label{fig:clear_ceq}
\end{center}
\end{figure*}

\renewcommand{\thefigure}{\arabic{figure}}
\addtocounter{figure}{-1}
\begin{figure*}[t]
\begin{center}
\includegraphics[width=6.1in]{title_Clear_CEQ.pdf}
\includegraphics[width=6.1in]{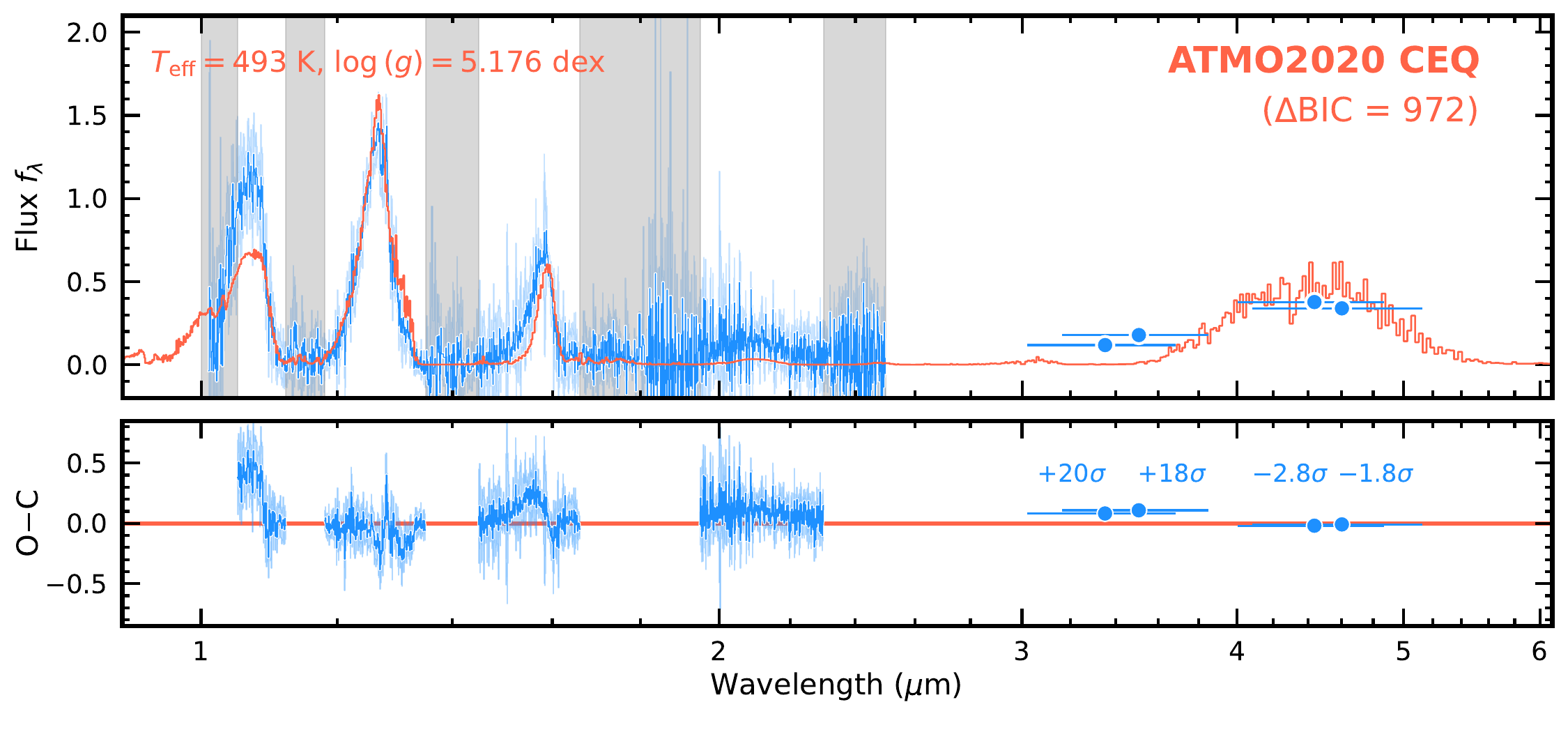}
\includegraphics[width=6.1in]{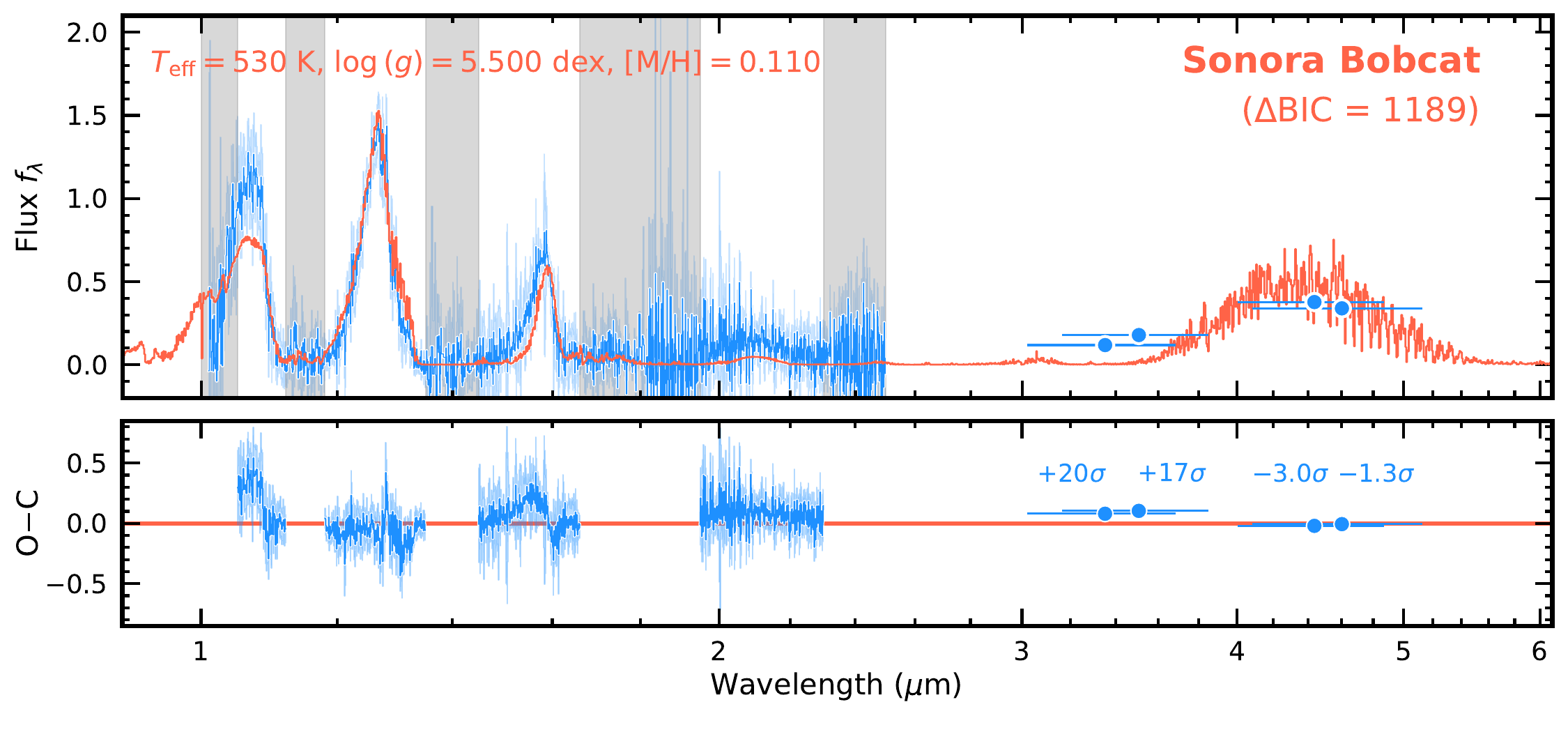}
\caption{Continued. }
\label{fig:clear_ceq}
\end{center}
\end{figure*}

\subsection{Methodology} 
\label{subsec:methods}

We constrain the physical properties of COCONUTS-2b by comparing each model grid to its Gemini/F2 spectrum and $W1$, $W2$, [3.6], and [4.5] photometry. For the analysis presented in this section, we exclude the $W3$ and $W4$ (upper limit) magnitudes, as the \texttt{Sonora Elf Owl} models do not cover the filter response curves of these two bands. As discussed in Section~\ref{subsec:different_input}, our analysis results and conclusions are not significantly altered when $W3$ is included. Upcoming JWST NIRSPEC and MIRI spectroscopy (GO 2124, GO 3514, GO 6463) will provide more insights into this object's appearance and properties in the mid-infrared. Additionally, our analysis uses the Gemini/F2 spectrum in the wavelength ranges of 1.05--1.12~$\mu$m, 1.18--1.35~$\mu$m, 1.45--1.66~$\mu$m, and 1.95--2.30~$\mu$m (see Figure~\ref{fig:c2b}). The lower wavelength cut of 1.05~$\mu$m is determined considering the flux drop of the Gemini/F2 data when compared with spectral templates (Section~\ref{sec:spt}).

We perform the forward-modeling analysis using the nested sampling algorithm \texttt{PyMultiNest} \citep{2008MNRAS.384..449F, 2009MNRAS.398.1601F, 2019OJAp....2E..10F, 2014A&A...564A.125B}, with 4000 live points and a 0.8 sampling efficiency. {The fitting process terminates once the change in the natural logarithmic evidence between iterations drops below 0.5.}  The free parameters include the model grid parameters, radius ($R$), and a hyper-parameter accounting for any underestimated flux uncertainties of our Gemini/F2 spectrum, described in a logarithmic scale ($\log{(b)}$). Uniform priors are adopted for all these parameters. The grid parameters are constrained within the ranges described in Section~\ref{subsec:modelgrids}; $R$ is constrained between 0.5~R$_{\rm Jup}$ and 2.5~R$_{\rm Jup}$; and the boundaries of $\log{(b)}$ correspond to a range from $10\%$ of the minimum Gemini/F2 spectral flux uncertainty to 100 times the maximum flux uncertainty over the fitted wavelength range (i.e., $2.2 \times 10^{-20}$ -- $8.2 \times 10^{-16}$~erg~s$^{-1}$~cm$^{-2}$~\AA$^{-1}$). 

The model spectra are interpolated to a given set of grid parameters. This interpolation is performed in logarithmic scales for $T_{\rm eff}$, $K_{\rm zz}$, and fluxes, and in linear scales for $\log{(g)}$, [M/H], $f_{\rm sed}$, and C/O. The interpolated models are then scaled by $(R/d)^{2}$, with $d = 10.888$~pc fixed at the planet host star's distance \citep{2021AJ....161..147B}. Log-likelihood is evaluated via
\begin{equation}
\ln{\mathcal{L}} = \ln{\mathcal{L_{\rm spec}}} + \ln{\mathcal{L_{\rm phot}}} 
\end{equation}
where
\begin{equation}
\begin{aligned}
\ln{\mathcal{L_{\rm spec}}} =& -\frac{1}{2} \sum_{i}^{N_{\rm pix}}{\frac{(\mathcal{F}_{ {\rm obs}, i} - \mathcal{F}_{ {\rm mod}, i})^{2}}{\sigma_{ {\rm obs}, i}^2 + b^{2}}} - \frac{1}{2} \ln{\left[2\pi (\sigma_{ {\rm obs}, i}^2 + b^{2})\right]} \\
\ln{\mathcal{L_{\rm phot}}} =& -\frac{1}{2} \sum_{j}^{N_{\rm phot}}{\frac{(\mathcal{F}_{ {\rm obs}, j} - \mathcal{F}_{ {\rm mod}, j})^{2}}{\sigma_{ {\rm obs}, j}^2}} - \frac{1}{2} \ln{\left[2\pi (\sigma_{ {\rm obs}, j}^2)\right]}
\end{aligned}
\end{equation}
Here, $\mathcal{F}_{\rm obs}$ and $\sigma_{\rm obs}$ represent the observed flux and uncertainty of spectrophotometry; $N_{\rm pix}$ is the total number of spectral wavelength pixels, and $N_{\rm phot} = 4$ is the number of broadband photometry. For photometric data, the predicted model flux, $\mathcal{F}_{\rm mod}$, is calculated using the filter response curves \citep{2008PASP..120.1233H, 2011ApJ...735..112J}.

\begin{figure*}[t]
\begin{center}
\includegraphics[width=5.9in]{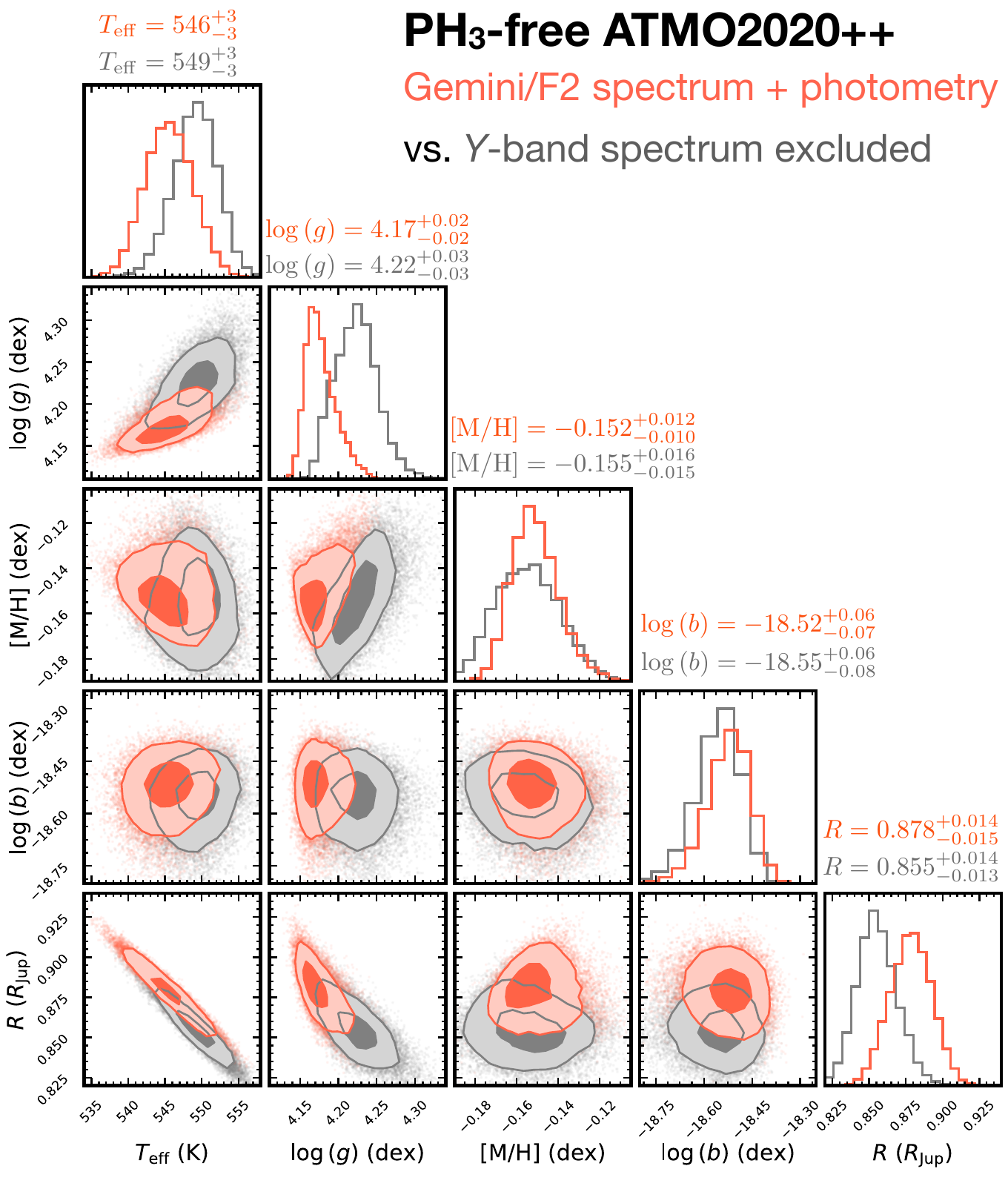}
\caption{{Posterior distributions of COCONUTS-2b's spectroscopically inferred parameters using the \texttt{PH$_{3}$-free ATMO2020++} models. The results derived with the $Y$-band spectral fluxes included and excluded are shown in orange and grey, respectively.} The $1\sigma$ and $2\sigma$ confidence intervals are highlighted.}
\label{fig:example_corner}
\end{center}
\end{figure*}

\subsection{Results} 
\label{subsec:results}

We perform forward-modeling analysis for COCONUTS-2b using the atmospheric model grids described in Section~\ref{subsec:modelgrids}. These models are categorized based on their assumptions:
\begin{enumerate}
\item[$\bullet$] Models with disequilibrium chemistry, diabatic atmospheric thermal structure, and no clouds: \texttt{ATMO2020++} and \texttt{PH$_{3}$-free ATMO2020++}.
\item[$\bullet$] Models with disequilibrium chemistry and clouds: \texttt{Exo-REM} and \texttt{LB23 Cloudy NEQ}.
\item[$\bullet$] Models with disequilibrium chemistry and no clouds: \texttt{ATMO2020 NEQ weak}, \texttt{ATMO2020 NEQ strong}, \texttt{Sonora Elf Owl}, and \texttt{LB23 Clear NEQ}.
\item[$\bullet$] Models with equilibrium chemistry and clouds: \texttt{Morley12}, \texttt{Morley14}, \texttt{Linder19 Cloudy}, and \texttt{LB23 Cloudy CEQ}.
\item[$\bullet$] Models with equilibrium chemistry and no clouds: \texttt{ATMO2020 CEQ}, \texttt{Sonora Bobcat}, \texttt{Linder19 Clear}, and \texttt{LB23 Clear CEQ}.
\end{enumerate}

The bolometric luminosity of COCONUTS-2b is estimated using the inferred $T_{\rm eff}$ and $R$ posteriors:
\begin{equation} \label{eq:lbol}
L_{\rm bol} = 4\pi R^{2} \sigma_{\rm SB} T_{\rm eff}^{4} + 4\pi d^{2} \int_{\Lambda}{\left(\mathcal{F}_{\lambda, {\rm obs}} - \langle \mathcal{F}_{\lambda, {\rm mod}} \rangle\right) d\lambda}
\end{equation}
{Here, $\sigma_{\rm SB}$ is the Stefan-Boltzmann constant, $d$ is the distance of COCONUTS-2b, and $\Lambda$ represents the wavelength ranges used in the forward-modeling analysis, i.e., 1.05--1.12~$\mu$m, 1.18--1.35~$\mu$m, 1.45--1.66~$\mu$m, and 1.95--2.30~$\mu$m.} The second term {in Equation~\ref{eq:lbol}} accounts for any discrepancies between the observed fluxes and the fitted model spectra. Specifically, $\left(\mathcal{F}_{\lambda, {\rm obs}} - \langle \mathcal{F}_{\lambda, {\rm mod}} \rangle \right)$ represents the residuals between the observed Gemini/F2 spectrum and the maximum-likelihood model spectrum. Our error propagation {of $L_{\rm bol}$} accounts for the observed flux uncertainties and the best-fitted hyper-parameter $b$ (Section~\ref{subsec:methods}).

The spectroscopically inferred mass of COCONUTS-2b can be computed for each model grid based on $\log{(g)}$ and $R$, but these values are not explicitly reported in this work. These spectroscopically inferred masses should be interpreted with caution, given the known challenges of accurately determining $\log{(g)}$ and $R$ from thermal emission spectra of gas-giant exoplanets and brown dwarfs \citep[e.g.,][]{2008ApJ...678.1372C, 2009ApJ...702..154S, 2011ApJ...740..108L, 2017ApJ...848...83L, 2019ApJ...877...24Z, 2020ApJ...905...46G, 2022ApJ...930..136L, 2021ApJ...916...53Z, 2021ApJ...921...95Z, 2023AJ....166..198Z, 2023ApJ...953..170H}. 

We compute the Bayesian Information Criterion \citep[BIC;][]{1978AnSta...6..461S, 2007MNRAS.377L..74L} for each model grid as follows:
\begin{equation}
{\rm BIC} \equiv -2\ln{\mathcal{L}_{\rm max}} + k \ln{N}
\end{equation}
where $\mathcal{L}_{\rm max}$ is the maximum likelihood, $k$ is the number of free parameters, and $N$ is the number of data points. {As proposed by \cite{Kass1995}, a model with a BIC that is lower by more than $6$ compared to another model is strongly preferred for interpreting the observations. If the $\Delta$BIC between the two models is less than 2, the preference is not considered significant.} With this metric as guidance, we visually examine data-model discrepancies to evaluate the various model assumptions. 

Figures~\ref{fig:best_neq}--\ref{fig:clear_ceq} present our results, with confidence intervals of the inferred parameters summarized in Table~\ref{tab:fm_params_key_runs}. We compute a $\Delta$BIC value for each model grid by comparing its BIC with that of the most preferred grid, \texttt{PH$_{3}$-free ATMO2020++}; a corner plot of the inferred parameters from this grid is shown in Figure~\ref{fig:example_corner}. Results for the \texttt{LB23 Cloudy CEQ} and \texttt{NEQ} models are skipped in this section because their inferred $T_{\rm eff}$ posteriors are pinned at the upper boundary of 400~K that is too cold for COCONUTS-2b. In Appendix~\ref{app:lb23_cloudy}, we present the $\chi^{2}$-based spectral fitting analyses of COCONUTS-2b using these models.

\section{Discussion} 
\label{sec:discussion}

\subsection{Disequilibrium Chemistry, Diabatic Thermal Structure, and Clouds of COCONUTS-2\lowercase{b}}
\label{subsec:properties}

Among all the model grids, \texttt{PH$_{3}$-free ATMO2020++}, \texttt{ATMO2020++}, and \texttt{Exo-REM} are preferred for matching the observed spectrophotometry of COCONUTS-2b, as they yield the lowest BIC values (Figures~\ref{fig:best_neq}--\ref{fig:clouds_neq}). These grids incorporate disequilibrium chemistry and additional atmospheric processes --- diabatic thermal structure or clouds --- that are not the assumptions of the remaining models.

The  \texttt{PH$_{3}$-free ATMO2020++} and \texttt{ATMO2020++} model grids are expansions of \texttt{ATMO2020 NEQ strong}, with a modified adiabatic index (Section~\ref{subsec:atmo2020plusplus}). This modification was motivated and guided by discrepancies between \texttt{ATMO2020 NEQ strong} and the observed spectrum of UGPS~J072227.51$-$054031.2 (the fifth best-matched single spectral template of COCONUTS-2b; Figure~\ref{fig:spt_single}) in the $J$, $K$, and [3.6] bands \citep[see Section~5.1 of][]{2021ApJ...918...11L}. Therefore, it is not surprising that these models are preferred against other standard disequilibrium cloud-free models. With a modified $\gamma = 1.25$, the atmospheric thermal structure becomes slightly hotter than the standard case in the upper atmosphere, where the $W1$, $W2$, [3.6], and [4.5] fluxes are emitted, and slightly colder in the deep atmosphere, where the $Y$, $J$, and $H$ fluxes are emitted \citep[e.g.,][]{2021ApJ...918...11L}. This is equivalent to a reduced temperature gradient of the temperature-pressure profile. The resulting emission spectra appear slightly redder in the near-infrared wavelengths, similar to the effects of cloud opacities, even though these models do not include clouds. This prescribed $\gamma$ modification may be interpreted in the context of the thermo-compositional convection triggered by the transitions of CO $\rightleftharpoons$ CH$_{4}$ and N$_{2} \rightleftharpoons$ NH$_{3}$, the breaking of gravity waves, chromospheric activity, latent heating during condensation, or non-1D dynamics in the atmospheres \citep[e.g.,][]{2014MNRAS.440.3675S, 2015ApJ...804L..17T, 2019ApJ...876..144T, 2016Natur.536..190O, 2021ApJ...918...11L, 2021ApJ...922...26T}. The diabatic atmospheric thermal structure has also been suggested in the atmospheres of other late-T and Y dwarfs based on forward models or atmospheric retrievals \citep[e.g.,][]{2016ApJ...824....2L, 2017ApJ...842..118L, 2019ApJ...882..117L, 2021ApJ...918...11L, 2019ApJ...877...24Z, 2023ApJ...959...86L, 2024RNAAS...8...13L, 2023AJ....166...57M, 2024AJ....167....5L}. 

\begin{figure*}[t]
\begin{center}
\includegraphics[width=7.in]{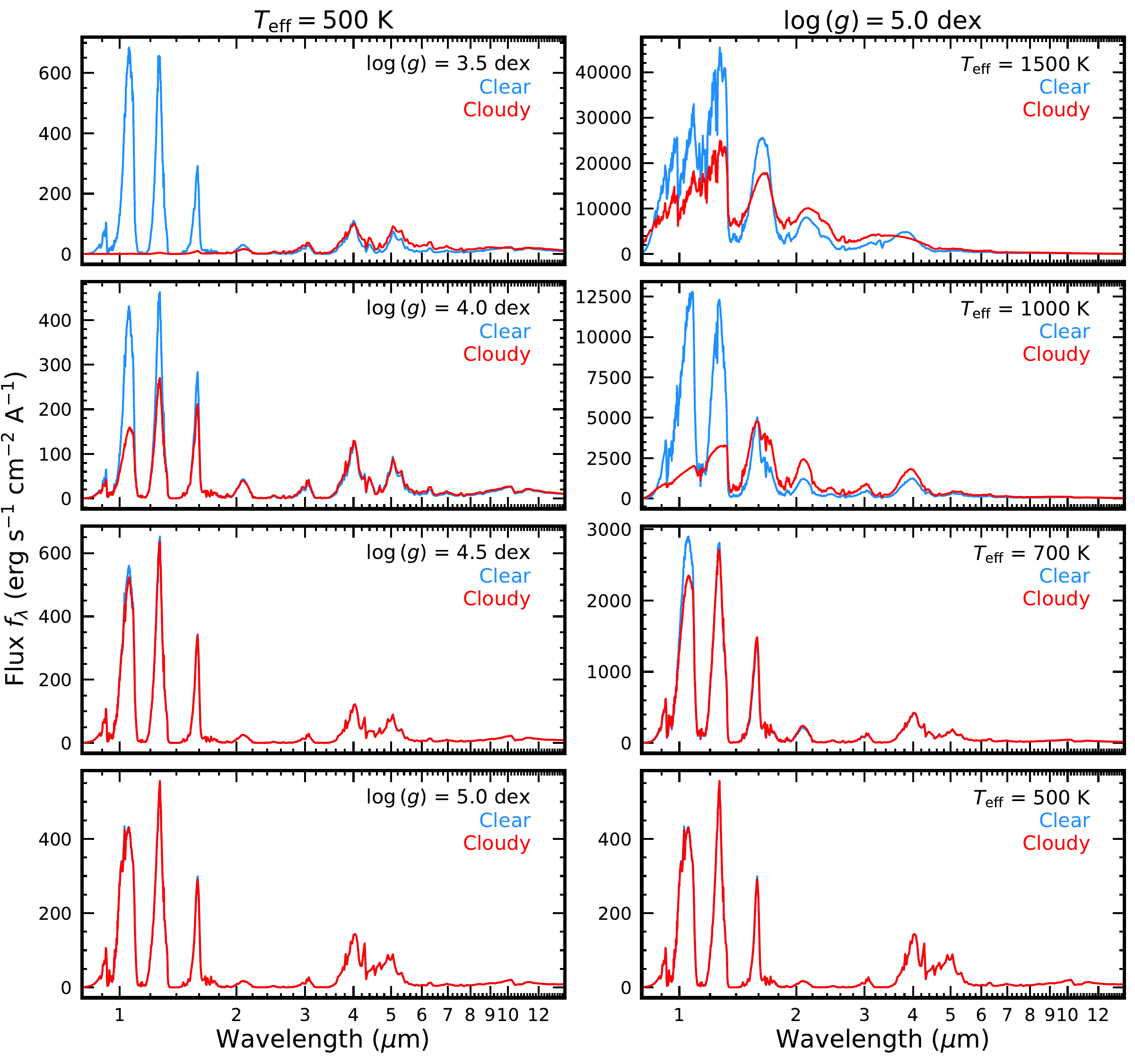}
\caption{The \texttt{Exo-REM} $R \sim 500$ model spectra computed with cloud-free (blue) and cloudy (red) atmospheres. The left column shows spectra with $T_{\rm eff} = 500$~K and varying $\log{(g)}$; the integrated fluxes of these spectra over 0.34--250~$\mu$m are consistent within 0.004~dex. The right column shows spectra with $\log{(g)} = 5.0$~dex and varying $T_{\rm eff}$. All spectra correspond to [M/H]$= 0$~dex and C/O$=0.4$, consistent with the inferred values for COCONUTS-2b based on \texttt{Exo-REM} (Table~\ref{tab:fm_params_key_runs}). }
\label{fig:exorem_clear_vs_cloudy}
\end{center}
\end{figure*}

{The \texttt{Exo-REM} model grid yields a $\Delta$BIC of 111 (Table~\ref{tab:fm_params_key_runs}), meaning that it is disfavored compared to the \texttt{PH$_{3}$-free ATMO2020++} and \texttt{ATMO2020++} models. Nevertheless, \texttt{Exo-REM} is preferred among all the model grids with the standard $\gamma$. The \texttt{Exo-REM} models used in this study incorporate iron and silicate clouds \citep{2018ApJ...854..172C}.} These cloud species are crucial for gas-giant planets and brown dwarfs near the L/T transition \citep[e.g.,][]{2006asup.book....1L, 2015ARA&A..53..279M}. In colder atmospheres of late-T and Y dwarfs, iron and silicate clouds condense below or near the bottom of the near-infrared photospheres. The primary cloud species at the T/Y transition include KCl, Na$_{2}$S, and water \citep[e.g.,][]{2012ApJ...756..172M, 2014ApJ...787...78M}, which are not included by the public \texttt{Exo-REM} models. However, within the \texttt{Exo-REM} framework, the iron and silicate clouds still impact the emergent spectra at low effective temperatures when surface gravities are low. Figure~\ref{fig:exorem_clear_vs_cloudy} compares \texttt{Exo-REM} model spectra with and without clouds; the spectral resolutions of these models ($R \sim 500$) are lower than those used in our forward-modeling analysis ($R \sim 2 \times 10^{4}$). Although the spectral effects of iron and silicate clouds weaken at colder effective temperatures, they remain significant at the characteristic $T_{\rm eff}$ values of T/Y dwarfs (e.g., 500~K) when $\log{(g)} \lesssim 4.0$~dex. Our spectroscopically inferred $T_{\rm eff} \approx 450$~K and $\log{(g)} = 3.348^{+0.004}_{-0.006}$~dex (Table~\ref{tab:fm_params_key_runs}) indicate that cloud effects are important within the \texttt{Exo-REM} framework. Therefore, our analysis suggests the need for cloud opacities for interpreting COCONUTS-2b's spectrophotometry {when the standard atmospheric adiabatic index is assumed.} However, we cannot conclude whether iron and silicate clouds are the predominant cloud species {for COCONUTS-2b}. 

These findings suggest dynamic processes in the atmosphere of COCONUTS-2b, including disequilibrium chemistry, a diabatic thermal structure, and {potential clouds} \citep[also see][]{2021ApJ...916L..11Z}. Such processes are likely prevalent in the atmospheres of late-T and Y dwarfs, potentially explaining the observed diversity in their spectrophotometry \citep[e.g.,][]{2017ApJ...842..118L, 2020AJ....159..257B, 2021ApJS..253....7K, 2023ApJ...959...63S}. These processes may also lead to top-of-atmosphere inhomogeneity, such as hot/cold spots and patchy clouds \citep[e.g.,][]{2014ApJ...788L...6Z, 2013ApJ...776...85S, 2019ApJ...883....4S, 2020SSRv..216..139S, 2019ApJ...874..111T, 2021MNRAS.502.2198T, 2020A&A...643A..23T}, resulting in rotationally-modulated or irregular time-evolving variabilities that can be probed via spectrophotometric monitoring \citep[e.g.,][]{2009ApJ...701.1534A, 2014ApJ...793...75R, 2016ApJ...830..141L, 2016ApJ...818..176Z, 2022AJ....164..239Z, 2017Sci...357..683A, 2019ApJ...875L..15M, 2020ApJ...893L..30B, 2020AJ....159..125L, 2022ApJ...924...68V, 2024MNRAS.527.6624L}. {To date, variability measurements for T/Y dwarfs are less extensive than those for objects near the L/T transition. Only a handful of T8--Y2 objects have exhibited variability signals, with amplitudes reaching up to $2.6\%$ in the $J$ band \citep[e.g.,][]{2019ApJ...875L..15M} or $7.5\%$ near 4.5~$\mu$m \citep[e.g.,][]{2016ApJ...823..152C, 2016ApJ...832...58E, 2016ApJ...830..141L}. Variability monitoring of COCONUTS-2b, however, has yet to be reported, especially considering the challenges posed by its faintness ($J = 19.342$~mag) in ground-based observations. Space-based observations by, e.g., JWST, offer a promising approach for investigating the atmospheric variability of this object.  }

\subsection{Major Discrepancies between Data and Models}
\label{subsec:data_model_offsets}

\subsubsection{All Model Grids: Under-predicted Fluxes in the $Y$ band}
\label{subsubsec:offsets_all_models}

As shown in Figures~\ref{fig:best_neq}--\ref{fig:clear_ceq}, all model grids predict $Y$-band fluxes that are fainter than the observations by $1-2\sigma$, regardless of their underlying assumptions. Similar offsets have been observed in forward modeling analyses of other late-T and Y dwarfs \citep[e.g.,][]{2011AJ....142..169B, 2015ApJ...804...92S, 2016ApJ...824....2L, 2020A&A...637A..38P, 2021ApJ...918...11L, 2024ApJ...961..210P, 2021ApJ...921...95Z}. It has been suggested in the literature that incorporating disequilibrium chemistry may alleviate this offset, as vertical mixing can reduce the abundance of NH$_{3}$, which has an absorption feature in the $Y$ band (Figure~\ref{fig:c2b}) based on theoretical predictions \citep[e.g.,][]{2012ApJ...750...74S}. Indeed, this alone cannot fully explain the $Y$-band discrepancies observed for COCONUTS-2b, as some models used in our analysis already include disequilibrium chemistry {(Section~\ref{subsec:modelgrids})}.

The $Y$-band fluxes of late-T and Y dwarfs are influenced by the pressure-broadened wings of alkali resonance lines in the optical wavelengths (Figure~\ref{fig:c2b}). \cite{2020A&A...637A..38P} found that the ATMO2020 models, which incorporate the latest opacity database of K \citep{2016A&A...589A..21A}, yield better data-model consistencies in the $Y$ band spectrum of the late-T dwarf GJ~570D \citep[$T_{\rm eff} \approx 786$~K;][]{2021ApJ...916...53Z} compared to models with previous line profiles \citep{2003ApJ...583..985B, 2003A&A...411L.473A}. Despite this, several model grids in our analysis already incorporate the latest opacities of Na \citep{2019A&A...628A.120A} and/or K \citep{2016A&A...589A..21A}, suggesting that the alkali opacities are likely not the primary source of the persistent $Y$-band offsets across the model grids. 

The $Y$-band data-model offsets may be attributed to uncertainties in alkali chemistry models, including the condensation and rainout of Na and K. \cite{2020A&A...637A..38P} noted that reducing the K abundance by nearly an order of magnitude in the \texttt{ATMO2020} models could resolve the $Y$-band data-model differences seen in other Y dwarfs. Additionally, alkali rainout was also suggested by atmospheric retrievals performed for large ensembles of late-T and Y dwarfs \citep[e.g.,][]{2017ApJ...848...83L, 2019ApJ...877...24Z, 2022ApJ...936...44Z}.

\renewcommand{\arraystretch}{1.18} 
{ 
\begin{deluxetable*}{lcccccccccc}
\setlength{\tabcolsep}{1.3pt} 
\tablecaption{Atmospheric properties of COCONUTS-2b ($Y$-band spectrum excluded)} \label{tab:fm_params_key_runs_noY} 
\tablehead{ \multicolumn{1}{l}{Atmospheric Model Grid}   &   \multicolumn{1}{c}{$\Delta$BIC}\tablenotemark{\scriptsize a}   &   \multicolumn{1}{c}{$T_{\rm eff}$}   &   \multicolumn{1}{c}{$\log{(g)}$}   &   \multicolumn{1}{c}{[M/H]\tablenotemark{\scriptsize b}}   &   \multicolumn{1}{c}{C/O\tablenotemark{\scriptsize c}}   &   \multicolumn{1}{c}{$f_{\rm sed}$\tablenotemark{\scriptsize b}}   &   \multicolumn{1}{c}{$\log{(K_{\rm zz})}$\tablenotemark{\scriptsize b}}   &   \multicolumn{1}{c}{$R$}   &   \multicolumn{1}{c}{$\log{(b)}$}   &   \multicolumn{1}{c}{$\log{(L_{\rm bol}/L_{\odot})}$\tablenotemark{\scriptsize c}}    \\ 
\multicolumn{1}{l}{}   &   \multicolumn{1}{c}{}   &   \multicolumn{1}{c}{(K)}   &   \multicolumn{1}{c}{(dex)}   &   \multicolumn{1}{c}{(dex)}   &   \multicolumn{1}{c}{}   &   \multicolumn{1}{c}{}   &   \multicolumn{1}{c}{(cm$^{2}/$s)}   &   \multicolumn{1}{c}{(R$_{\rm Jup}$)}   &   \multicolumn{1}{c}{(erg s$^{-1}$ cm$^{-2}$ \AA$^{-1}$)}   &   \multicolumn{1}{c}{(dex)}  } 
\startdata 
\multicolumn{11}{c}{disequilibrium chemistry + diabatic thermal structure + no clouds}   \\ 
\hline   
\texttt{ATMO2020++} (M1)\tablenotemark{\scriptsize d} & $-34$ & $563.1^{+2.8}_{-2.9}$ & $4.611^{+0.015}_{-0.014}$ & $-0.105^{+0.012}_{-0.012}$ & ($0.55$) & $\dots$ & ($5.78^{+0.03}_{-0.03}$) & $0.795^{+0.011}_{-0.011}$ & $-18.60^{+0.07}_{-0.09}$ & $-6.215^{+0.005}_{-0.005}$  \\ 
\texttt{ATMO2020++} (M2)\tablenotemark{\scriptsize d} & $-33$ & $555.0^{+2.4}_{-2.5}$ & $4.682^{+0.018}_{-0.014}$ & $0.024^{+0.013}_{-0.012}$ & ($0.55$) & $\dots$ & ($5.64^{+0.03}_{-0.04}$) & $0.826^{+0.011}_{-0.011}$ & $-18.68^{+0.09}_{-0.14}$ & $-6.207^{+0.005}_{-0.005}$  \\ 
\texttt{PH$_{3}$-free ATMO2020++} & $0$ & $549.3^{+2.8}_{-3.1}$ & $4.223^{+0.027}_{-0.028}$ & $-0.155^{+0.016}_{-0.015}$ & ($0.55$) & $\dots$ & ($6.55^{+0.06}_{-0.05}$) & $0.855^{+0.014}_{-0.013}$ & $-18.55^{+0.06}_{-0.08}$ & $-6.194^{+0.005}_{-0.006}$  \\ 
\hline   
\multicolumn{11}{c}{disequilibrium chemistry + clouds}   \\ 
\hline   
\texttt{Exo-REM} (M1)\tablenotemark{\scriptsize d} & $209$ & $452.9^{+1.8}_{-1.9}$ & $3.332^{+0.011}_{-0.010}$ & $-0.132^{+0.010}_{-0.009}$ & $0.435^{+0.002}_{-0.003}$ & (microphys.) & (profile) & $1.887^{+0.025}_{-0.024}$ & $-18.38^{+0.04}_{-0.05}$ & $-5.845^{+0.005}_{-0.006}$  \\ 
\texttt{Exo-REM} (M2)\tablenotemark{\scriptsize d} & $214$ & $450.0^{+0.7}_{-0.9}$ & $3.340^{+0.005}_{-0.008}$ & $-0.011^{+0.010}_{-0.018}$ & $0.400^{+0.001}_{-0.000}$ & (microphys.) & (profile) & $1.904^{+0.014}_{-0.012}$ & $-18.41^{+0.04}_{-0.05}$ & $-5.848^{+0.005}_{-0.005}$  \\ 
\hline   
\multicolumn{11}{c}{disequilibrium chemistry + no clouds}   \\ 
\hline   
\texttt{Sonora Elf Owl} & $605$ & $469.2^{+2.7}_{-2.7}$ & $3.243^{+0.021}_{-0.009}$ & $-0.276^{+0.027}_{-0.028}$ & $0.503^{+0.005}_{-0.003}$ & $\dots$ & $8.98^{+0.01}_{-0.03}$ & $1.116^{+0.021}_{-0.020}$ & $-18.40^{+0.05}_{-0.05}$ & $-6.233^{+0.006}_{-0.007}$  \\ 
\texttt{LB23 Clear NEQ} & $638$ & $464.7^{+1.1}_{-1.2}$ & $3.791^{+0.013}_{-0.020}$ & $-0.320^{+0.021}_{-0.022}$ & (0.537) & $\dots$ & (profile) & $1.131^{+0.010}_{-0.010}$ & $-18.31^{+0.04}_{-0.04}$ & $-6.237^{+0.004}_{-0.004}$  \\ 
\texttt{ATMO2020 NEQ strong} (M1)\tablenotemark{\scriptsize d} & $903$ & $465.7^{+2.7}_{-2.7}$ & $5.130^{+0.033}_{-0.033}$ & (0) & ($0.55$) & $\dots$ & ($4.74^{+0.07}_{-0.07}$) & $1.000^{+0.014}_{-0.014}$ & $-18.32^{+0.04}_{-0.04}$ & $-6.337^{+0.004}_{-0.004}$  \\ 
\texttt{ATMO2020 NEQ strong} (M2)\tablenotemark{\scriptsize d} & $908$ & $449.1^{+2.3}_{-2.3}$ & $4.896^{+0.031}_{-0.032}$ & (0) & ($0.55$) & $\dots$ & ($5.21^{+0.06}_{-0.06}$) & $1.101^{+0.020}_{-0.017}$ & $-18.40^{+0.04}_{-0.05}$ & $-6.317^{+0.007}_{-0.006}$  \\ 
\texttt{ATMO2020 NEQ weak} (M1)\tablenotemark{\scriptsize d} & $942$ & $476.4^{+2.1}_{-2.2}$ & $5.139^{+0.033}_{-0.035}$ & (0) & ($0.55$) & $\dots$ & ($2.72^{+0.07}_{-0.07}$) & $0.933^{+0.011}_{-0.011}$ & $-18.32^{+0.04}_{-0.04}$ & $-6.357^{+0.004}_{-0.004}$  \\ 
\texttt{ATMO2020 NEQ weak} (M2)\tablenotemark{\scriptsize d} & $951$ & $463.8^{+2.4}_{-2.4}$ & $4.900^{+0.039}_{-0.039}$ & (0) & ($0.55$) & $\dots$ & ($3.20^{+0.08}_{-0.08}$) & $0.999^{+0.015}_{-0.016}$ & $-18.41^{+0.05}_{-0.05}$ & $-6.345^{+0.004}_{-0.005}$  \\ 
\hline   
\multicolumn{11}{c}{equilibrium chemistry + clouds}   \\ 
\hline   
\texttt{Linder19 Cloudy} & $504$ & $640.4^{+3.2}_{-3.7}$ & $3.597^{+0.040}_{-0.042}$ & $-0.395^{+0.007}_{-0.003}$ & ($0.55$) & $0.90^{+0.06}_{-0.04}$ & $\dots$ & $0.593^{+0.005}_{-0.005}$ & $-18.17^{+0.03}_{-0.03}$ & $-6.253^{+0.009}_{-0.011}$  \\ 
\texttt{Morley12} & $1044$ & $580.9^{+4.7}_{-4.6}$ & $4.944^{+0.050}_{-0.054}$ & (0) & ($0.497$) & $4.93^{+0.05}_{-0.10}$ & $\dots$ & $0.599^{+0.009}_{-0.009}$ & $-18.01^{+0.02}_{-0.02}$ & $-6.410^{+0.005}_{-0.004}$  \\ 
\texttt{Morley14}\tablenotemark{\scriptsize e} & $1687$ & $449.9^{+0.1}_{-0.1}$ & $4.394^{+0.048}_{-0.041}$ & (0) & ($0.497$) & (5) & $\dots$ & $1.012^{+0.005}_{-0.004}$ & $-17.99^{+0.02}_{-0.02}$ & $-6.372^{+0.004}_{-0.004}$  \\ 
\hline   
\multicolumn{11}{c}{equilibrium chemistry + no clouds}   \\ 
\hline   
\texttt{LB23 Clear CEQ} (M1)\tablenotemark{\scriptsize d} & $987$ & $550.1^{+1.3}_{-1.1}$ & $4.995^{+0.004}_{-0.009}$ & $-0.167^{+0.014}_{-0.015}$ & (0.537) & $\dots$ & $\dots$ & $0.650^{+0.004}_{-0.004}$ & $-18.08^{+0.02}_{-0.02}$ & $-6.420^{+0.004}_{-0.004}$  \\ 
\texttt{LB23 Clear CEQ} (M2)\tablenotemark{\scriptsize d} & $990$ & $559.1^{+1.4}_{-2.6}$ & $4.667^{+0.028}_{-0.019}$ & $-0.279^{+0.020}_{-0.016}$ & (0.537) & $\dots$ & $\dots$ & $0.634^{+0.006}_{-0.004}$ & $-18.10^{+0.02}_{-0.02}$ & $-6.415^{+0.004}_{-0.005}$  \\ 
\texttt{ATMO2020 CEQ} & $1047$ & $495.6^{+2.0}_{-2.0}$ & $5.351^{+0.046}_{-0.045}$ & (0) & ($0.55$) & $\dots$ & $\dots$ & $0.822^{+0.009}_{-0.009}$ & $-18.15^{+0.03}_{-0.03}$ & $-6.396^{+0.004}_{-0.004}$  \\ 
\texttt{Linder19 Clear} & $1051$ & $483.0^{+3.1}_{-2.8}$ & $3.539^{+0.064}_{-0.081}$ & $-0.389^{+0.019}_{-0.008}$ & ($0.55$) & $\dots$ & $\dots$ & $0.876^{+0.013}_{-0.014}$ & $-18.35^{+0.04}_{-0.05}$ & $-6.386^{+0.004}_{-0.004}$  \\ 
\texttt{Sonora Bobcat} & $1340$ & $547.3^{+3.6}_{-3.5}$ & $5.322^{+0.061}_{-0.062}$ & $-0.123^{+0.027}_{-0.027}$ & ($0.458$) & $\dots$ & $\dots$ & $0.665^{+0.008}_{-0.008}$ & $-18.11^{+0.03}_{-0.03}$ & $-6.410^{+0.005}_{-0.005}$  \\ 
\enddata 
\tablenotetext{a}{This column lists the BIC value of each model grid relative to that of the \texttt{PH$_{3}$-free ATMO2020++} models.}  
\tablenotetext{b}{Values inside parentheses are fixed by the model grids. In particular, the $\log{(K_{\rm zz})}$ values of the \texttt{ATMO2020}, \texttt{ATMO2020++}, and \texttt{PH$_{3}$-free ATMO2020++} modeling results are linearly interpolated from the inferred $\log{(g)}$ based on Figure~1 of Phillips et al. (2021). }  
\tablenotetext{c}{The bolometric luminosity is computed based on Equation~\ref{eq:lbol} that accounts for the offsets between observed fluxes and the fitted models}  
\tablenotetext{d}{The parameter posteriors inferred by these model grids exhibit two modes, with the corresponding results labeled as ``M1'' and ``M2''.  }  
\tablenotetext{e}{The inferred $T_{\rm eff}$ based on \texttt{Morley14} models is pinned at the maximum grid value at 450~K.}  
\end{deluxetable*} 
} 
 \renewcommand{\arraystretch}{1.25} 
{ 
\begin{deluxetable*}{lcccccccccc}
\setlength{\tabcolsep}{1.3pt} 
\tablecaption{Atmospheric properties of COCONUTS-2b ($W3$ photometry included)} \label{tab:fm_params_key_runs_w3} 
\tablehead{ \multicolumn{1}{l}{Atmospheric Model Grid\tablenotemark{\scriptsize a}}   &   \multicolumn{1}{c}{$\Delta$BIC}\tablenotemark{\scriptsize b}   &   \multicolumn{1}{c}{$T_{\rm eff}$}   &   \multicolumn{1}{c}{$\log{(g)}$}   &   \multicolumn{1}{c}{[M/H]\tablenotemark{\scriptsize c}}   &   \multicolumn{1}{c}{C/O\tablenotemark{\scriptsize c}}   &   \multicolumn{1}{c}{$f_{\rm sed}$\tablenotemark{\scriptsize c}}   &   \multicolumn{1}{c}{$\log{(K_{\rm zz})}$\tablenotemark{\scriptsize c}}   &   \multicolumn{1}{c}{$R$}   &   \multicolumn{1}{c}{$\log{(b)}$}   &   \multicolumn{1}{c}{$\log{(L_{\rm bol}/L_{\odot})}$\tablenotemark{\scriptsize d}}    \\ 
\multicolumn{1}{l}{}   &   \multicolumn{1}{c}{}   &   \multicolumn{1}{c}{(K)}   &   \multicolumn{1}{c}{(dex)}   &   \multicolumn{1}{c}{(dex)}   &   \multicolumn{1}{c}{}   &   \multicolumn{1}{c}{}   &   \multicolumn{1}{c}{(cm$^{2}/$s)}   &   \multicolumn{1}{c}{(R$_{\rm Jup}$)}   &   \multicolumn{1}{c}{(erg s$^{-1}$ cm$^{-2}$ \AA$^{-1}$)}   &   \multicolumn{1}{c}{(dex)}  } 
\startdata 
\multicolumn{10}{c}{disequilibrium chemistry + diabatic thermal structure + no clouds}   \\ 
\hline   
\texttt{PH$_{3}$-free ATMO2020++} & $0$ & $544.6^{+3.3}_{-3.2}$ & $4.170^{+0.022}_{-0.014}$ & $-0.151^{+0.012}_{-0.010}$ & ($0.55$) & $\dots$ & ($6.66^{+0.03}_{-0.04}$) & $0.883^{+0.014}_{-0.015}$ & $-18.52^{+0.06}_{-0.07}$ & $-6.177^{+0.005}_{-0.005}$  \\ 
\texttt{ATMO2020++} & $4$ & $533.2^{+3.9}_{-3.9}$ & $4.212^{+0.034}_{-0.041}$ & $-0.208^{+0.024}_{-0.021}$ & ($0.55$) & $\dots$ & ($6.58^{+0.08}_{-0.07}$) & $0.940^{+0.020}_{-0.020}$ & $-18.54^{+0.06}_{-0.08}$ & $-6.159^{+0.007}_{-0.006}$  \\ 
\hline   
\multicolumn{10}{c}{disequilibrium chemistry + clouds}   \\ 
\hline   
\texttt{Exo-REM} & $105$ & $449.9^{+0.4}_{-0.6}$ & $3.349^{+0.004}_{-0.006}$ & $-0.001^{+0.008}_{-0.009}$ & $0.400^{+0.000}_{-0.000}$ & (microphys.) & (profile) & $1.899^{+0.012}_{-0.011}$ & $-18.34^{+0.04}_{-0.05}$ & $-5.850^{+0.004}_{-0.004}$  \\ 
\hline   
\multicolumn{10}{c}{disequilibrium chemistry + no clouds}   \\ 
\hline   
\texttt{LB23 Clear NEQ} & $562$ & $463.1^{+1.0}_{-1.0}$ & $3.759^{+0.012}_{-0.006}$ & $-0.304^{+0.021}_{-0.021}$ & (0.537) & $\dots$ & (profile) & $1.151^{+0.009}_{-0.009}$ & $-18.28^{+0.04}_{-0.04}$ & $-6.225^{+0.004}_{-0.004}$  \\ 
\texttt{ATMO2020 NEQ strong} & $842$ & $444.9^{+2.6}_{-2.6}$ & $4.799^{+0.036}_{-0.036}$ & (0) & ($0.55$) & $\dots$ & ($5.40^{+0.07}_{-0.07}$) & $1.152^{+0.022}_{-0.022}$ & $-18.38^{+0.04}_{-0.05}$ & $-6.291^{+0.007}_{-0.007}$  \\ 
\texttt{ATMO2020 NEQ weak} & $899$ & $461.1^{+2.4}_{-2.5}$ & $4.799^{+0.038}_{-0.039}$ & (0) & ($0.55$) & $\dots$ & ($3.40^{+0.08}_{-0.08}$) & $1.027^{+0.017}_{-0.016}$ & $-18.41^{+0.05}_{-0.05}$ & $-6.326^{+0.005}_{-0.005}$  \\ 
\hline   
\multicolumn{10}{c}{equilibrium chemistry + clouds}   \\ 
\hline   
\texttt{Linder19 Cloudy} & $390$ & $676.1^{+2.8}_{-3.2}$ & $3.251^{+0.027}_{-0.036}$ & $-0.396^{+0.006}_{-0.003}$ & ($0.55$) & $0.78^{+0.03}_{-0.03}$ & $\dots$ & $0.575^{+0.005}_{-0.005}$ & $-18.08^{+0.02}_{-0.02}$ & $-6.186^{+0.006}_{-0.006}$  \\ 
\texttt{Morley12} & $1015$ & $570.0^{+4.8}_{-4.9}$ & $4.718^{+0.053}_{-0.055}$ & (0) & ($0.497$) & $4.97^{+0.02}_{-0.03}$ & $\dots$ & $0.622^{+0.010}_{-0.010}$ & $-18.02^{+0.02}_{-0.02}$ & $-6.405^{+0.005}_{-0.005}$  \\ 
\texttt{Morley14}\tablenotemark{\scriptsize e} & $1538$ & $449.9^{+0.1}_{-0.1}$ & $4.204^{+0.034}_{-0.035}$ & (0) & ($0.497$) & (5) & $\dots$ & $1.024^{+0.004}_{-0.004}$ & $-17.99^{+0.02}_{-0.02}$ & $-6.360^{+0.004}_{-0.004}$  \\ 
\hline   
\multicolumn{10}{c}{equilibrium chemistry + no clouds}   \\ 
\hline   
\texttt{LB23 Clear CEQ} (M1)\tablenotemark{\scriptsize f} & $906$ & $561.7^{+0.7}_{-0.8}$ & $4.630^{+0.013}_{-0.016}$ & $-0.262^{+0.012}_{-0.012}$ & (0.537) & $\dots$ & $\dots$ & $0.631^{+0.004}_{-0.004}$ & $-18.10^{+0.02}_{-0.02}$ & $-6.408^{+0.004}_{-0.004}$  \\ 
\texttt{LB23 Clear CEQ} (M2)\tablenotemark{\scriptsize f} & $910$ & $559.6^{+0.7}_{-0.7}$ & $4.415^{+0.019}_{-0.012}$ & $-0.312^{+0.012}_{-0.013}$ & (0.537) & $\dots$ & $\dots$ & $0.633^{+0.004}_{-0.004}$ & $-18.12^{+0.03}_{-0.02}$ & $-6.411^{+0.004}_{-0.004}$  \\ 
\texttt{Linder19 Clear} & $935$ & $487.5^{+1.6}_{-1.6}$ & $3.218^{+0.052}_{-0.050}$ & $-0.394^{+0.010}_{-0.005}$ & ($0.55$) & $\dots$ & $\dots$ & $0.864^{+0.008}_{-0.009}$ & $-18.37^{+0.04}_{-0.05}$ & $-6.378^{+0.004}_{-0.004}$  \\ 
\texttt{ATMO2020 CEQ} & $993$ & $493.0^{+2.0}_{-1.9}$ & $5.179^{+0.042}_{-0.041}$ & (0) & ($0.55$) & $\dots$ & $\dots$ & $0.835^{+0.008}_{-0.008}$ & $-18.18^{+0.03}_{-0.03}$ & $-6.387^{+0.004}_{-0.004}$  \\ 
\texttt{Sonora Bobcat} & $1298$ & $530.3^{+2.6}_{-2.7}$ & $5.493^{+0.005}_{-0.010}$ & $0.102^{+0.020}_{-0.021}$ & ($0.458$) & $\dots$ & $\dots$ & $0.701^{+0.008}_{-0.008}$ & $-18.19^{+0.03}_{-0.03}$ & $-6.414^{+0.004}_{-0.004}$  \\ 
\enddata 
\tablenotetext{a}{The Sonora Elf Owl models are not used because the synthetic spectra (with a wavelength cutoff at 15~$\mu$m) do not cover the full wavelength of the $W3$ filter response.}  
\tablenotetext{b}{This column lists the BIC value of each model grid relative to that of the \texttt{PH$_{3}$-free ATMO2020++} models.}  
\tablenotetext{c}{Values inside parentheses are fixed by the model grids. In particular, the $\log{(K_{\rm zz})}$ values of the \texttt{ATMO2020}, \texttt{ATMO2020++}, and \texttt{PH$_{3}$-free ATMO2020++} modeling results are linearly interpolated from the inferred $\log{(g)}$ based on Figure~1 of Phillips et al. (2021). }  
\tablenotetext{d}{The bolometric luminosity is computed based on Equation~\ref{eq:lbol} that accounts for the offsets between observed fluxes and the fitted models}  
\tablenotetext{e}{The inferred $T_{\rm eff}$ based on \texttt{Morley14} models is pinned at the maximum grid value at 450~K.}  
\tablenotetext{f}{The parameter posteriors inferred by the \texttt{LB23 Clear CEQ} grid exhibit two modes, with the corresponding results labeled as ``M1'' and ``M2''.   }  
\end{deluxetable*} 
}

\subsubsection{Most Model Grids: Under-predicted Fluxes near 3.6~$\mu$m}
\label{subsubsec:offsets_most_models_36um}

All model grids, except \texttt{Exo-REM}, predict fainter fluxes than the observations at the wavelengths of $W1$ and [3.6] photometry (Figures~\ref{fig:best_neq}--\ref{fig:clear_ceq}), with significance values of $4\sigma-22\sigma$. These data-model offsets near 3.6~$\mu$m, as influenced by the CH$_{4}$ absorption, are sometimes connected with departures from equilibrium chemistry. The disequilibrium chemistry occurs when the timescale for vertical mixing {is shorter than} that of the CO $\rightarrow$ CH$_{4}$ chemical reaction \citep[e.g.,][]{1977Sci...198.1031P, 1996ApJ...472L..37F, 2000ApJ...541..374S, 2014ApJ...797...41Z, 2015ApJ...804L..17T, 2020AJ....160...63M, 2021ApJ...923..269K, 2023ApJ...942...71M, 2024ApJ...963...73M}. Vertical mixing quenches the CH$_{4}$ abundance in the deep atmosphere, leading to an overabundance of CO relative to CH$_{4}$. However, the offsets near 3.6~$\mu$m are present in both chemical equilibrium and disequilibrium models, indicating that additional factors may be involved.

\cite{2021ApJ...918...11L} proposed that upper-atmospheric heating could enhance the model-predicted fluxes near 3.6~$\mu$m and improve the data-model consistency for late-T and Y dwarfs. With a modified adiabatic index of $\gamma = 1.25$, the \texttt{ATMO2020++} and \texttt{PH$_{3}$-free ATMO2020++} models have slightly warmer atmospheres in the photospheres of $W1$ and [3.6] photometry as compared to models with the standard $\gamma$. However, our analysis suggests that such a $\gamma$, at least with this magnitude of modification, remains insufficient to match the observations (Figure~\ref{fig:best_neq}). Other studies have also investigated atmospheric heating \citep[e.g.,][]{2014MNRAS.440.3675S, 2017MNRAS.470.1177B}, including \cite{2018ApJ...858...97M}, who found no direct evidence for this effect in the atmosphere of WISE~J085510.83$-$071442.5.

Interestingly, the \texttt{Exo-REM} models predict $W1$ and [3.6] values that are consistent with observations within $3\sigma$. These models incorporate cloud opacities that are not included by most other grids, including \texttt{ATMO2020++} and \texttt{PH$_{3}$-free ATMO2020++}. This suggests that cloud formation may be important in resolving the data-model offsets near 3.6~$\mu$m. {Moreover, the improved data-model consistency of the \texttt{Exo-REM} models may also stem from other factors, such as the opacities of key atmospheric absorbers (e.g., CH$_{4}$, collision-induced absorption by H$_{2}$) and the specific treatment of vertical mixing processes.}

\subsubsection{Models with Equilibrium Chemistry and No Clouds: Under-predicted Fluxes in the $H$ band}
\label{subsubsec:offsets_clear_ceq}

For model grids with equilibrium chemistry and no clouds, the predicted fluxes are too faint in the blue wing of the $H$ band {(Figure~\ref{fig:clear_ceq})}. This offset indicates the need for disequilibrium chemistry, which not only impacts the abundances of CH$_{4}$ and CO (Section~\ref{subsubsec:offsets_most_models_36um}) but also reduces the abundance of NH$_{3}$ \citep[e.g.,][]{2006ApJ...647..552S, 2014ApJ...797...41Z}. Incorporating disequilibrium chemistry into these models can lead to a depletion of NH$_{3}$, which would enhance the model-predicted fluxes near 1.55~$\mu$m and help resolve the data-model discrepancies.

{ 
\begin{deluxetable*}{lccccccccccc}
\setlength{\tabcolsep}{2pt} 
\tablecaption{Evolution-based properties of COCONUTS-2b based on $L_{\rm bol}$ and age} \label{tab:evoparams} 
\tablehead{ \multicolumn{1}{l}{Evolution Models} &  \multicolumn{2}{c}{$T_{\rm eff}$} &  \multicolumn{1}{c}{} &  \multicolumn{2}{c}{$R$} &  \multicolumn{1}{c}{} &  \multicolumn{2}{c}{$\log{g}$} &  \multicolumn{1}{c}{} &  \multicolumn{2}{c}{$M$} \\ 
\cline{2-3} \cline{5-6} \cline{8-9} \cline{11-12} 
\multicolumn{1}{l}{}  &  \multicolumn{1}{c}{Median$\pm 1\sigma$} &  \multicolumn{1}{c}{$3\sigma$ C. I.} &  \multicolumn{1}{c}{}  &  \multicolumn{1}{c}{Median$\pm 1\sigma$} &  \multicolumn{1}{c}{$3\sigma$ C. I.} &  \multicolumn{1}{c}{}  &  \multicolumn{1}{c}{Median$\pm 1\sigma$} &  \multicolumn{1}{c}{$3\sigma$ C. I.} &  \multicolumn{1}{c}{}  &  \multicolumn{1}{c}{Median$\pm 1\sigma$} &  \multicolumn{1}{c}{$3\sigma$ C. I.} \\ 
\multicolumn{1}{l}{}  &  \multicolumn{1}{c}{(K)} &  \multicolumn{1}{c}{(K)} &  \multicolumn{1}{c}{}  &  \multicolumn{1}{c}{(R$_{\rm Jup}$)} &  \multicolumn{1}{c}{(R$_{\rm Jup}$)} &  \multicolumn{1}{c}{}  &  \multicolumn{1}{c}{(dex)} &  \multicolumn{1}{c}{(dex)} &  \multicolumn{1}{c}{}  &  \multicolumn{1}{c}{(M$_{\rm Jup}$)} &  \multicolumn{1}{c}{(M$_{\rm Jup}$)}  } 
\startdata 
hot-start ATMO2020  &  $490^{+47}_{-52}$  &  ($416$, $577$)  &  &  $1.09^{+0.03}_{-0.04}$  &  ($1.05$, $1.16$)  &    &  $4.21^{+0.18}_{-0.14}$  &   ($3.8$, $4.5$)   &  &  $8^{+2}_{-2}$  &   ($3$, $12$)  \\ 
hot-start Sonora Bobcat [M/H]$=+0.5$  &  $488^{+47}_{-52}$  &  ($413$, $575$)  &  &  $1.10^{+0.03}_{-0.04}$  &  ($1.05$, $1.17$)  &    &  $4.20^{+0.19}_{-0.14}$  &   ($3.8$, $4.4$)   &  &  $8^{+2}_{-2}$  &   ($3$, $12$)  \\ 
hot-start Sonora Bobcat [M/H]$=0$  &  $485^{+47}_{-52}$  &  ($411$, $570$)  &  &  $1.12^{+0.03}_{-0.04}$  &  ($1.07$, $1.19$)  &    &  $4.18^{+0.18}_{-0.14}$  &   ($3.8$, $4.4$)   &  &  $8^{+2}_{-2}$  &   ($3$, $12$)  \\ 
hot-start Sonora Bobcat [M/H]$=-0.5$  &  $484^{+46}_{-51}$  &  ($410$, $569$)  &  &  $1.12^{+0.03}_{-0.04}$  &  ($1.07$, $1.19$)  &    &  $4.16^{+0.19}_{-0.13}$  &   ($3.7$, $4.4$)   &  &  $7^{+2}_{-2}$  &   ($3$, $11$)  \\ 
cold-start Sonora Bobcat [M/H]$=0$  &  $466^{+33}_{-47}$  &  ($411$, $561$)  &  &  $1.12^{+0.03}_{-0.04}$  &  ($1.07$, $1.18$)  &    &  $4.17^{+0.17}_{-0.10}$  &   ($3.8$, $4.3$)   &  &  $7^{+2}_{-2}$  &   ($3$, $10$)  \\ 
\hline 
{\bf Adopted}\tablenotemark{\scriptsize a}  &  $483^{+44}_{-53}$  &  ($411$, $575$)  &  &  $1.11^{+0.03}_{-0.04}$  &  ($1.05$, $1.19$)  &    &  $4.19^{+0.18}_{-0.13}$  &   ($3.7$, $4.4$)   &  &  $8^{+2}_{-2}$  &   ($3$, $12$)  \\ 
\enddata 
\tablenotetext{a}{The adopted parameters are inferred from the concatenated chains of all the evolution models (Section~\ref{subsubsec:tgrm}).}  
\end{deluxetable*} 
}

\subsubsection{Models with Equilibrium Chemistry and Clouds: Offsets in the $J$ band}
\label{subsubsec:offsets_cloudy_ceq}

For \texttt{Linder19 Cloudy} and \texttt{Morley12} models that incorporate clouds into equilibrium chemistry, the predicted spectral shapes in the $J$ band are too broad (Figure~\ref{fig:cloudy_ceq}). These models also yield high $T_{\rm eff}$ values (570--655~K) that are 100--200~K hotter than predictions from most other atmospheric model grids (Table~\ref{tab:fm_params_key_runs}) and thermal evolution models (Section~\ref{subsubsec:tgrm}); their predicted radii (0.58--0.62~$R_{\rm Jup}$) are unphysically small, {compared to the evolution-based value of $1.11^{+0.03}_{-0.04}$~R$_{\rm Jup}$ (see Section~\ref{subsubsec:tgrm})}. These chemical equilibrium cloudy models attempt to use clouds and hotter $T_{\rm eff}$ to compensate for the lack of disequilibrium chemistry and avoid the $H$-band offsets discussed in the previous section. With $T_{\rm eff}$ around 600~K, sulfide clouds form at a few bars, attenuating the emergent fluxes in the $Y$ and $J$ bands and redistributing the fluxes toward longer wavelengths (e.g., $K$ band and beyond), while not significantly altering fluxes in the $H$ band \citep[e.g., Figure~4 of][]{2012ApJ...756..172M}. Such a hotter atmosphere has a relatively lower chemical equilibrium abundance of NH$_{3}$ and CH$_{4}$ in the near-infrared photospheres, leading to a broader $H$-band shape (compared to predictions of chemical equilibrium models) that matches the observations. However, the hotter cloudy atmosphere also results in a broader $J$-band shape with a fainter peak flux, causing the data-model offsets in the $J$ band. 

These explanations are supported by the analysis results with \texttt{Morley14} models. The inferred $T_{\rm eff}$ posterior is pinned at the maximum grid value of 450~K (Table~\ref{tab:fm_params_key_runs}), but this insufficiently high temperature leaves the $H$-band data-model offsets unresolved (Figure~\ref{fig:cloudy_ceq}).

\subsubsection{Models with Disequilibrium Chemistry and No Clouds: Under-predicted Fluxes in the $K$ band}
\label{subsubsec:offsets_clear_neq}

For model grids with disequilibrium chemistry and no clouds, the predicted fluxes are too faint in the $K$ band (Figure~\ref{fig:clear_neq}). This offset suggests the need for a modified adiabatic index or clouds. A slightly reduced adiabatic index (Section~\ref{subsec:properties}) can lead to hotter temperatures in the $K$-band photospheres, while clouds can redistribute the short-wavelength fluxes emergent from deep atmospheres to the $K$ band and longer wavelengths (Section~\ref{subsubsec:offsets_cloudy_ceq}). Both effects can enhance the model-predicted fluxes in the $K$ band, justifying why \texttt{PH$_{3}$-free ATMO2020++}, \texttt{ATMO2020++}, and \texttt{Exo-REM} are the preferred grids for interpreting the spectrophotometry of COCONUTS-2b.

\subsection{Impact of the Input Spectrophotometry}
\label{subsec:different_input}

We repeat the forward-modeling analysis with variations in COCONUTS-2b's spectrophotometry to assess their impacts on our results. 

First, we perform the analysis with the $Y$-band spectrum fully masked, focusing on fitting the spectrum over 1.18--1.35~$\mu$m, 1.45--1.66~$\mu$m, and 1.95--2.30~$\mu$m, along with $W1$, $W2$, [3.6], and [4.5] photometry. This approach is motivated by the persistent $Y$-band data-model offsets among all model grids in our analysis (Section~\ref{subsubsec:offsets_all_models}). {The \texttt{PH$_{3}$-free ATMO2020++}, \texttt{ATMO2020++} and \texttt{Exo-REM} models remain the most preferred grids for interpreting the spectrophotometry of COCONUTS-2b. Notably, several model grids --- including \texttt{ATMO2020++}, \texttt{Exo-REM}, \texttt{ATMO2020 NEQ strong}, \texttt{ATMO2020 NEQ weak}, and \texttt{LB23 Clear CEQ} --- exhibit bimodal parameter posteriors (Table~\ref{tab:fm_params_key_runs_noY}). In contrast, when the $Y$-band spectrum is included in spectral fits, only the \texttt{LB23 Clear CEQ} model grid yields bimodal posteriors. 

For the three preferred model grids, we also assess the changes in the inferred parameters between fits with and without the $Y$-band spectrum. For \texttt{PH$_{3}$-free ATMO2020++} and \texttt{Exo-REM}, most physical parameters remain consistent within $1.5\sigma$. The only significant deviation is found in the [M/H] predicted by \texttt{Exo-REM}, where including the $Y$-band spectrum leads to a [M/H] value that is $10\sigma$ higher than one of the modes inferred without $Y$ band. Additionally, parameter shifts in \texttt{ATMO2020++} become more significant, ranging from $4\sigma-12\sigma$, with absolute differences of $20-30$~K in $T_{\rm eff}$, $0.4-0.5$~dex in $\log{(g)}$, $0.05-0.23$~dex in [M/H], and $0.08-0.10$~R$_{\rm Jup}$ in radius.}

In the second experiment, we keep the same spectral wavelength ranges as in the main analysis (Section~\ref{sec:fm}) but additionally include the $W3$ photometry. No significant changes are found in the results. The inferred parameters of these two experiments are summarized in Tables~\ref{tab:fm_params_key_runs_noY} and \ref{tab:fm_params_key_runs_w3}. Figures presenting data-model comparisons are presented in Appendix~\ref{app:different_input}.

\begin{figure*}[t]
\begin{center}
\includegraphics[width=6.3in]{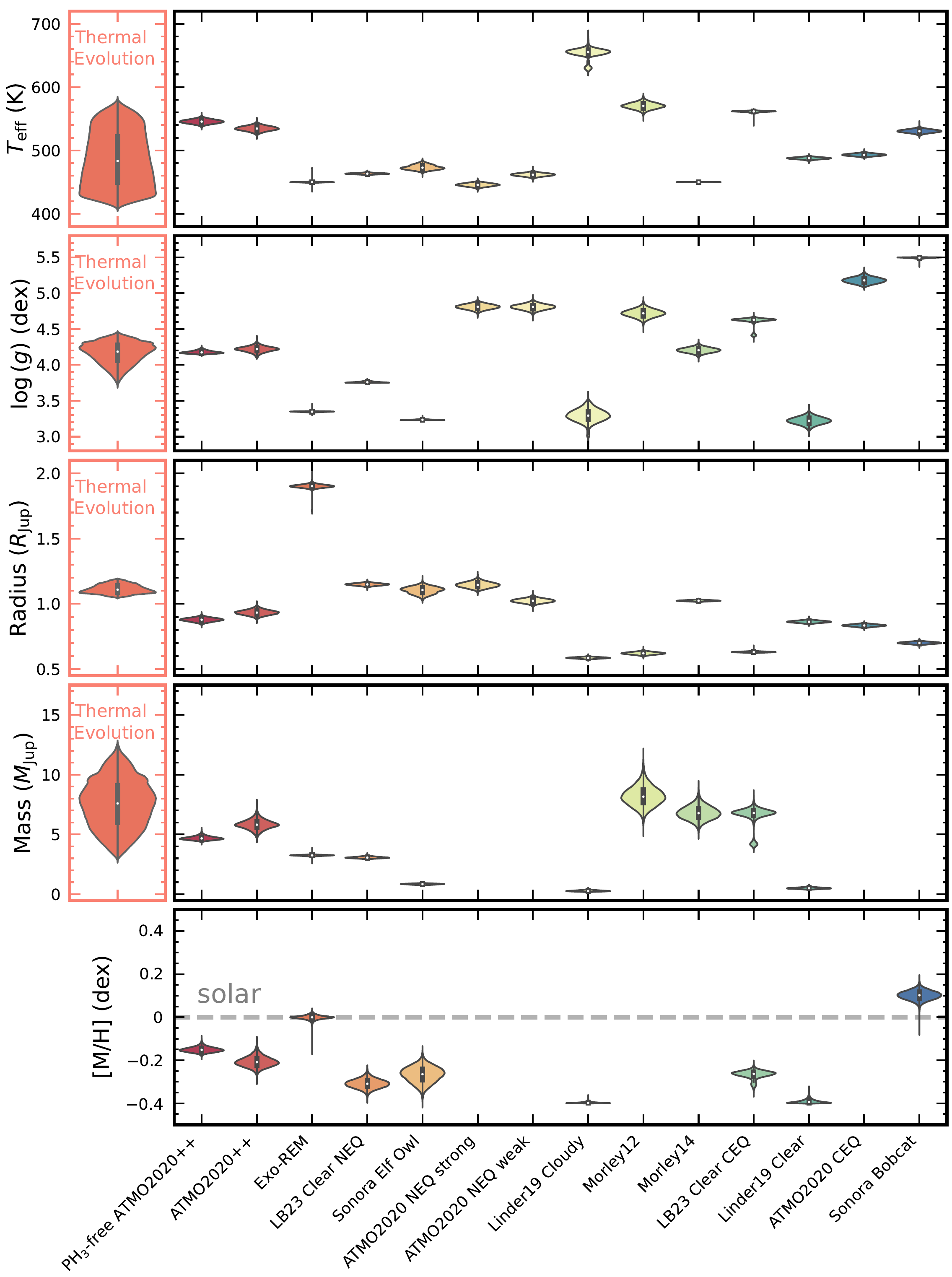}
\caption{Violin plot for the physical properties of COCONUTS-2b determined from thermal evolution models (the left column; Section~\ref{subsubsec:tgrm}) and {various atmospheric models using the Gemini/F2 spectrum and $W1/W2$[3.6]/[4.5] photometry} (the right column; Section~\ref{sec:fm}). The evolution-based parameter posteriors are obtained by concatenating the results from several hot-start and cold-start evolution models (Table~\ref{tab:evoparams}). The plotted atmospheric models are ordered by their corresponding BIC values (low to high from left to right). The spectroscopically inferred masses of the following atmospheric models are between 20--70~M$_{\rm Jup}$ and thus outside the y-axis range: \texttt{ATMO2020 NEQ strong}, \texttt{ATMO2020 NEQ weak}, \texttt{ATMO2020 CEQ}, and \texttt{Sonora Bobcat}. }
\label{fig:evo_vs_atm}
\end{center}
\end{figure*}

\subsection{A Recommended Set of Physical Properties for COCONUTS-2\lowercase{b}}
\label{subsec:rec_params}

The forward-modeling analyses provide valuable insights into the atmospheric processes of COCONUTS-2b, yet the resulting physical properties of different model grids differ noticeably. These spectroscopically inferred physical parameters may carry systematic errors \citep[e.g.,][]{2021ApJ...916...53Z, 2021ApJ...921...95Z, 2024ApJ...961..121H}, stemming from uncertainties in opacity databases and assumptions about exoplanet atmospheres. Thermal evolution models offer a complementary perspective on the fundamental properties of self-luminous exoplanets, brown dwarfs, and stars, as these models are less affected by some of these systematic effects (e.g., Section~5.1 of \citealt{2023AJ....166..198Z} and references therein). 

In this section, we summarize the physical properties of COCONUTS-2b by combining the results of our forward-modeling analysis with predictions from various thermal evolution models. We focus on bolometric luminosity (Section~\ref{subsubsec:lbol}), effective temperature, surface gravity, radius, and mass (Section~\ref{subsubsec:tgrm}).

\subsubsection{Bolometric Luminosity}
\label{subsubsec:lbol}

Table~\ref{tab:fm_params_key_runs} lists the inferred bolometric luminosities based on each model grid. These $L_{\rm bol}$ values are computed with data-model discrepancies already incorporated (Equation~\ref{eq:lbol}). Therefore, the variations in $L_{\rm bol}$ among model grids primarily reflect the different assumptions of each model rather than the quality of the spectral fits. 

We adopt the bolometric luminosity value from the most preferred grid, which is $\log{(L_{\rm bol} / L_{\odot})} = -6.18$~dex, along with a 0.5~dex-wide range of $[-6.43, -5.93]$~dex to encompass the scatter of predicted values across different model grids.\footnote{\cite{2021ApJ...916L..11Z} estimated $\log{(L_{\rm bol} / L_{\odot})} = -6.384 \pm 0.028$~dex by combining $J_{\rm MKO}$, $K_{\rm MKO}$, [3.6], and [4.5] broadband photometry. They used the ``super-magnitude'' bolometric correction, which is based on the \texttt{Sonora Bobcat} and \texttt{ATMO2020} models \citep[see][]{2024ApJ...967..115B}. Our computed bolometric luminosities from these two model grids, as shown in Tables~\ref{tab:fm_params_key_runs}--\ref{tab:fm_params_key_runs_w3}, yield consistent values with \cite{2021ApJ...916L..11Z}.} {The bolometric luminosity of COCONUTS-2b will be more precisely constrained by the upcoming JWST NIRSPEC and MIRI observations.}

\subsubsection{Effective Temperature, Surface Gravity, Radius, and Mass}
\label{subsubsec:tgrm}
We derive the physical properties of COCONUTS-2b using several thermal evolution models (Table~\ref{tab:evoparams}), including the hot-start ATMO2020 models \citep{2020A&A...637A..38P} and both the hot-start and cold-start Sonora Bobcat models \citep{2021ApJ...920...85M}. We assume uniform distributions for $\log{(L_{\rm bol} / L_{\odot})}$ and age, ranging from $-6.43$~dex to $-5.93$~dex (Section~\ref{subsubsec:lbol}) and from 150~Myr to 800~Myr \citep{2021ApJ...916L..11Z}, respectively. The evolution models are interpolated on a linear scale for $\log{(g)}$ and logarithmic scales for $T_{\rm eff}$, $L_{\rm bol}$, age, $R$, and $M$. 

The inferred parameters are summarized in Table~\ref{tab:evoparams}. Given the consistency of predictions across different evolution models, we concatenate the chains of a given parameter derived from all models and adopt the median values and confidence intervals. Based on all these evolution models, COCONUTS-2b has $T_{\rm eff} = 483^{+44}_{-53}$~K, $\log{(g)} = 4.19^{+0.18}_{-0.13}$~dex, $R = 1.11^{+0.03}_{-0.04}$~R$_{\rm Jup}$, and $M = 8 \pm 2$~M$_{\rm Jup}$. Considering the mass of COCONUTS-2A \citep[$0.37\pm0.02$~M$_{\odot}$;][]{2021ApJ...916L..11Z}, the companion-to-host mass ratio of this system is $0.021 \pm 0.005$. 

Figure~\ref{fig:evo_vs_atm} compares these evolution-based parameters with those inferred from atmospheric models through spectral fitting (Section~\ref{sec:fm}). {For the three preferred model grids, \texttt{PH$_{3}$-free ATMO2020++}, \texttt{ATMO2020++}, and \texttt{Exo-REM}, their inferred effective temperatures align well with predictions from the thermal evolution models. The $\log{(g)}$ and $M$ values inferred by \texttt{PH$_{3}$-free ATMO2020++} and \texttt{ATMO2020++} also match the evolution-based parameters, while \texttt{Exo-REM} predicts lower values. However, all three model grids predict different radii with evolution-based values, with \texttt{Exo-REM} showing significantly larger inconsistencies than \texttt{PH$_{3}$-free ATMO2020++} and \texttt{ATMO2020++}. Beyond these three model grids, several other grids predict unphysically small radii, near 0.6~R$_{\rm Jup}$ (Table~\ref{tab:fm_params_key_runs}), which contradict the fundamental mass-radius relationship for brown dwarfs \citep{1989ApJ...345..939B, 1996ApJ...460..993S, 2021ApJ...920...85M}. These discrepancies between atmospheric and evolution model predictions have been reported for other gas-giant exoplanets and brown dwarfs \citep[e.g.,][]{2020ApJ...905...46G, 2020ApJ...891..171Z, 2021ApJ...916...53Z, 2021ApJ...921...95Z, 2022ApJ...936...44Z, 2023ApJ...953..170H, 2024ApJ...961..210P, 2024ApJ...961..121H}. These differences may arise from uncertainties in the opacities adopted by the atmospheric models, assumptions within atmospheric and evolution models, or unknown factors such as the binarity or age uncertainties of the objects (see also Section 5.1 of \citealt{2023AJ....166..198Z}). 

Additionally, the uncertainties in parameters inferred from atmospheric models are significantly smaller than those derived from evolution models. Similar trends of small statistical uncertainties from atmospheric models have been also observed in previous forward-modeling studies \citep[e.g.,][]{2007ApJ...667..537L, 2010ApJS..186...63R, 2020ApJ...891..171Z, 2024ApJ...961..121H, 2024ApJ...966L..11P, 2024ApJ...961..210P}. A more sophisticated likelihood function that accounts for model interpolation uncertainties and correlations in data-model residuals across adjacent wavelengths may provide more realistic parameter uncertainties \citep[e.g.,][]{2015ApJ...812..128C, 2021ApJ...916...53Z, 2021ApJ...921...95Z}.   

}

\subsection{The Sub-/Near-solar [M/H] and C/O in the Atmosphere of COCONUTS-2b}
\label{subsubsec:mhcokzz}

{Most atmospheric models predict a sub-solar metallicity for COCONUTS-2b, regardless of variations in the input spectrophotometry used for spectral fits (Figure~\ref{fig:evo_vs_atm}). However, exceptions exist where the inferred metallicities are closer to the solar value (Tables~\ref{tab:fm_params_key_runs}--\ref{tab:fm_params_key_runs_w3}). Specifically, when $Y$-band fluxes are included in the spectral fits, the \texttt{Exo-REM} grid yields [M/H]$= -0.001^{+0.008}_{-0.009}$~dex; also, the least preferred grid, \texttt{Sonora Bobcat}, gives [M/H]$=0.102^{+0.020}_{-0.021}$~dex. When the $Y$-band fluxes are excluded, the \texttt{ATMO2020++} and \texttt{Exo-REM} grids show bimodal posterior distributions. In both grids, the mode with higher log-likelihood values suggests a sub-solar metallicity, while the other mode indicates a near-solar value (Table~\ref{tab:fm_params_key_runs_noY}). Furthermore}, the only two model grids with non-solar C/O axes, \texttt{Exo-REM} and \texttt{Sonora Elf Owl}, suggest C/O of $0.4-0.5$. These values are lower than or comparable with the solar C/O range of 0.45--0.55 as reported by, e.g., \cite{2003ApJ...591.1220L}, \cite{2006CoAst.147...76A}, \cite{2009ARA&A..47..481A}, \cite{2010fee..book..157L}, and \cite{2011SoPh..268..255C}. {These metallicity and C/O estimates for COCONUTS-2b should be validated with upcoming JWST observations, which will cover a broader wavelength range. As shown by \cite{2024ApJ...966L..11P}, the spectroscopically inferred [M/H] and C/O ratios for a L-type brown dwarf can be sensitive to the spectral wavelength range used in the fit. }

{In summary, our analysis indicates that COCONUTS-2b's atmosphere has sub- or near-solar [M/H] and C/O.} Similar findings were also reported for WASP-77~Ab \citep{2021Natur.598..580L, 2023ApJ...953L..24A}. A sub-solar atmospheric [M/H] suggests that COCONUTS-2b was not influenced by core erosion or planetesimal bombardment, processes that can lead to metal enrichment in the atmospheres \citep[e.g.,][]{2005ApJ...622L.145A, 2011ApJ...736L..29M, 2019ApJ...874L..31T, 2019ARA&A..57..617M, 2021A&A...654A..72S, 2023AJ....166..198Z, 2024RvMG...90..411K}. The sub-/near-solar C/O may be explained if COCONUTS-2b initially assembled its atmosphere at very close orbital separations from its parent star, accreting oxygen-rich and carbon-depleted gas from within ice lines \citep[e.g.,][]{2021Natur.598..580L, 2023ApJ...953L..24A}. This process might have been followed by outward migration after disk dissipation \citep[driven by, for example, planet-planet interactions and stellar fly-bys;][]{2024ApJ...967..147M}, bringing COCONUTS-2b to its current wide orbit. 

{Additionally, our most preferred grid, \texttt{PH$_{3}$-free ATMO2020++}, estimates a metallicity of [M/H]$=-0.152^{+0.012}_{-0.010}$~dex for COCONUTS-2b, which is a factor of $0.71^{+0.14}_{-0.12}$ relative to the metallicity of its parent star, COCONUTS-2A ([M/H]$_{A}$ = $0.00 \pm 0.08$~dex; \citealt{2019A&A...629A..80H}). This $2\sigma$ difference in [M/H] between COCONUTS-2b and 2A lowers the likelihood of an in-situ, stellar-binary-like formation scenario. However, our third-preferred grid, \texttt{Exo-REM}, suggests that COCONUTS-2b and 2A have consistent atmospheric metallicities, which supports a binary-like formation pathway. }

\section{Summary}
\label{sec:summary}

We have studied COCONUTS-2b's atmospheric properties based on our newly observed 1.0--2.5~$\mu$m spectroscopy from Gemini/FLAMINGOS-2. The main findings are summarized below.

\begin{enumerate}
\item[$\bullet$] COCONUTS-2b straddles the T/Y transition with a spectral type of T9.5$\pm 0.5$, as determined by comparing our Gemini/F2 spectrum to the near-infrared spectra of T0--Y1 dwarf templates. 

\item[$\bullet$] We compare the near-infrared spectra and $W1$, $W2$, [3.6] and [4.5] broad-band photometry of COCONUTS-2b with sixteen state-of-the-art atmospheric model grids of T/Y dwarfs to investigate what model assumptions can better approximate the particular atmosphere of COCONUTS-2b. The most preferred model grids are \texttt{PH$_{3}$-free ATMO2020++}, \texttt{ATMO2020++}, and \texttt{Exo-REM}, suggesting the presence of dis-equilibrium chemistry, along with a diabatic thermal structure and/or clouds, in the atmosphere of COCONUTS-2b. 

\item[$\bullet$] Offsets between data and fitted models are discussed. We find that the predicted $Y$-band fluxes by all these models are much fainter than the observations, highlighting the uncertainties in the {alkali chemistry models and opacities}. Also, all the atmospheric model grids, except for \texttt{Exo-REM}, predict fainter fluxes than the observations near 3.6~$\mu$m; the upper-atmospheric heating and/or clouds may play important roles in resolving these offsets. Additionally, we find that (1) the models with chemical equilibrium and no clouds under-predict COCONUTS-2b's fluxes in the $H$ band, (2) the models with chemical equilibrium and clouds yield a broad and faint $J$-band spectral shape, and (3) the models with disequilibrium chemistry and no clouds under-predict the observed fluxes in the $K$ band. These offsets justify the need for dis-equilibrium chemistry, modifications to the atmospheric adiabatic index, and clouds in the model assumptions.

\item[$\bullet$] The forward-modeling analysis is repeated with the $Y$ band excluded or with the $W3$ photometry included. These variations of the input spectrophotometry can lead to changes in the inferred parameters but do not alter our discussions and conclusions. 

\item[$\bullet$] The bolometric luminosity of COCONUTS-2b is computed using spectroscopically inferred $T_{\rm eff}$ and $R$, with the residuals between observations and fitted models propagated. This leads to $\log{(L_{\rm bol} / L_{\odot})} = -6.18$~dex, with a $0.5$~dex-wide range of $[-6.43, -5.93]$~dex. This range accounts for the scatter of predicted values across models with different assumptions. 

\item[$\bullet$] Physical properties of COCONUTS-2b are derived based on various thermal evolution models using its bolometric luminosity ($\log{(L_{\rm bol}/L_{\odot})} = [-6.43, -5.93]$~dex) and its parent star's age (150--800~Myr), including $T_{\rm eff} = 483^{+44}_{-53}$~K, $\log{(g)} = 4.19^{+0.18}_{-0.13}$~dex, $R = 1.11^{+0.03}_{-0.04}$~R$_{\rm Jup}$, $M = 8 \pm 2$~M$_{\rm Jup}$, and a companion-to-host mass ratio of $0.021 \pm 0.005$. 

{
\item[$\bullet$] The spectroscopically inferred physical parameters from several model grids differ from those derived by thermal evolution models. For instance, \texttt{Exo-REM} predicts a $6.5\sigma$ lower $\log{(g)}$ and a $25\sigma$ higher radius than the evolution-based estimates. Such discrepancies between atmospheric and evolution models have been reported for other gas-giant exoplanets and brown dwarfs, likely arising from the systematics of model assumptions.
}

\item[$\bullet$] {Multiple atmospheric model grids consistently suggest that COCONUTS-2b's atmosphere has sub- or near-solar [M/H] and C/O ratio. These properties provide insights into its formation and potential outward migration;} they also suggest that COCONUTS-2b was likely not influenced by core erosion or planetesimal accretion. {Furthermore, according to our most preferred model grid, \texttt{PH$_{3}$-free ATMO2020++}, the metallicity of COCONUTS-2b ([M/H]$= -0.152^{+0.012}_{-0.010}$~dex) is $2\sigma$ lower than that of its parent star ([M/H]$=0.00 \pm 0.08$~dex), reducing the likelihood of an in-situ, binary-like formation for this planetary-mass companion. However, our third-preferred model grid, \texttt{Exo-REM}, suggests that COCONUTS-2b and 2A have consistent atmospheric metallicities, supporting a binary-like formation pathway. }
\end{enumerate}

{Looking forward, upcoming JWST observations of COCONUTS-2b (GO programs 2124, 3514, and 6463) will provide higher precision constraints on its atmospheric properties, offering new insights into its disequilibrium chemistry, thermal structure, cloud properties, and atmospheric compositions, further illuminating the formation history of COCONUTS-2b.}

\begin{acknowledgments}
Z.Z. thanks Benjamin Charnay, {Ga\"{e}l Chauvin}, Jackie Faherty, Brianna Lacy, and Eliot Vrijmoet for the helpful discussions. Figure~\ref{fig:context} is inspired by Ga\"{e}l Chauvin's presentation during the conference ``Open Problems in the Astrophysics of Gas Giants'' in Chile in December 2023. Z.Z. acknowledges support from the NASA Hubble Fellowship grant HST-HF2- 51522.001-A. {T.J.D. acknowledges support from UKRI STFC AGP grant ST/W001209/1.} We acknowledge the use of the lux supercomputer at UC Santa Cruz, funded by NSF MRI grant AST 1828315. {For open access, the author has applied a Creative Commons Attribution (CC BY) license to any Author Accepted Manuscript version arising from this submission.} This work is based on observations obtained at the international Gemini Observatory, a program of NSF NOIRLab, which is managed by the Association of Universities for Research in Astronomy (AURA) under a cooperative agreement with the U.S. National Science Foundation on behalf of the Gemini Observatory partnership: the U.S. National Science Foundation (United States), National Research Council (Canada), Agencia Nacional de Investigaci\'{o}n y Desarrollo (Chile), Ministerio de Ciencia, Tecnolog\'{i}a e Innovaci\'{o}n (Argentina), Minist\'{e}rio da Ci\^{e}ncia, Tecnologia, Inova\c{c}\~{o}es e Comunica\c{c}\~{o}es (Brazil), and Korea Astronomy and Space Science Institute (Republic of Korea). This work has benefitted from The UltracoolSheet,\footnote{\url{http://bit.ly/UltracoolSheet}} maintained by Will Best, Trent Dupuy, Michael Liu, Aniket Sanghi, Rob Siverd, and Zhoujian Zhang, and developed from compilations by \cite{2012ApJS..201...19D}, \cite{2013Sci...341.1492D}, \cite{2016ApJ...833...96L}, \cite{2018ApJS..234....1B}, \cite{2021AJ....161...42B}, \cite{2023ApJ...959...63S}, and \cite{2023AJ....166..103S}. This research has benefited from the SpeX Prism Library and the SpeX Prism Library Analysis Toolkit, maintained by Adam Burgasser.\footnote{\url{https://splat.physics.ucsd.edu/splat/}} {This research has made use of the NASA Exoplanet Archive, which is operated by the California Institute of Technology, under contract with the National Aeronautics and Space Administration under the Exoplanet Exploration Program.} This work used IRAF, as distributed by the National Optical Astronomy Observatory, which is operated by the Association of Universities for Research in Astronomy (AURA) under a cooperative agreement with the National Science Foundation. 
\end{acknowledgments}

\vspace{5mm}
\facilities{Gemini South (FLAMINGOS-2)}
\software{\texttt{PyMultiNest} \citep[][]{2014A&A...564A.125B}, \texttt{MultiNest} \citep[][]{2008MNRAS.384..449F, 2009MNRAS.398.1601F, 2019OJAp....2E..10F}, \texttt{corner.py} \citep[][]{corner}, \texttt{astropy} \citep{2013A&A...558A..33A, 2018AJ....156..123A, 2022ApJ...935..167A}, \texttt{ipython} \citep{PER-GRA:2007}, \texttt{numpy} \citep{2020NumPy-Array}, \texttt{scipy} \citep{2020SciPy-NMeth}, \texttt{matplotlib} \citep{Hunter:2007}.}

\appendix

\section{Comparison of the Gemini/F2 Spectroscopy and an Archival Spectrum}
\label{app:comp_archival}

\begin{figure*}[t]
\begin{center}
\includegraphics[width=7.in]{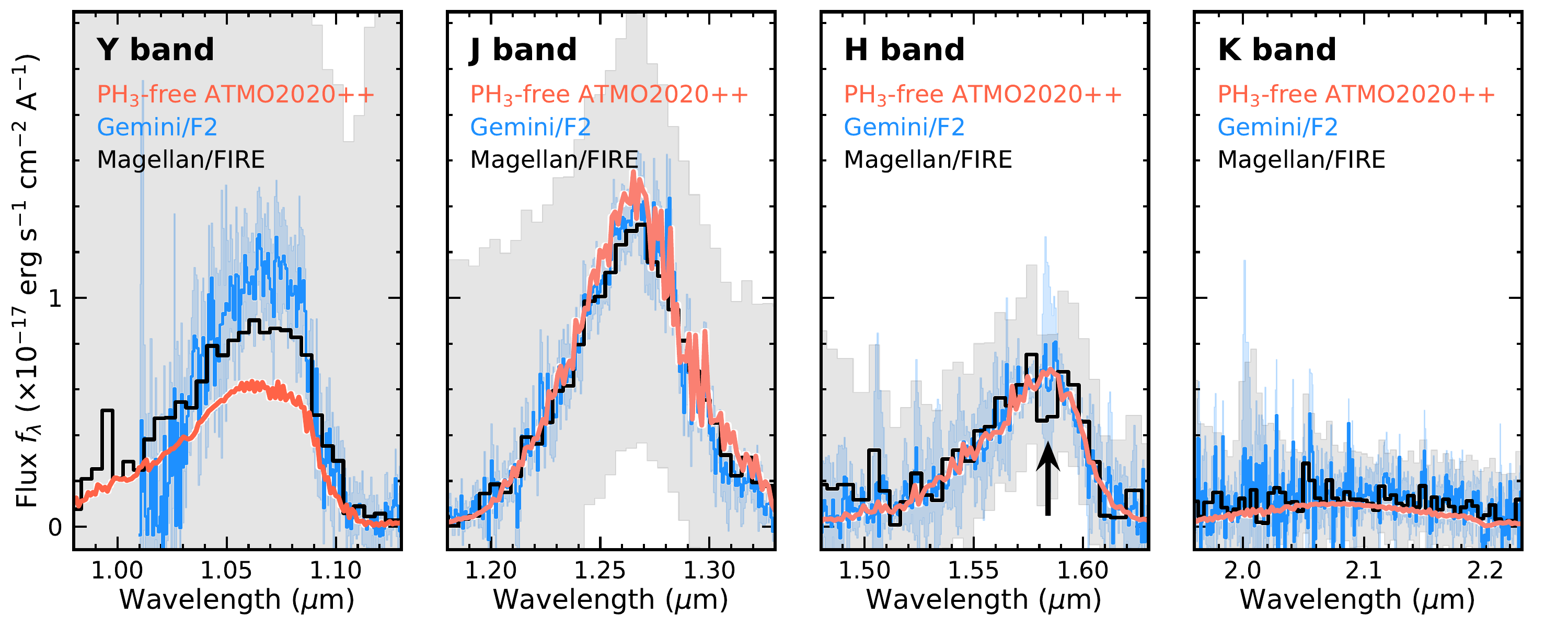}
\caption{Comparison of the Gemini/F2 spectrum of COCONUTS-2b (blue) with the archival Magellan/FIRE spectrum (black) across the $Y$, $J$, $H$, and $K$ bands, {with $1\sigma$ flux uncertainties highlighted by colored shades.} Both spectra are flux-calibrated using the same $J$-band magnitude (see Section~\ref{sec:observation}). The maximum-likelihood \texttt{PH$_{3}$-free ATMO2020++} model spectrum (orange), as shown in Figure~\ref{fig:best_neq}, is overlaid for comparison. {The Magellan/FIRE spectrum exhibits an absorption feature in the $H$ band (marked by an arrow symbol), which is absent in the Gemini/F2 spectrum with a higher S/N and spectral resolution.}  }
\label{fig:f2_fire}
\end{center}
\end{figure*}

We compare our Gemini/F2 spectrum with an archival near-infrared spectrum obtained by \cite{2011ApJS..197...19K} on 2010 April 3 UT, before \cite{2021ApJ...916L..11Z} discovered that this previously thought free-floating brown dwarf is instead a planetary-mass companion of the young star COCONUTS-2A. The archival spectrum ($R\sim300$) was collected using the Folded-port Infrared Echellette \citep[FIRE;][]{2008SPIE.7014E..0US} mounted on the 6.5~m Magellan Baade Telescope. {The Magellan/FIRE spectrum has a S/N of $\approx 0.5$, $\approx 1.4$, $\approx 2.0$, and $\approx 1.0$ per pixel near the peak of the $Y$, $J$, $H$, and $K$ band, respectively; these values are greatly lower than the quality of our Gemini/F2 spectrum, which has an S/N of $\approx 6$ in $Y$ band, $\approx 10$ in $J$ band, $\approx 8$ in $H$ band, and $\approx 2$ in $K$ band.} As shown in Figure~\ref{fig:f2_fire}, the Gemini/F2 and Magellan/FIRE spectra of COCONUTS-2b are generally consistent.

{When comparing our Gemini/F2 spectrum with the median spectral fluxes of Magellan/FIRE, we find that the latter spectrum shows} slightly fainter fluxes near the peak of the $Y$ band (1.05--1.09~$\mu$m); {this difference is not significant given the flux uncertainties of the Magellan/FIRE spectrum. We also note that} the $Y$-band fluxes of the Gemini/F2 spectrum over the wavelength range of 1.05--1.09~$\mu$m remain stable despite variations in data reduction setups. Extracting COCONUTS-2b's Gemini/F2 spectra using the object's own spectral trace (which is not recommended for our target; see Section~\ref{sec:observation} and \citealt{2016ApJ...824....2L}), instead of using the standard star as a reference (as done in Section~\ref{sec:observation}), results in fainter $Y$-band fluxes by only $\sim 0.5\sigma$ over a narrow wavelength range of 1.07--1.09~$\mu$m. However, at wavelengths below $\approx 1.03$~$\mu$m, the Gemini/F2 spectral fluxes are likely affected by the comprised accuracy of the sensitivity function near the detector edge. Additionally, in the $H$ band, the archival spectrum exhibits an absorption feature {near 1.58~$\mu$m that} is absent in the Gemini/F2 spectrum, which has a higher S/N and spectral resolution. {In summary, our Gemini/F2 spectrum provides a higher-quality set of near-infrared spectroscopy for COCONUTS-2b compared to the archival data. }

\section{$\chi^{2}$-based Spectral Fits with the \texttt{LB23 Cloudy} Models}
\label{app:lb23_cloudy}

We compare the Gemini/F2 spectrum of COCONUTS-2b with all the \texttt{LB23 Cloudy CEQ} and \texttt{NEQ} models over the wavelengths of 1.05--1.12~$\mu$m, 1.18--1.35~$\mu$m, 1.45--1.66~$\mu$m, and 1.95--2.30~$\mu$m (Section~\ref{subsec:methods}). A $\chi^{2}$ value of each model is computed by comparing the observed fluxes with the model spectrum scaled by a factor of $\alpha$, as defined in Equation~2 of \cite{2024ApJ...961..210P}. The scaling factor corresponding to the best-fit model is then used to derive the planet's radius as $R = d \sqrt{\alpha}$, with $d = 10.888$~pc. 

\begin{figure*}[t]
\begin{center}
\includegraphics[width=7.in]{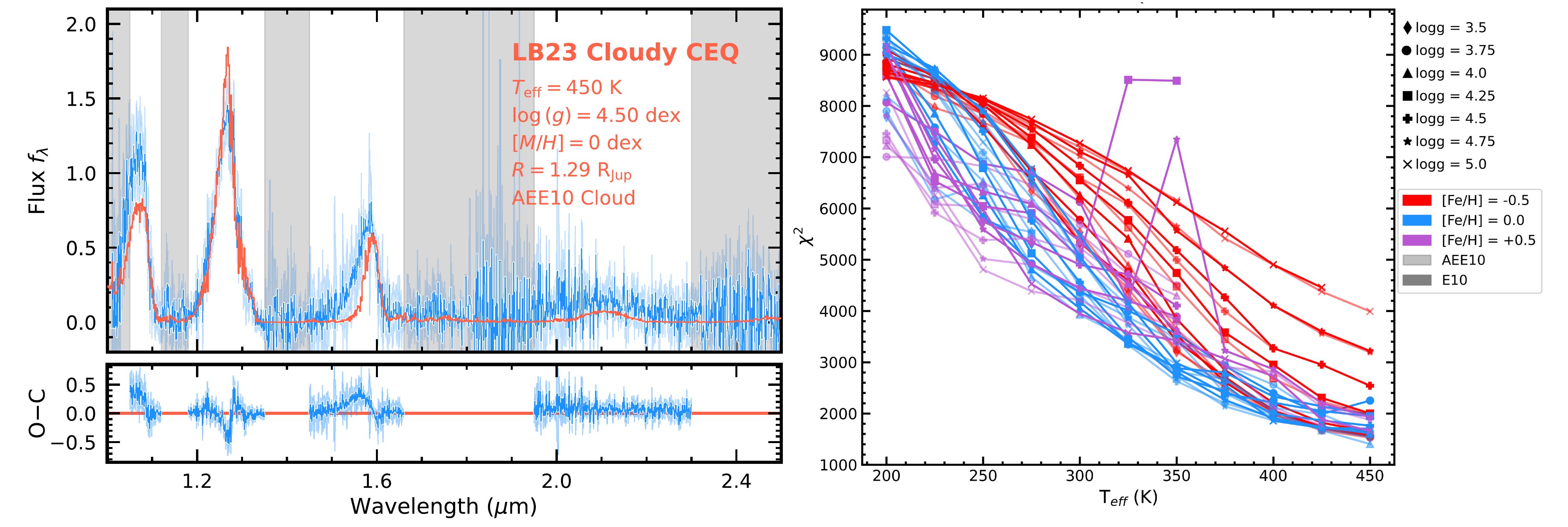}
\includegraphics[width=7.in]{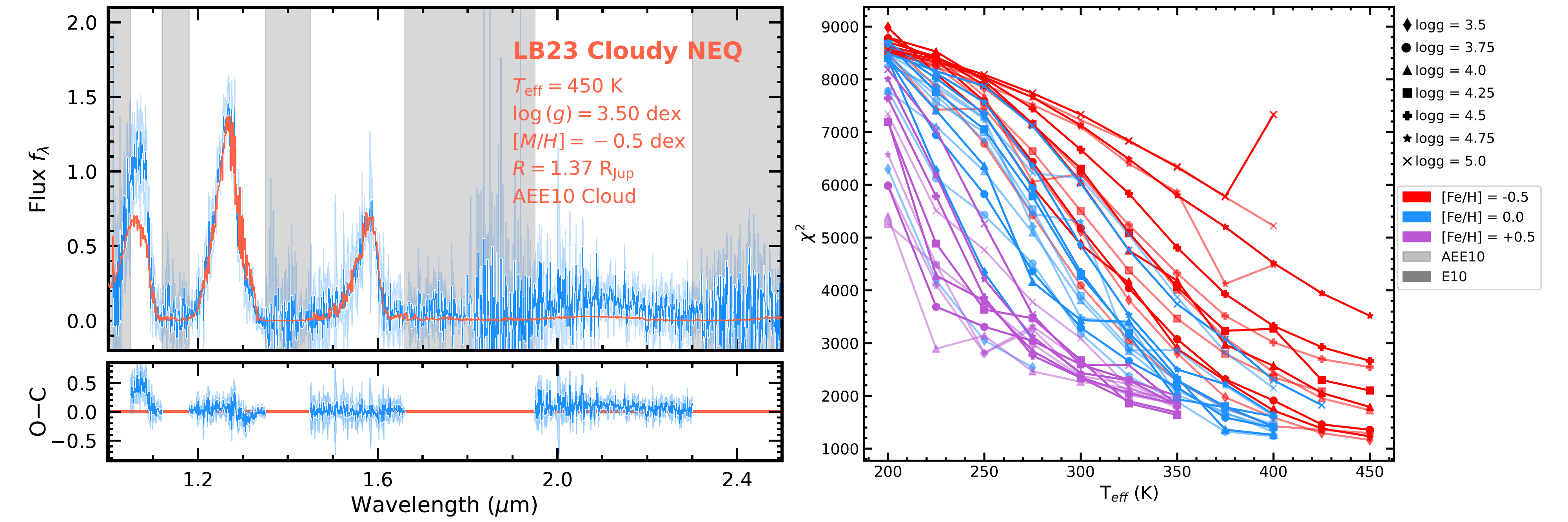}
\caption{{\it Top}: $\chi^{2}$-based spectral fitting results using the \texttt{LB23 Cloudy CEQ} grid. The top left panel presents the observed Gemini/F2 spectrum of COCONUTS-2b (blue) overlaid with the best-fit model spectrum (orange). The best-fitted physical parameters are labeled. The grey-shaded regions indicate the masked wavelength regions. The bottom left panel presents the data-model residuals. The right panel presents our computed $\chi^{2}$ for a given set of grid parameters, including $T_{\rm eff}$, $\log{(g)}$, [M/H], and cloud properties (AEE10 or E10). {\it Bottom}: Spectral fits with the \texttt{LB23 Cloudy NEQ} grid, with the same format as the top row. }
\label{fig:lb23_cloudy}
\end{center}
\end{figure*}

Figure~\ref{fig:lb23_cloudy} presents our results, including the best-fitted model spectra, data-model residuals, and the $\chi^{2}$ over the parameter space of each model grid. The $\chi^{2}$ values of these models decrease toward hotter effective temperatures, suggesting that COCONUTS-2b's $T_{\rm eff}$ is beyond the parameter space of these models. Additionally, the publicly available \texttt{LB23 Cloudy} models include two versions of clouds: ``AEE10'' and ``E10''. Both models assume a cloud particle size of 10~$\mu$m, but the former version assumes a more vertically extended cloud layer \citep[see Section~2.3 of][]{2023ApJ...950....8L}, which we find can better match the spectroscopy of COCONUTS-2b.

\section{Forward-modeling Analyses with Different Portions of the Spectrophotometry}
\label{app:different_input}

Figure~\ref{fig:noY} presents the results of our forward-modeling analysis with the $Y$-band Gemini/F2 spectrum fully masked. Figure~\ref{fig:w3} presents the results with the $W3$ photometry included. The confidence intervals of inferred parameters are summarized in Tables~\ref{tab:fm_params_key_runs_noY} and \ref{tab:fm_params_key_runs_w3}.

\begin{figure*}[t]
\begin{center}
\includegraphics[width=6.1in]{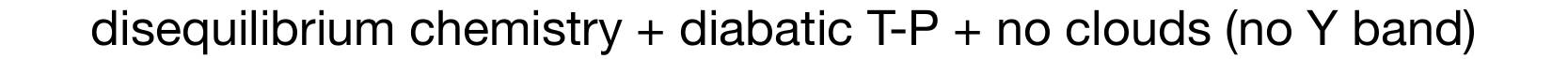}
\includegraphics[width=6.1in]{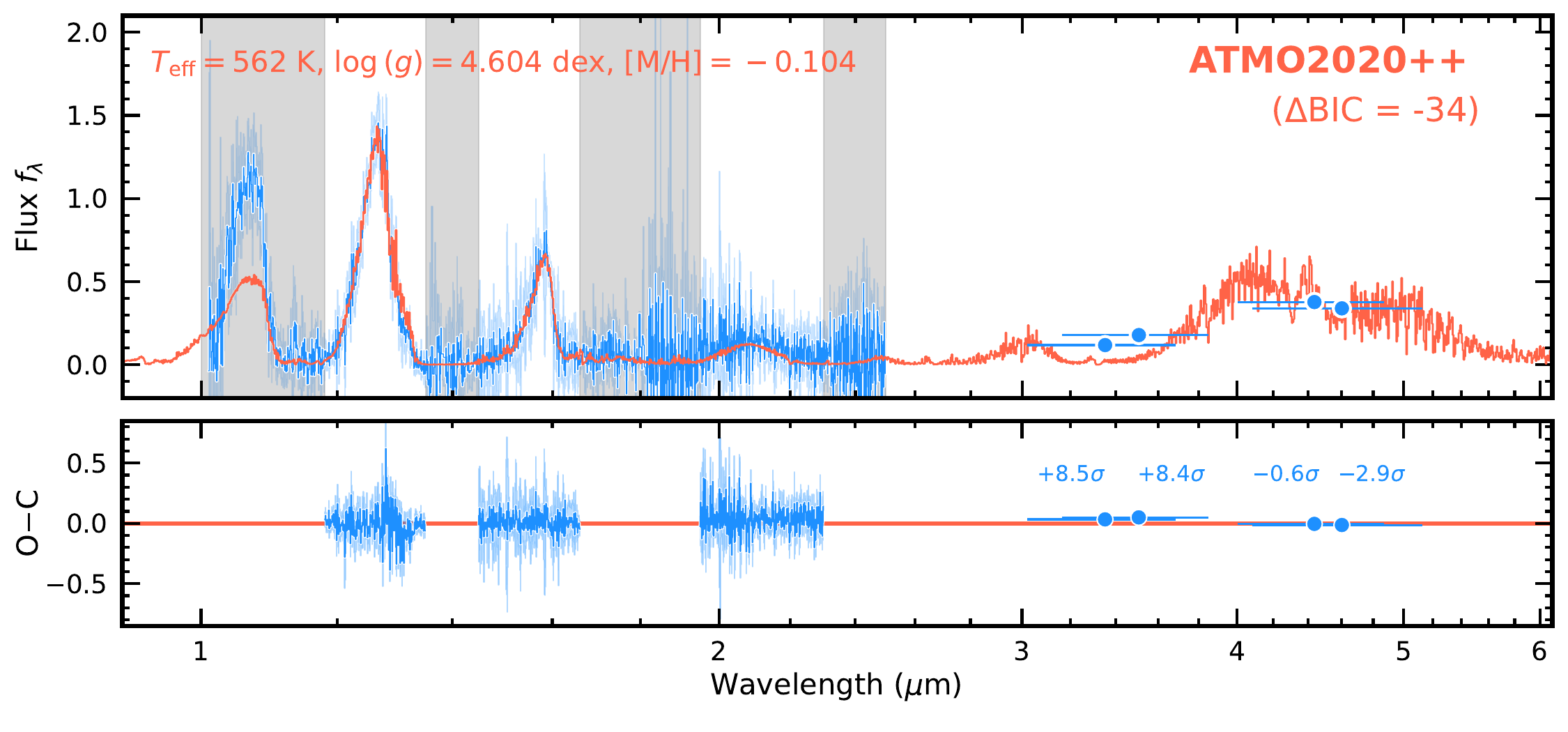}
\includegraphics[width=6.1in]{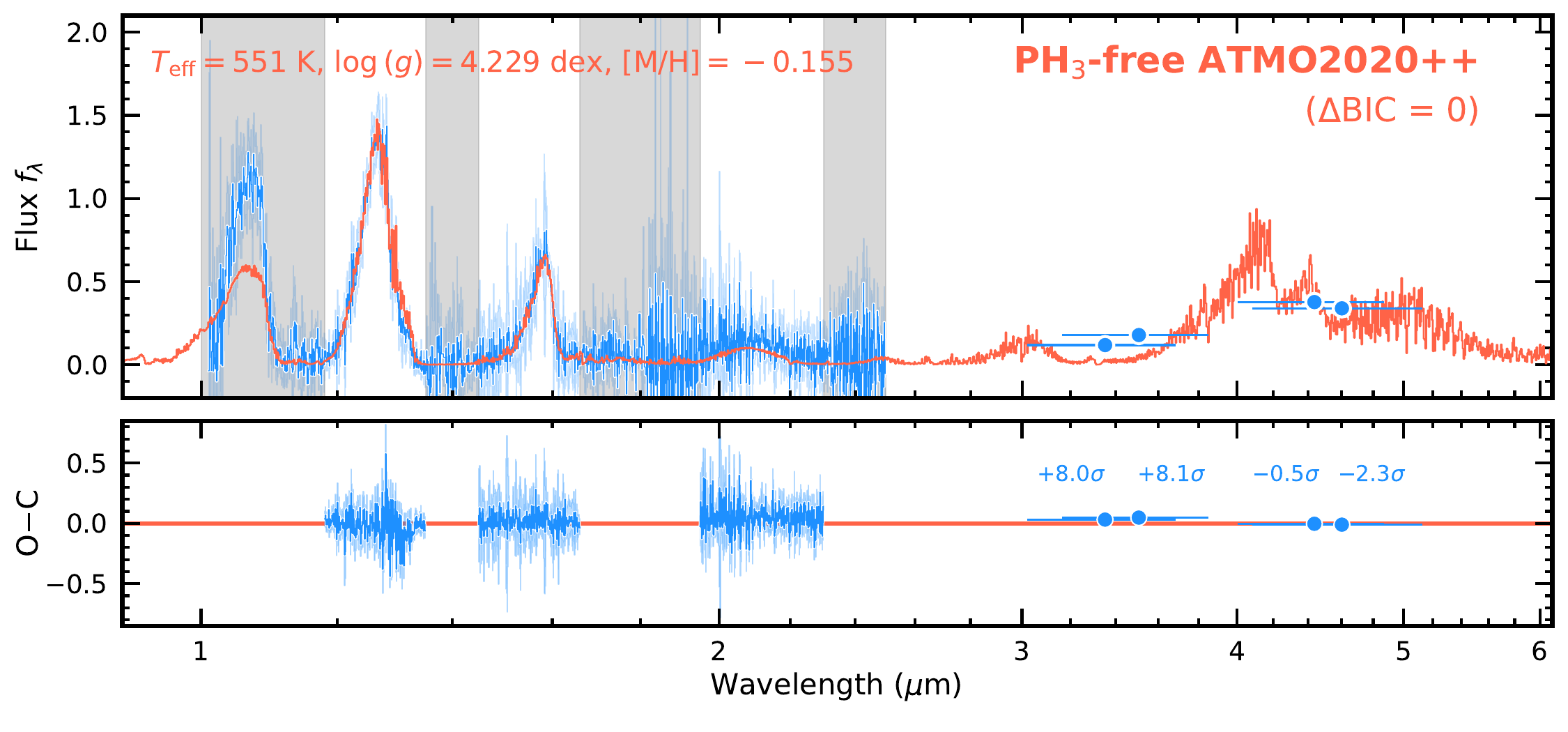}
\includegraphics[width=6.1in]{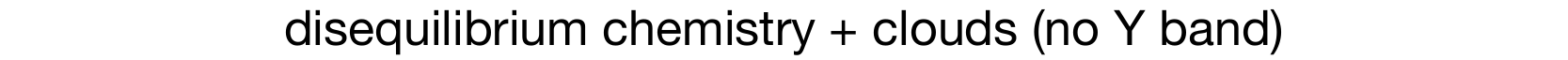}
\includegraphics[width=6.1in]{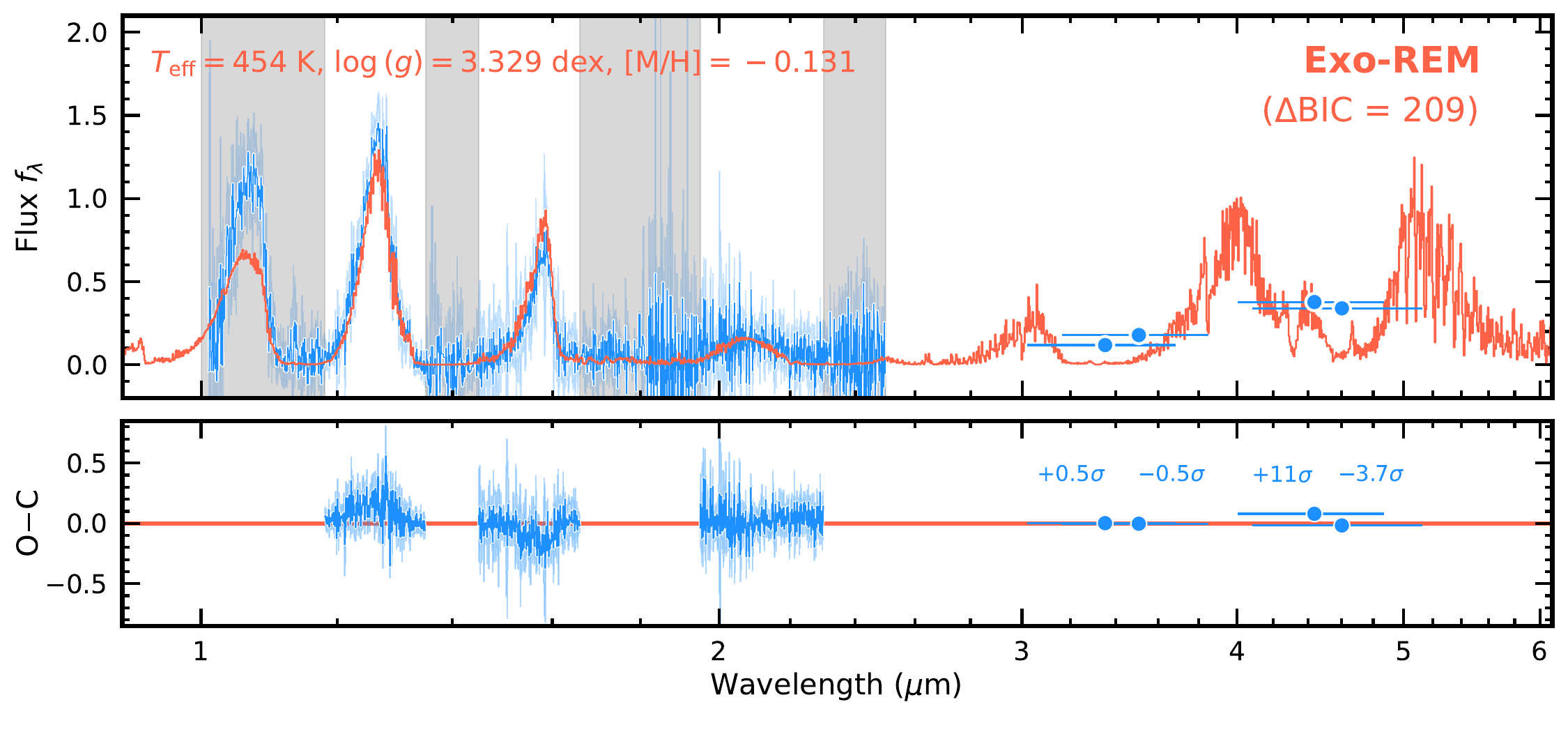}
\caption{Forward-modeling analysis results with the $Y$-band Gemini/F2 spectrum fully masked. The format is the same as Figure~\ref{fig:best_neq}. }
\label{fig:noY}
\end{center}
\end{figure*}

\renewcommand{\thefigure}{\arabic{figure}}
\addtocounter{figure}{-1}
\begin{figure*}[t]
\begin{center}
\includegraphics[width=6.1in]{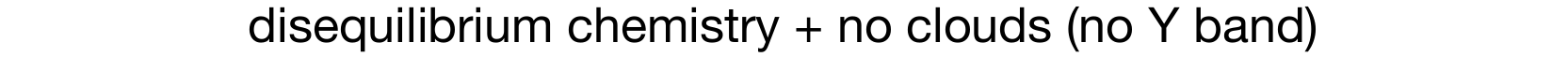}
\includegraphics[width=6.1in]{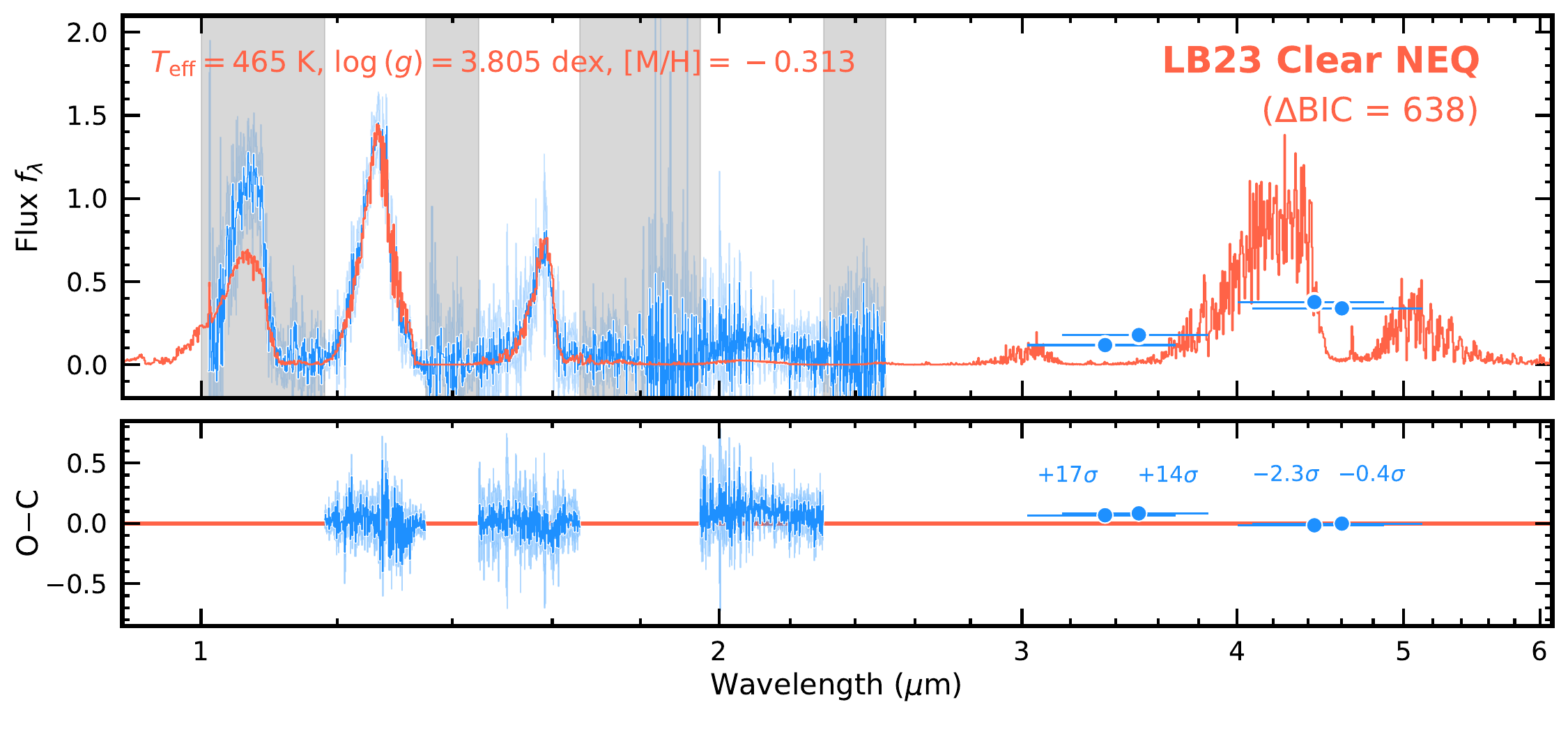}
\includegraphics[width=6.1in]{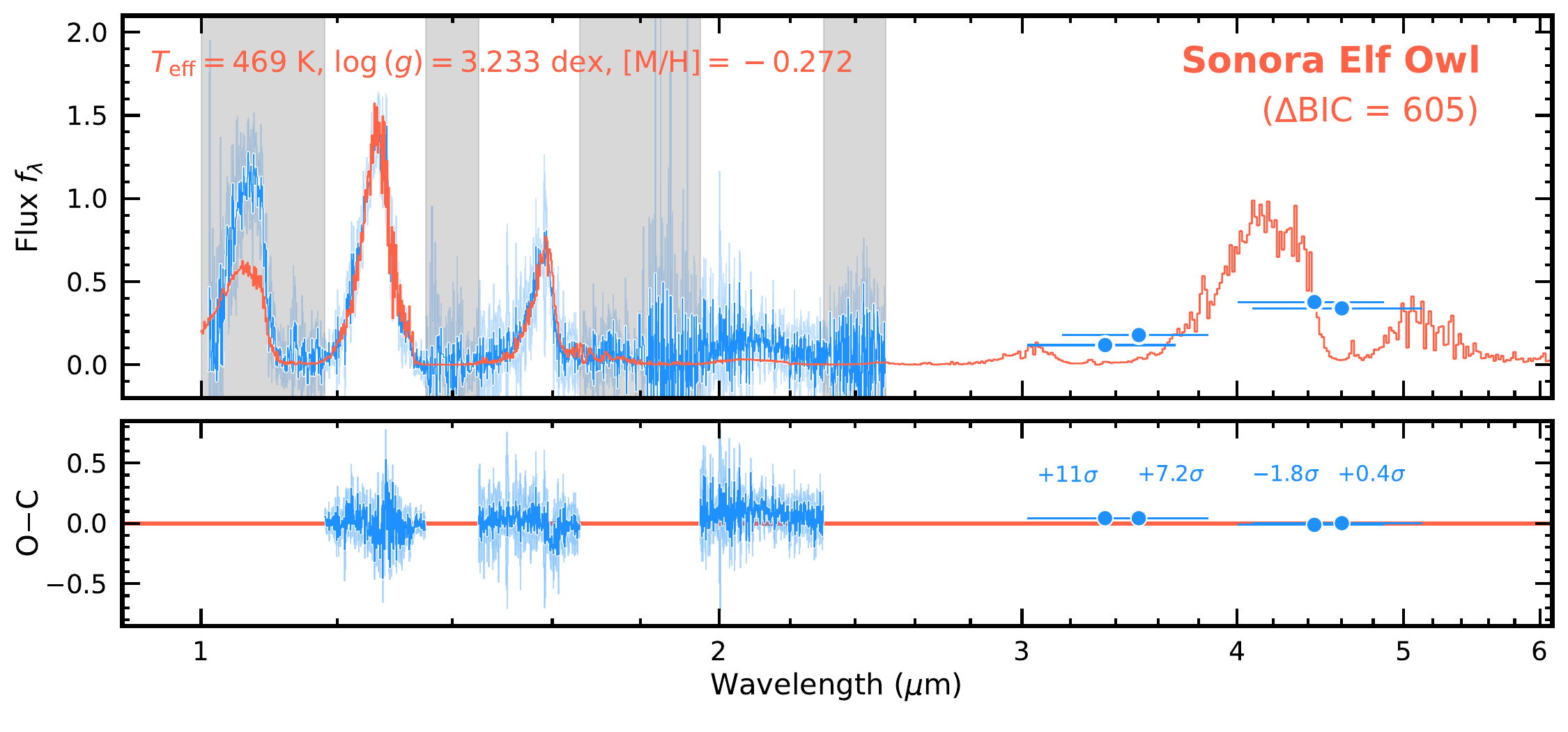}
\caption{Continued. }
\label{fig:noY}
\end{center}
\end{figure*}

\renewcommand{\thefigure}{\arabic{figure}}
\addtocounter{figure}{-1}
\begin{figure*}[t]
\begin{center}
\includegraphics[width=6.1in]{title_Clear_NEQ_noY.pdf}
\includegraphics[width=6.1in]{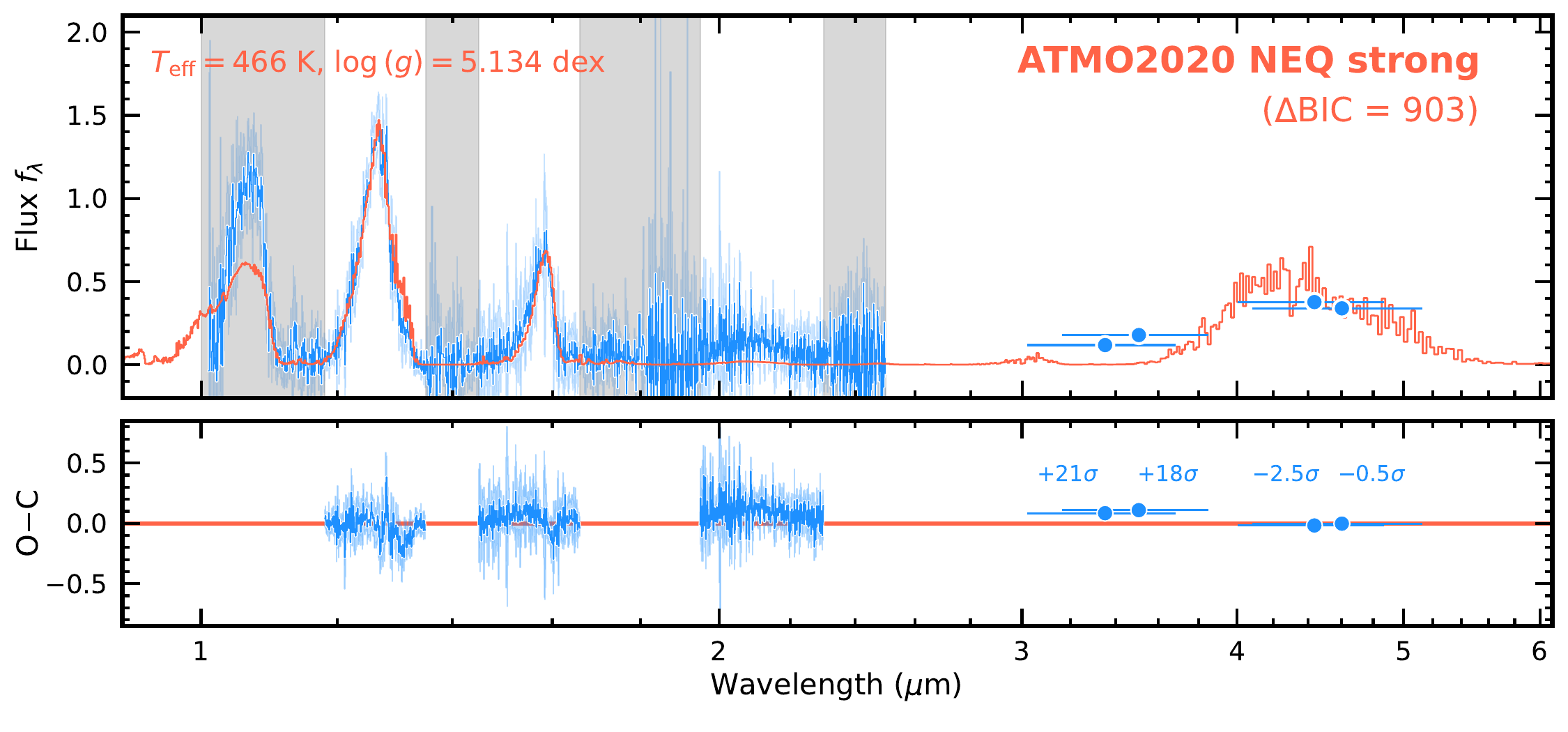}
\includegraphics[width=6.1in]{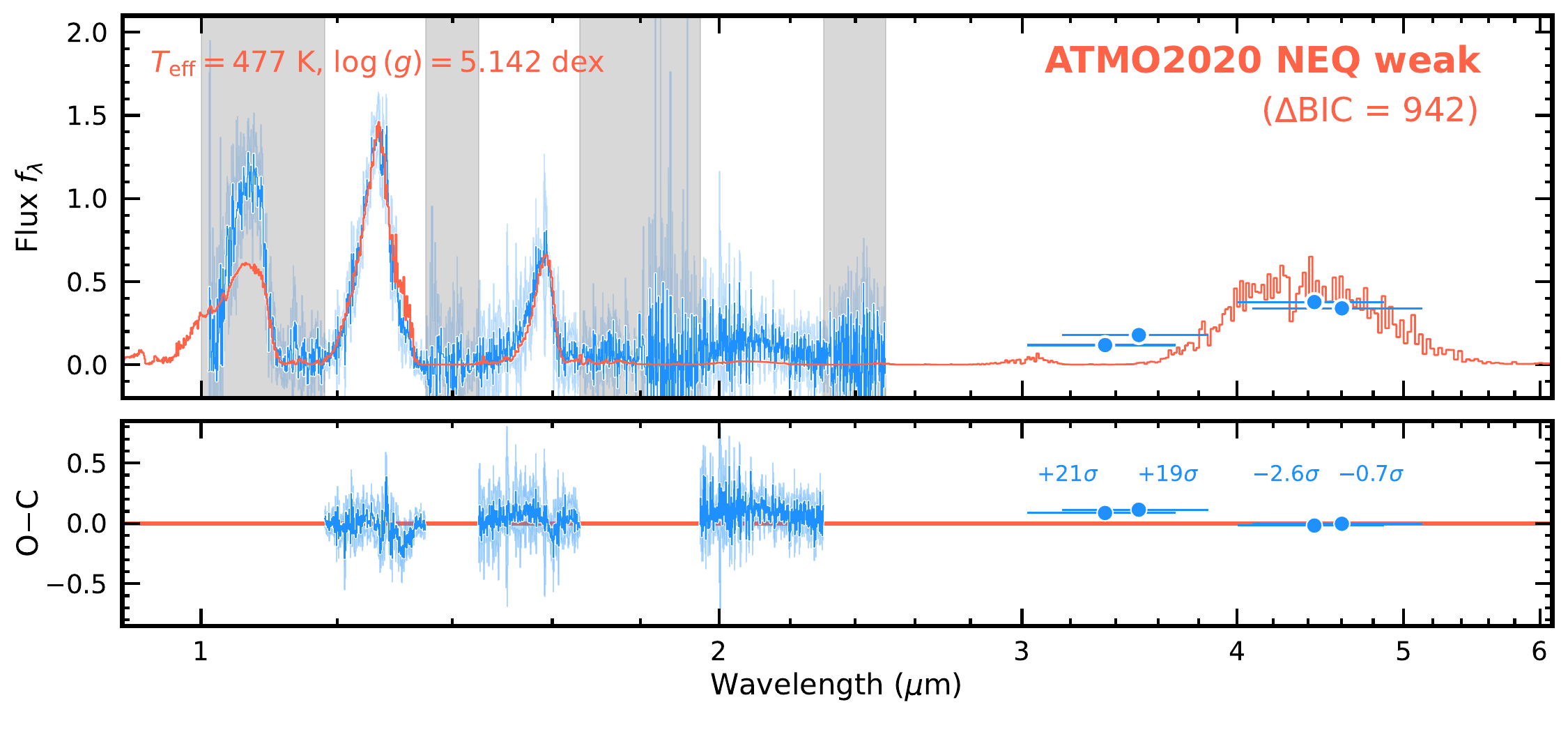}
\caption{Continued. }
\label{fig:noY}
\end{center}
\end{figure*}

\renewcommand{\thefigure}{\arabic{figure}}
\addtocounter{figure}{-1}
\begin{figure*}[t]
\begin{center}
\includegraphics[width=6.1in]{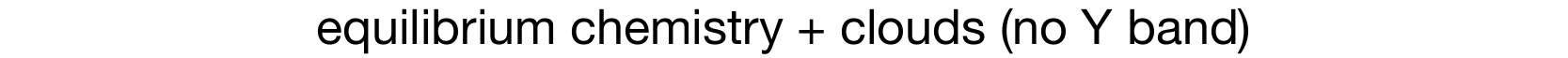}
\includegraphics[width=6.1in]{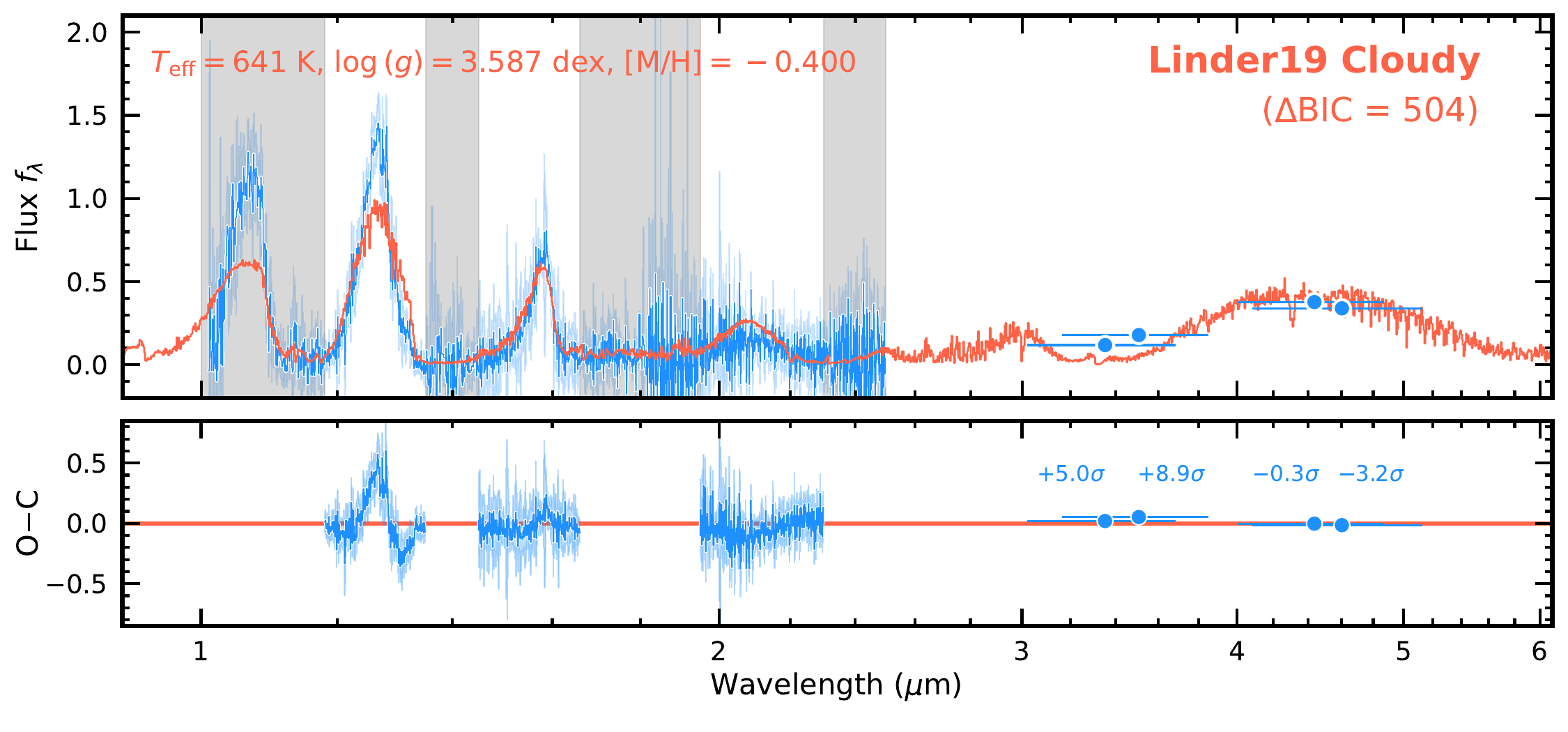}
\includegraphics[width=6.1in]{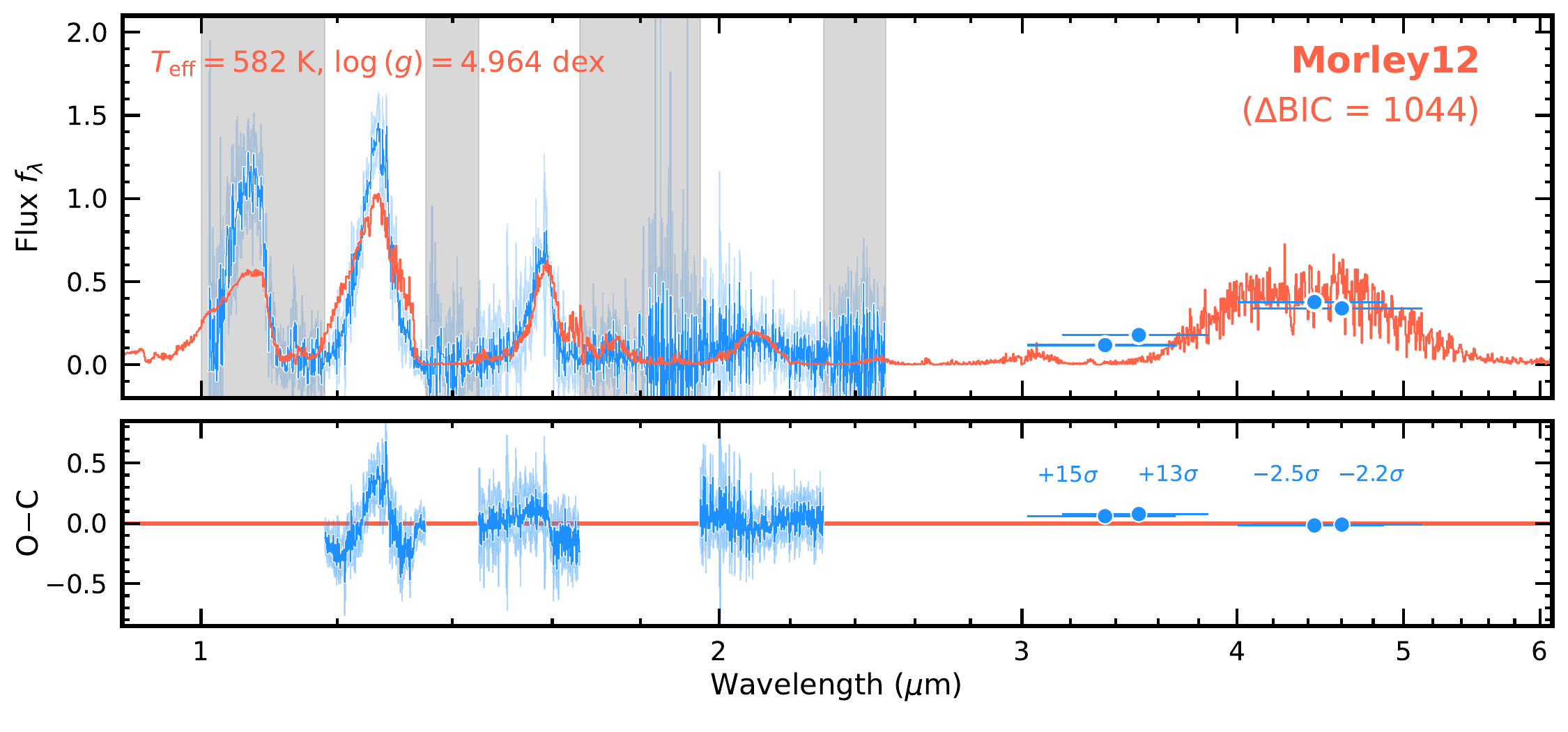}
\includegraphics[width=6.1in]{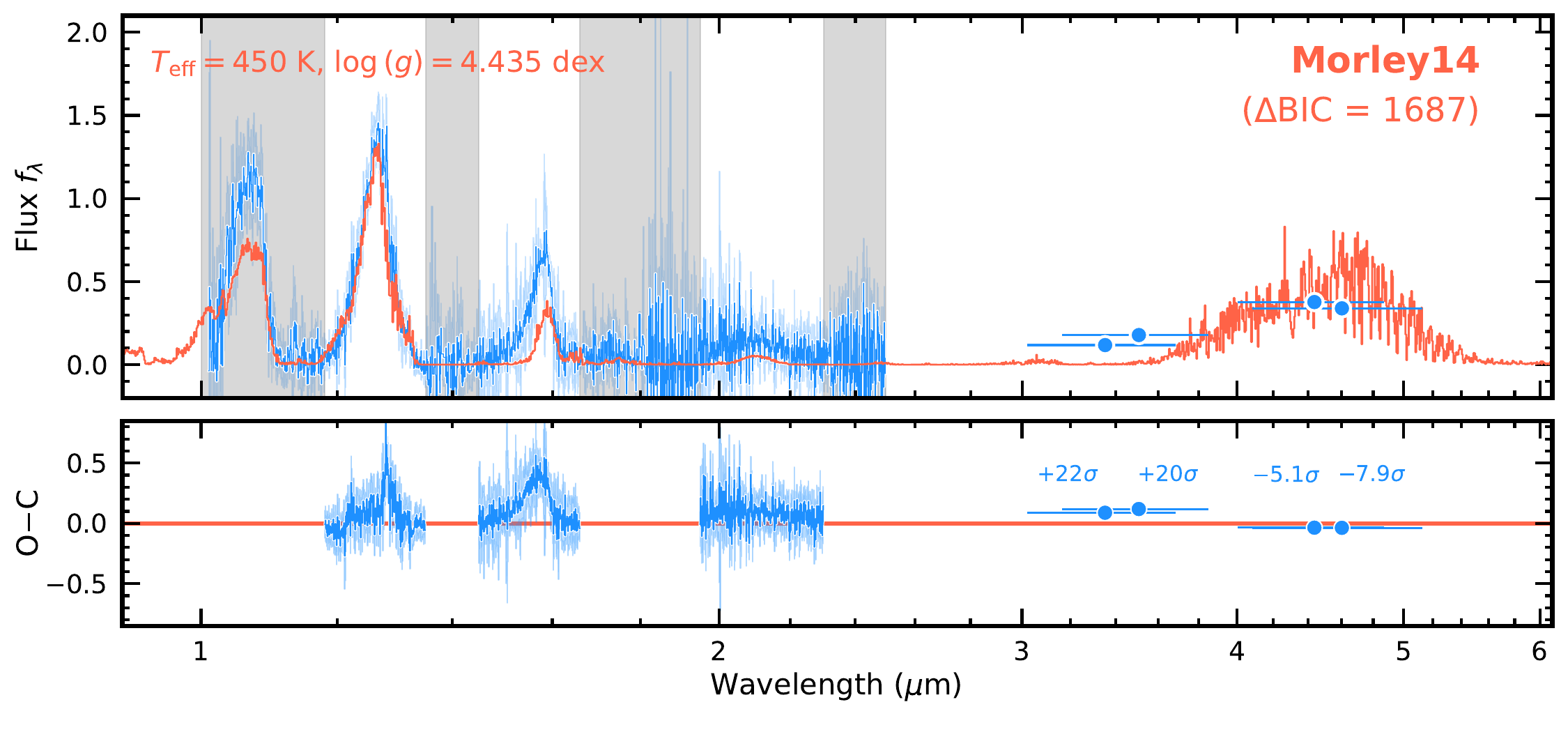}
\caption{Continued. }
\label{fig:noY}
\end{center}
\end{figure*}

\renewcommand{\thefigure}{\arabic{figure}}
\addtocounter{figure}{-1}
\begin{figure*}[t]
\begin{center}
\includegraphics[width=6.1in]{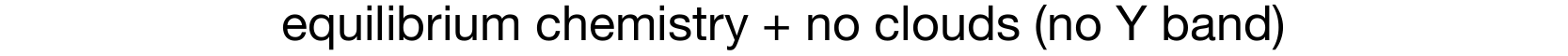}
\includegraphics[width=6.1in]{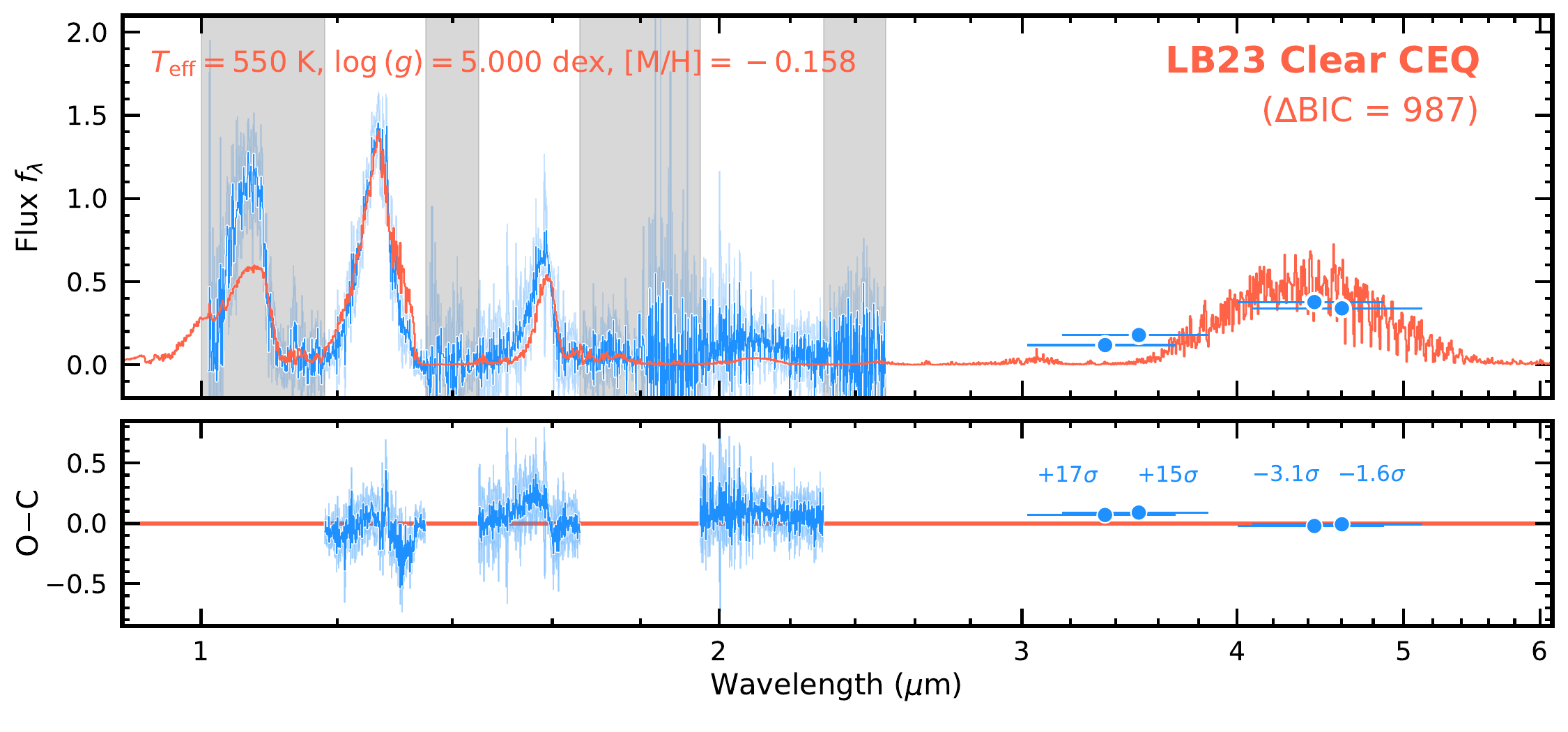}
\includegraphics[width=6.1in]{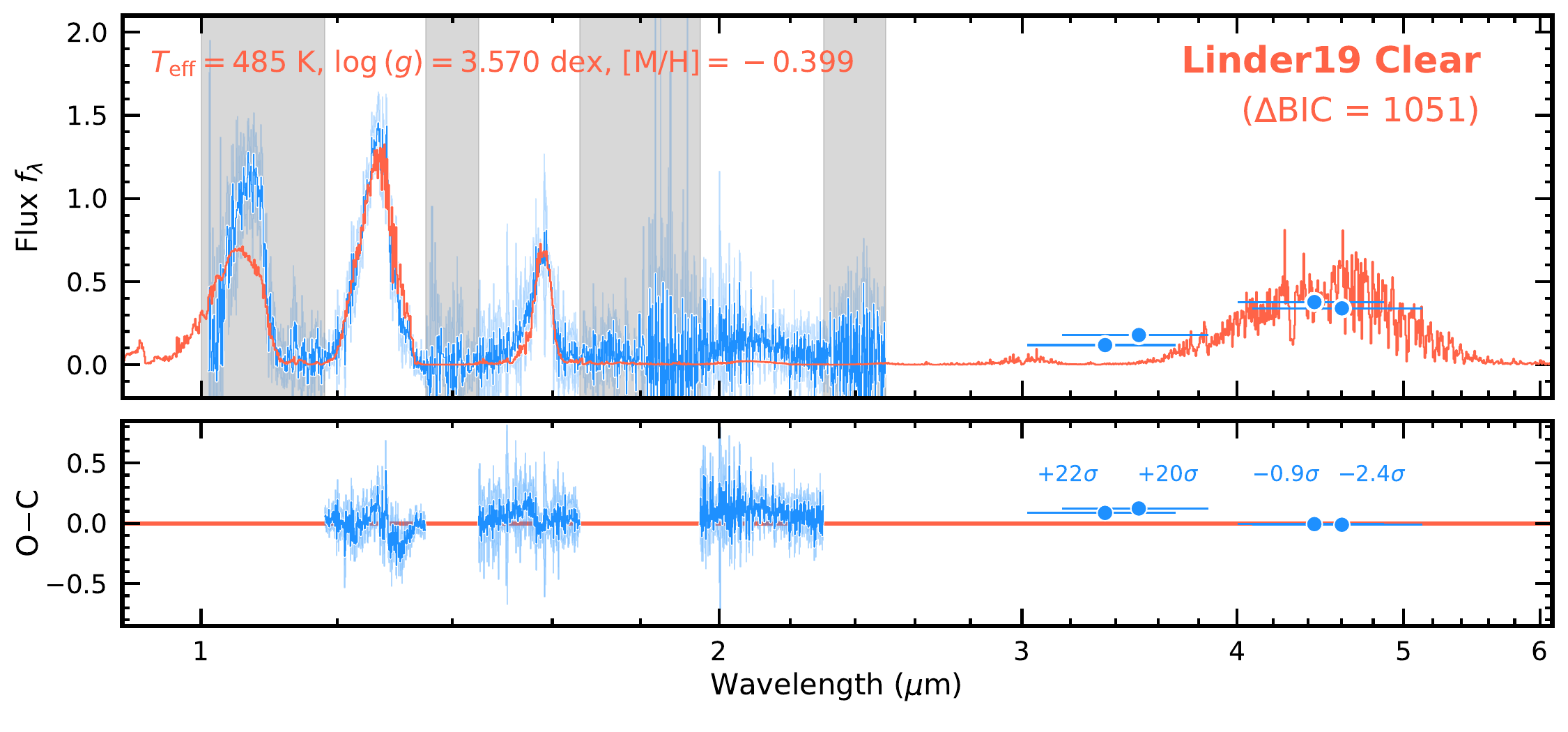}
\caption{Continued. }
\label{fig:noY}
\end{center}
\end{figure*}

\renewcommand{\thefigure}{\arabic{figure}}
\addtocounter{figure}{-1}
\begin{figure*}[t]
\begin{center}
\includegraphics[width=6.1in]{title_Clear_CEQ_noY.pdf}
\includegraphics[width=6.1in]{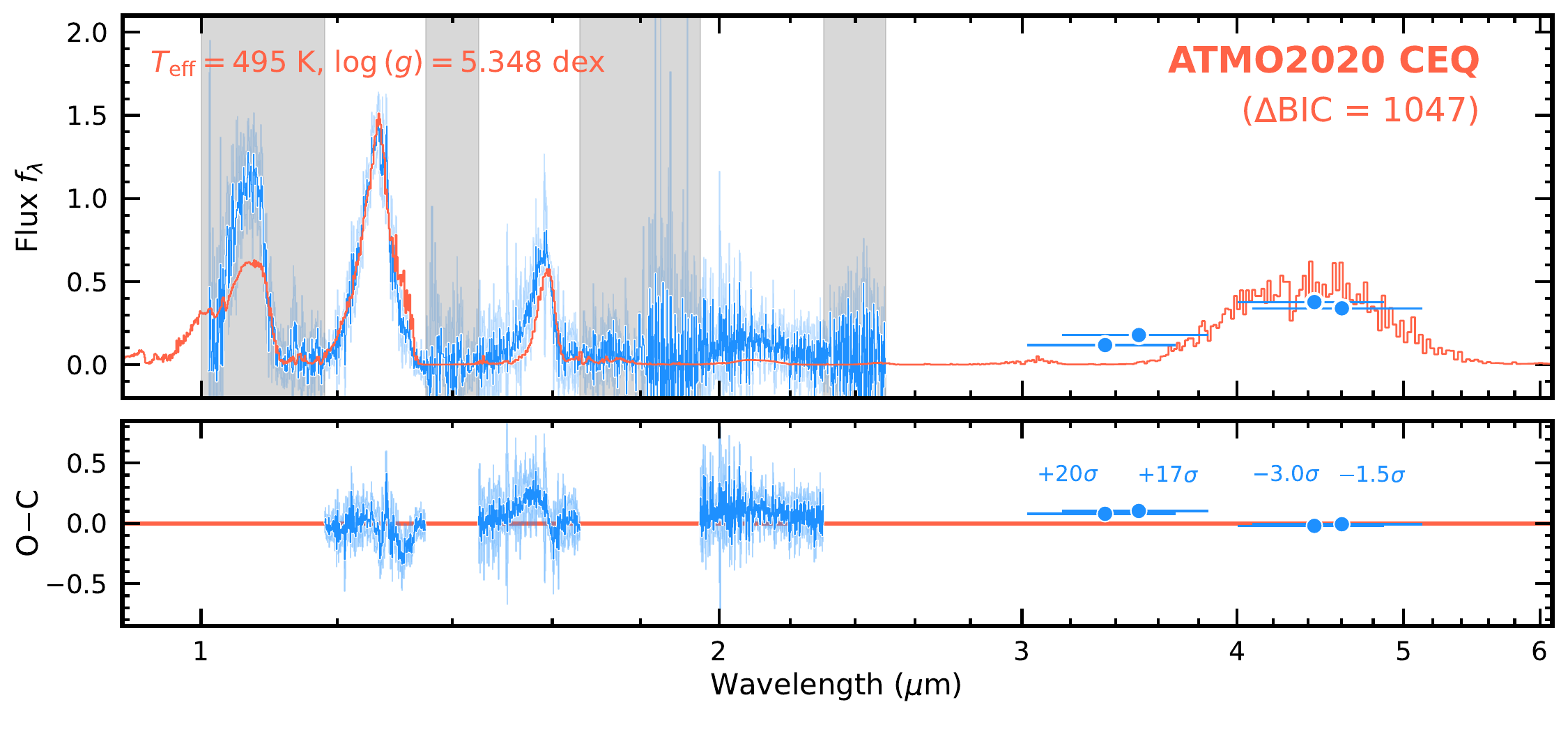}
\includegraphics[width=6.1in]{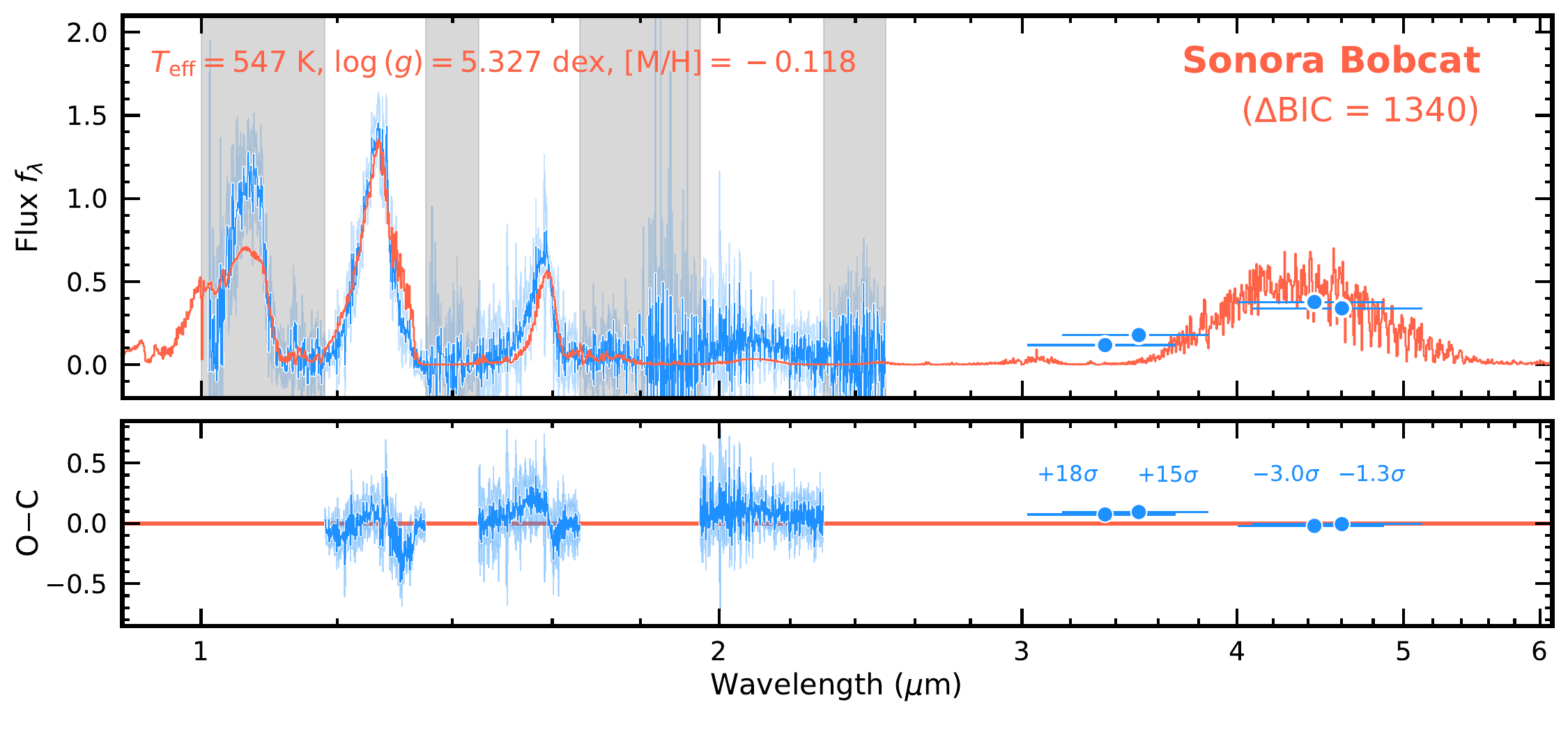}
\caption{Continued. }
\label{fig:noY}
\end{center}
\end{figure*}

\begin{figure*}[t]
\begin{center}
\includegraphics[width=6.1in]{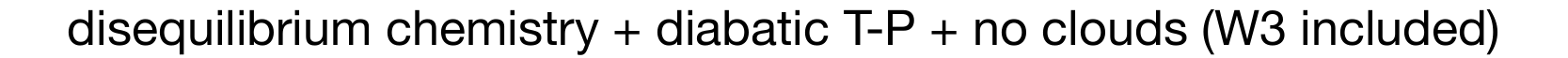}
\includegraphics[width=6.1in]{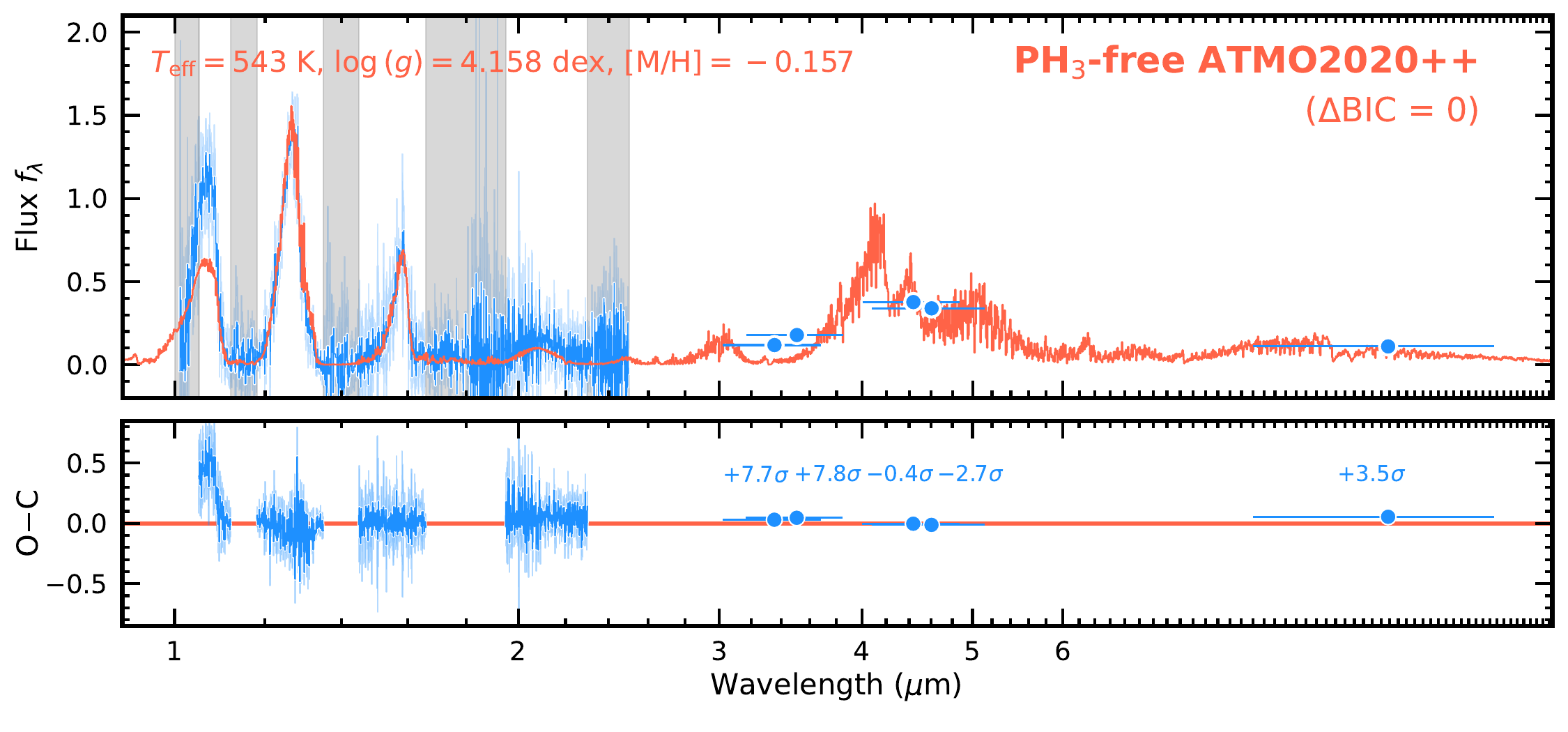}
\includegraphics[width=6.1in]{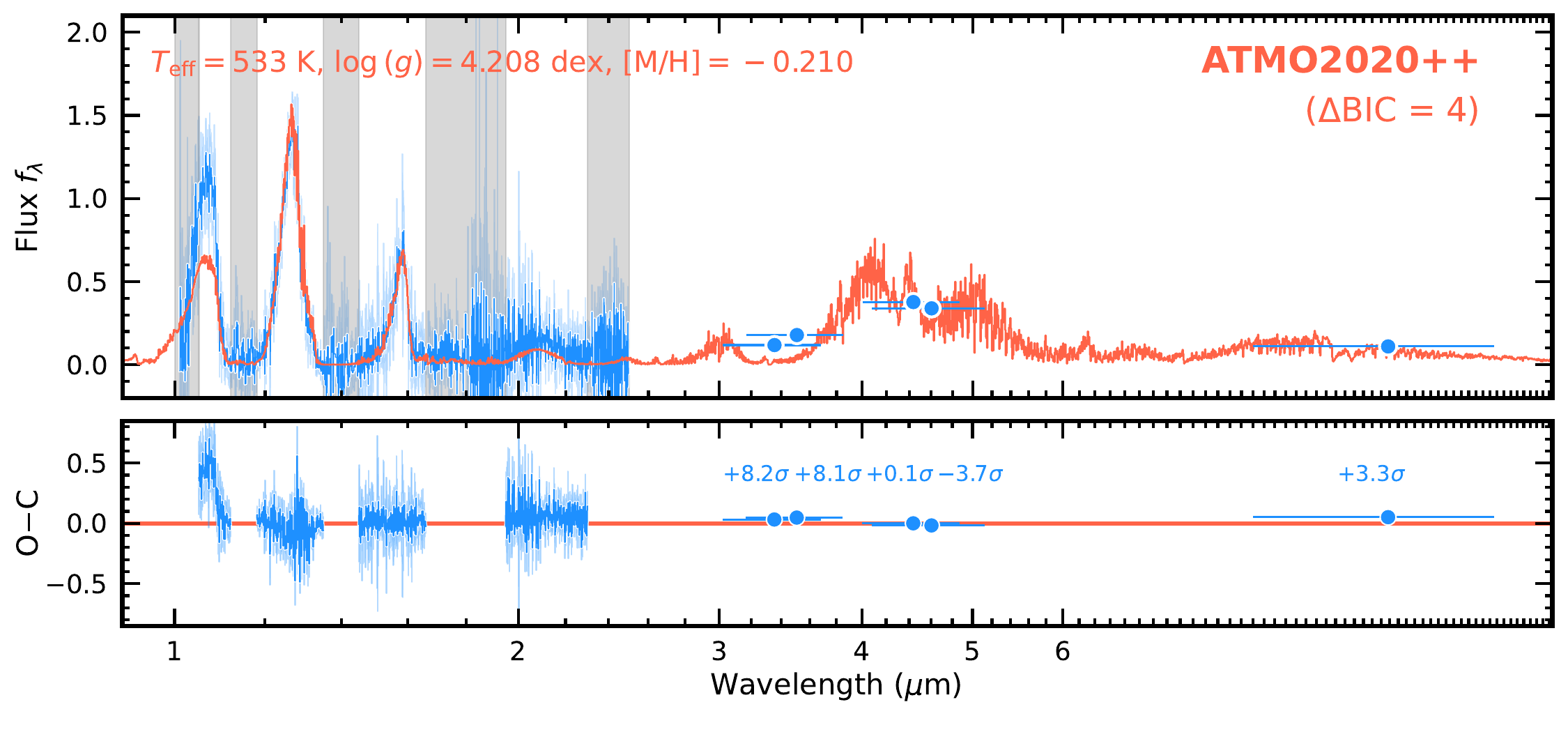}
\includegraphics[width=6.1in]{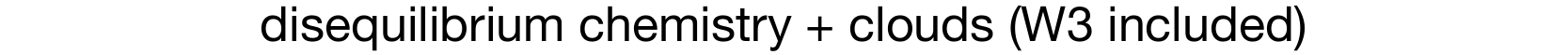}
\includegraphics[width=6.1in]{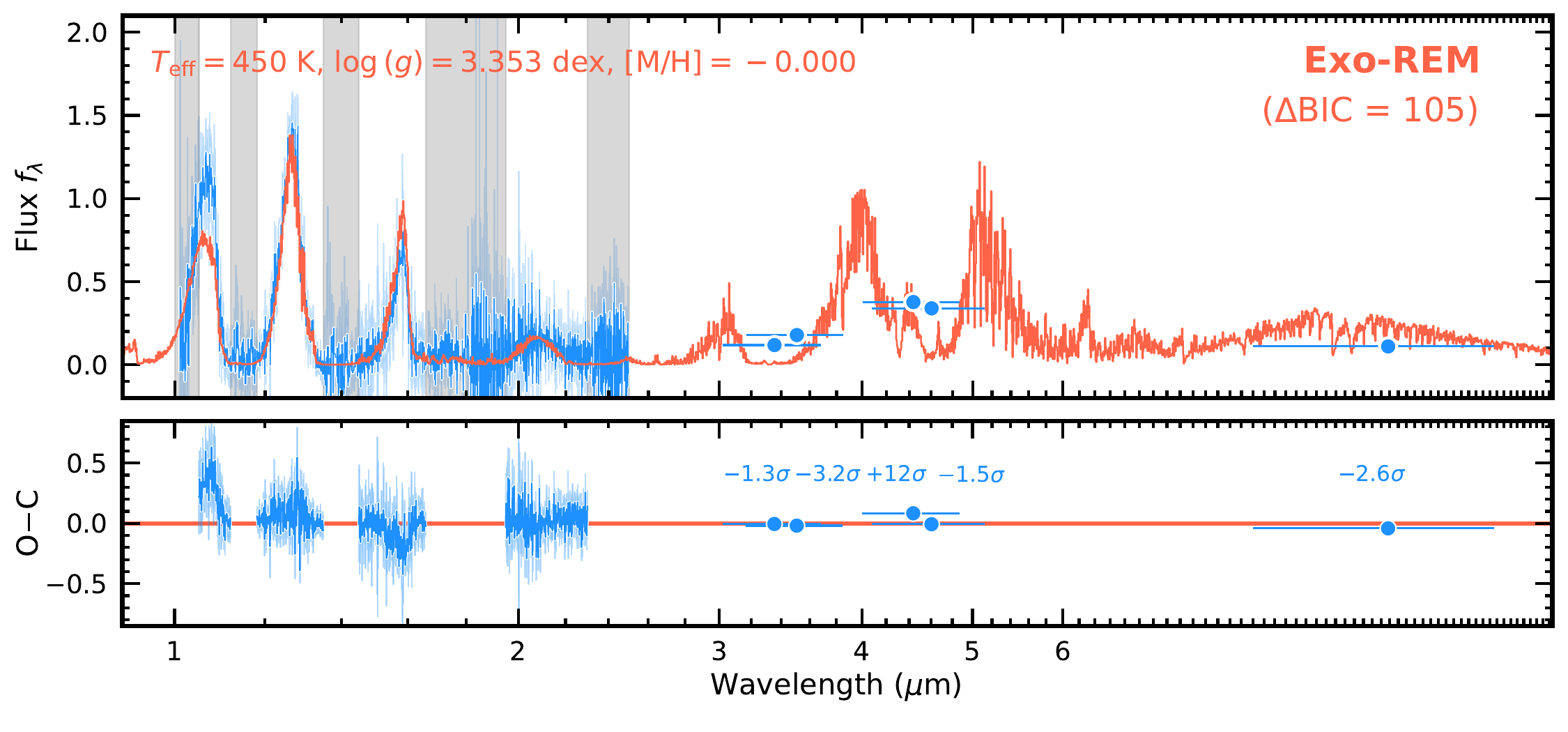}
\caption{Forward-modeling analysis results with the $W3$ photometry included. The format is the same as Figure~\ref{fig:best_neq}. }
\label{fig:w3}
\end{center}
\end{figure*}

\renewcommand{\thefigure}{\arabic{figure}}
\addtocounter{figure}{-1}
\begin{figure*}[t]
\begin{center}
\includegraphics[width=6.1in]{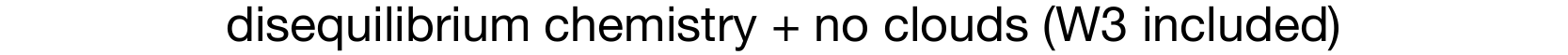}
\includegraphics[width=6.1in]{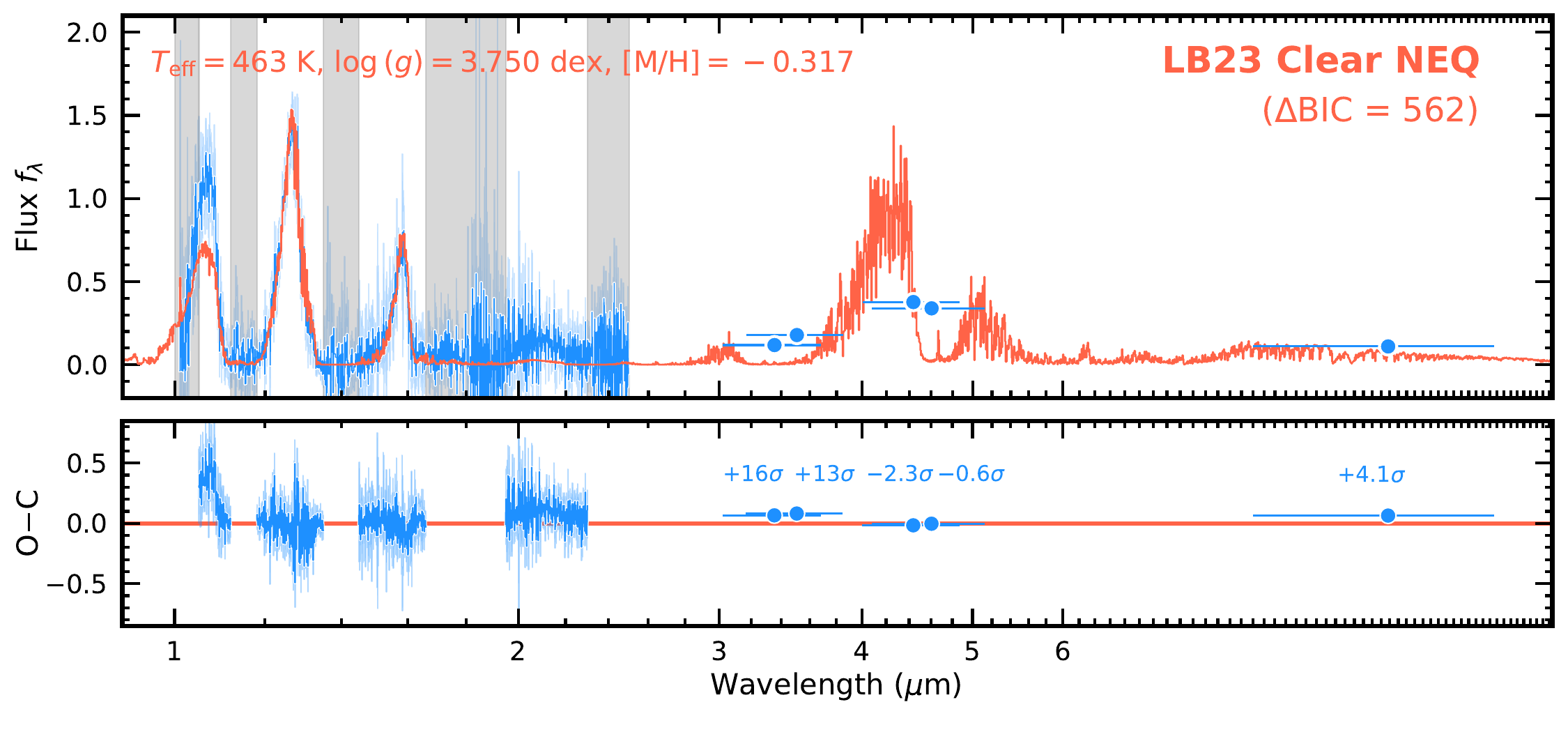}
\includegraphics[width=6.1in]{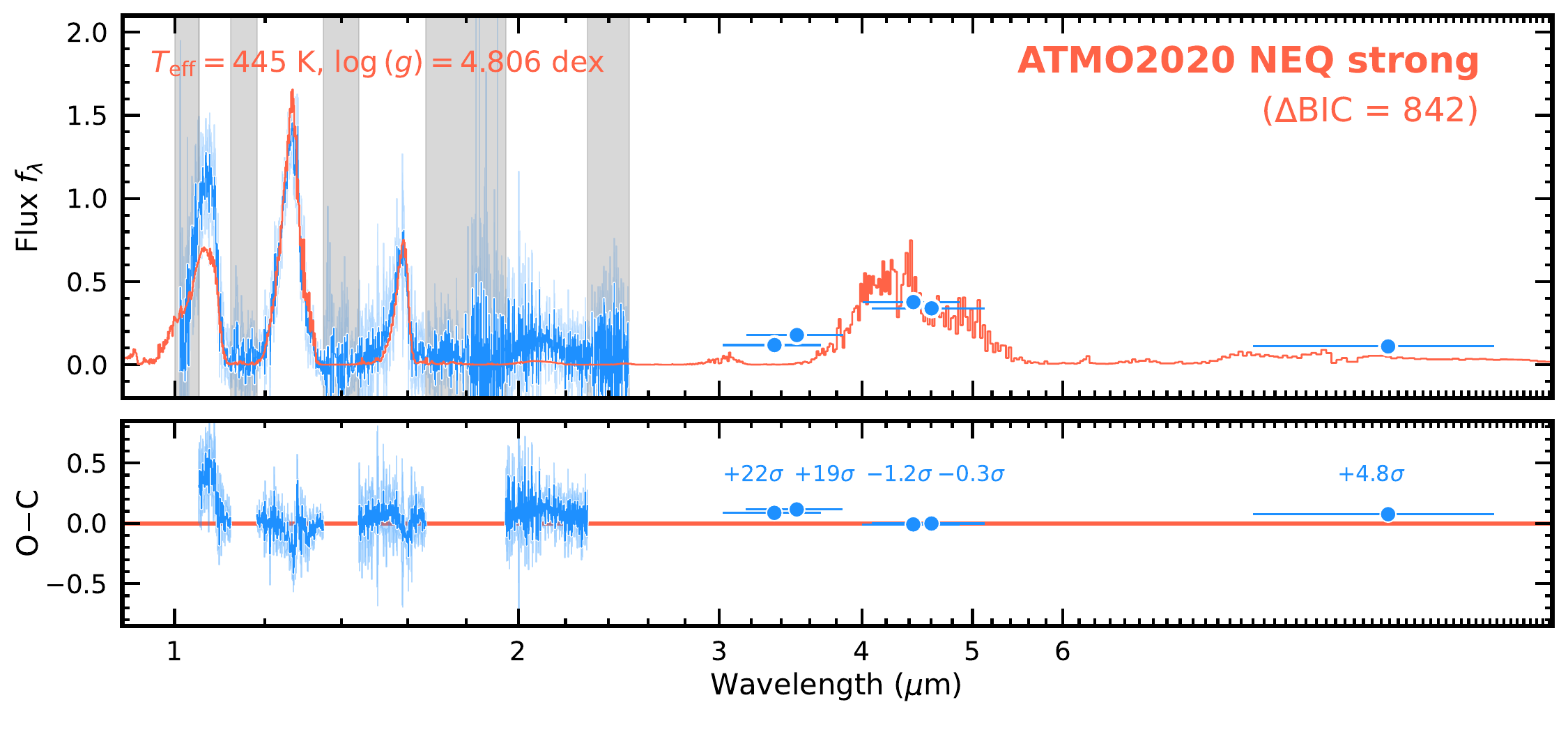}
\includegraphics[width=6.1in]{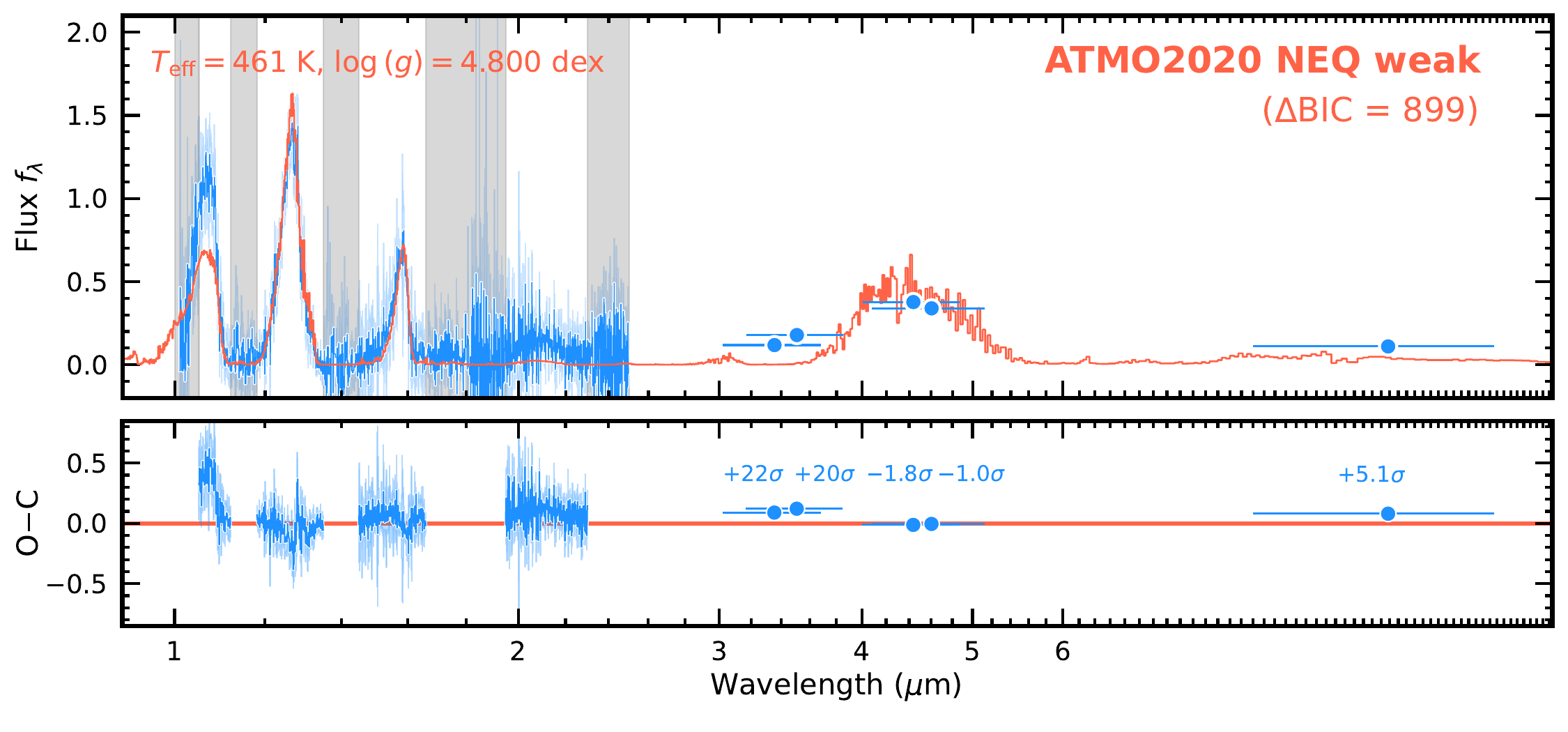}
\caption{Continued. }
\label{fig:w3}
\end{center}
\end{figure*}

\renewcommand{\thefigure}{\arabic{figure}}
\addtocounter{figure}{-1}
\begin{figure*}[t]
\begin{center}
\includegraphics[width=6.1in]{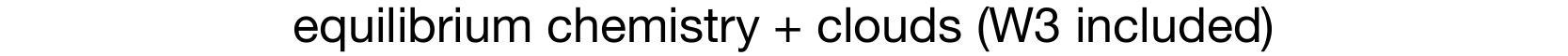}
\includegraphics[width=6.1in]{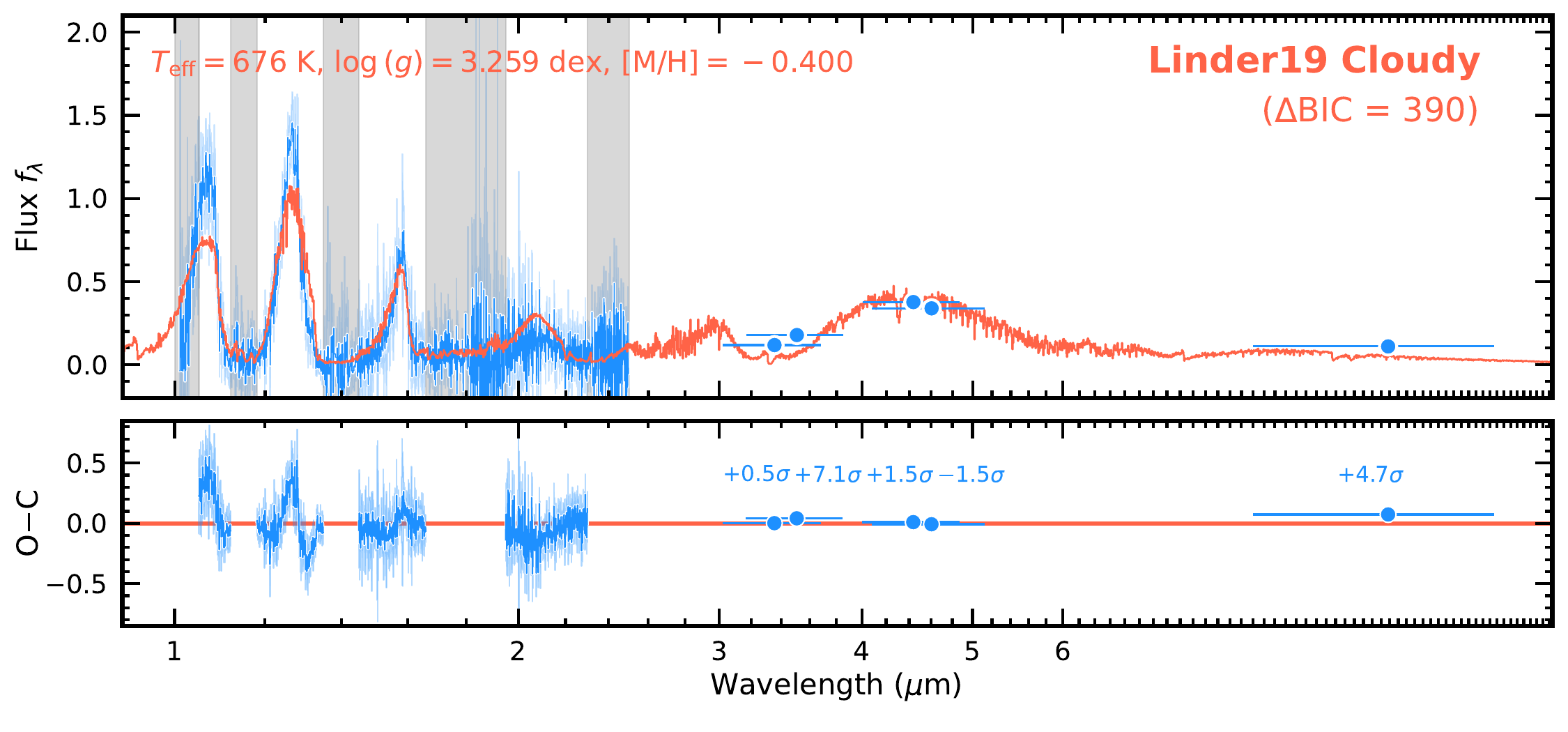}
\includegraphics[width=6.1in]{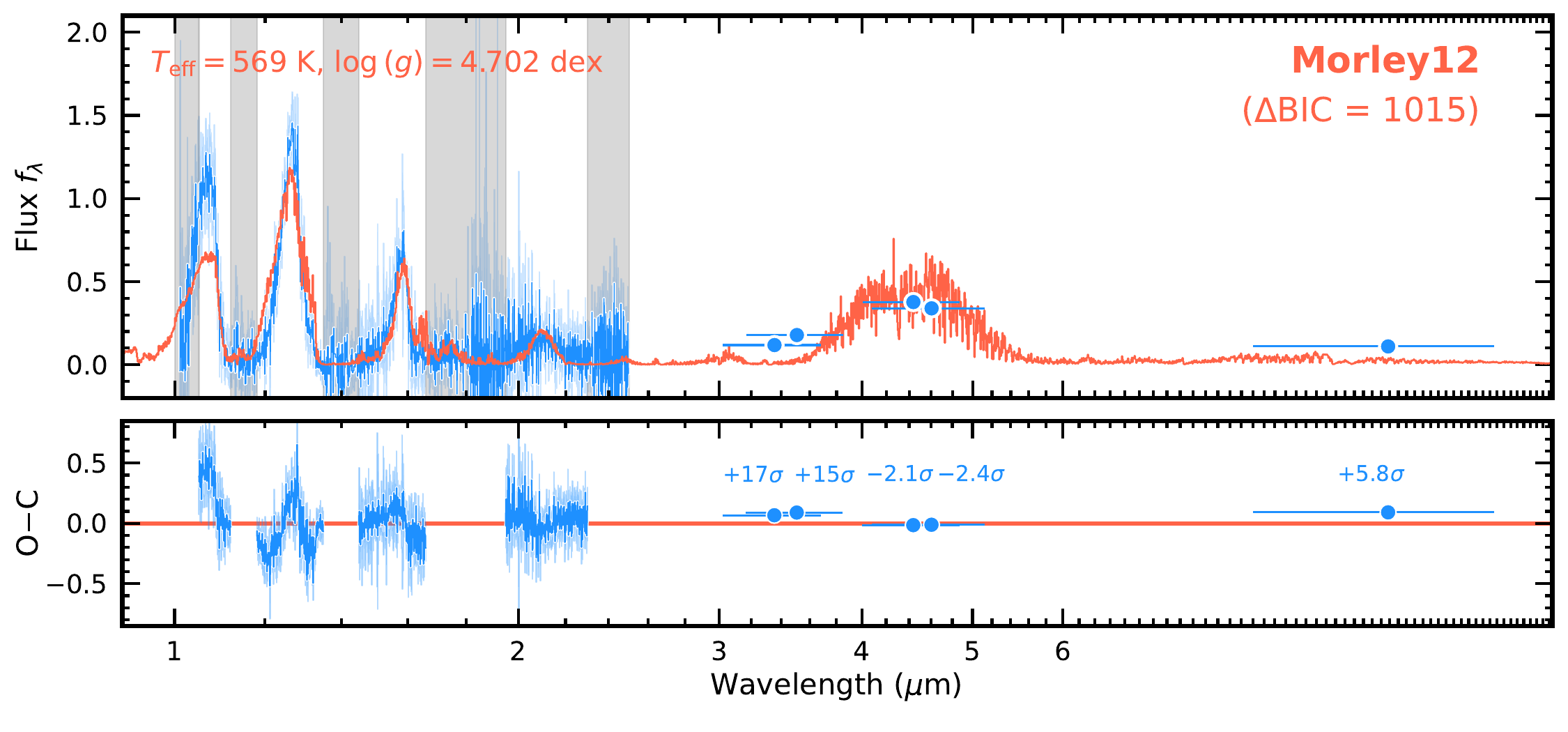}
\includegraphics[width=6.1in]{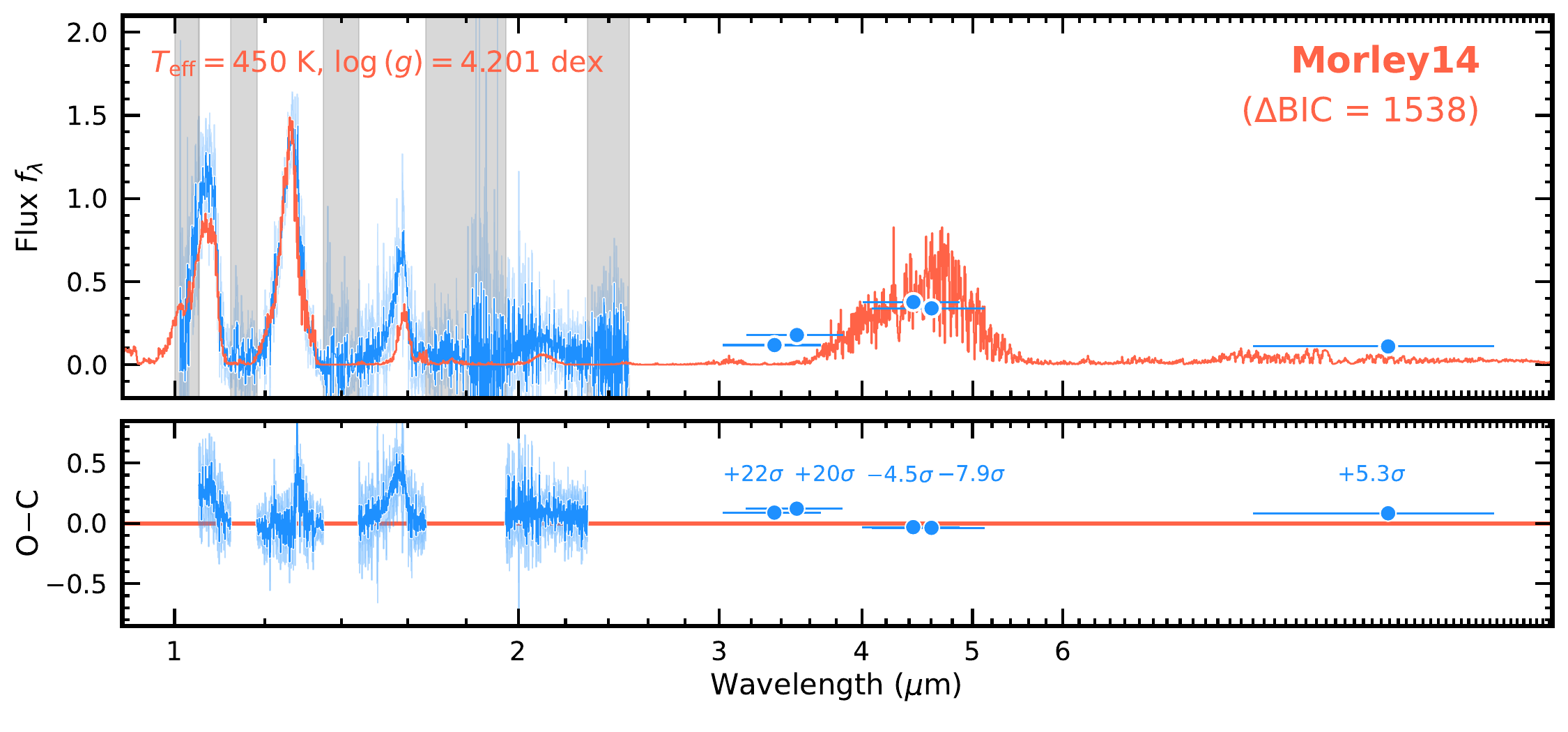}
\caption{Continued. }
\label{fig:w3}
\end{center}
\end{figure*}

\renewcommand{\thefigure}{\arabic{figure}}
\addtocounter{figure}{-1}
\begin{figure*}[t]
\begin{center}
\includegraphics[width=6.1in]{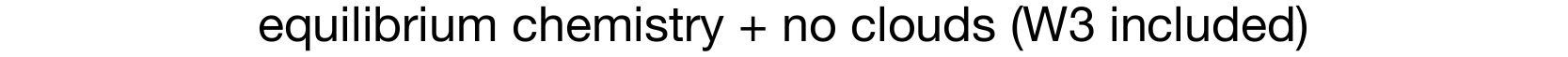}
\includegraphics[width=6.1in]{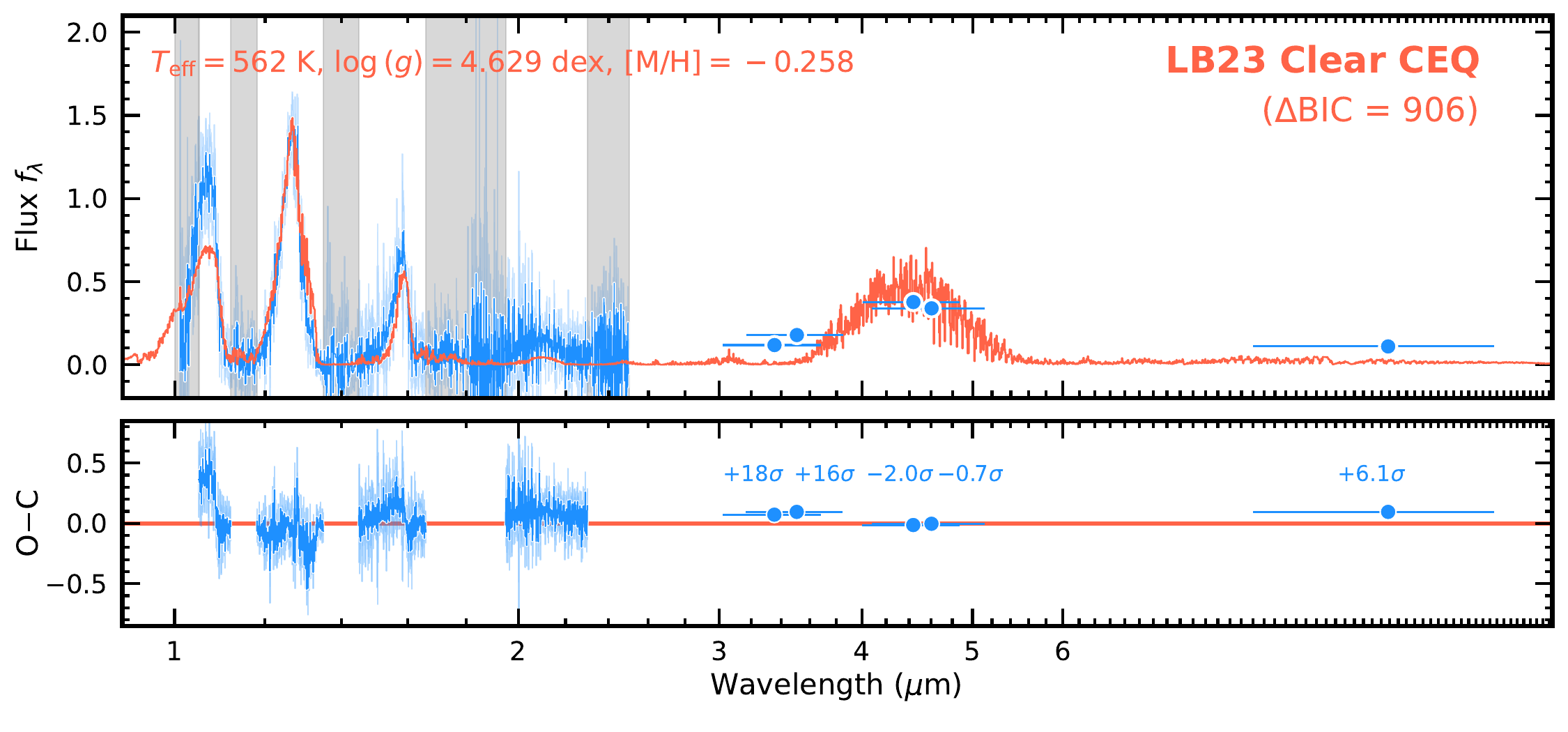}
\includegraphics[width=6.1in]{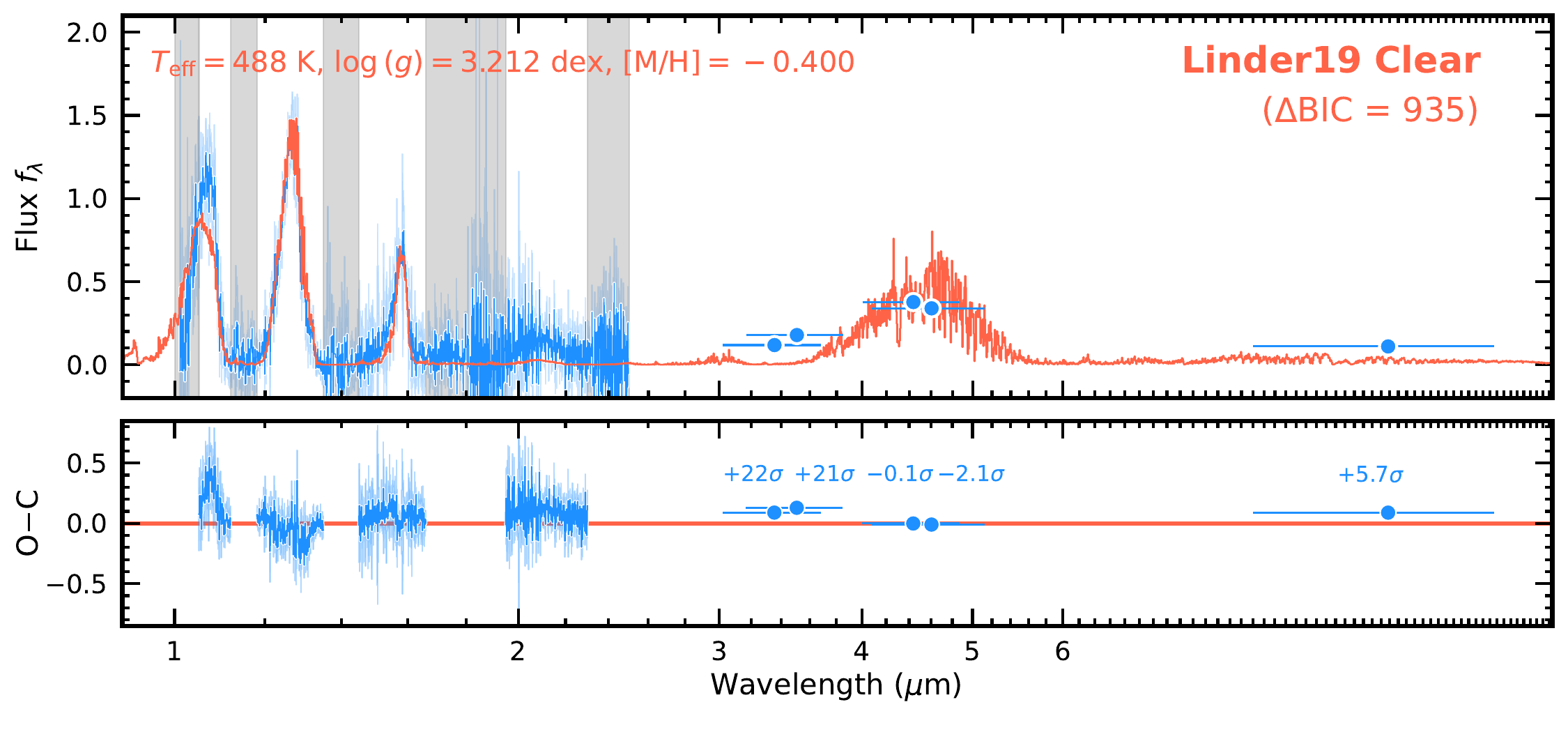}
\caption{Continued. }
\label{fig:w3}
\end{center}
\end{figure*}

\renewcommand{\thefigure}{\arabic{figure}}
\addtocounter{figure}{-1}
\begin{figure*}[t]
\begin{center}
\includegraphics[width=6.1in]{title_Clear_CEQ_w3.pdf}
\includegraphics[width=6.1in]{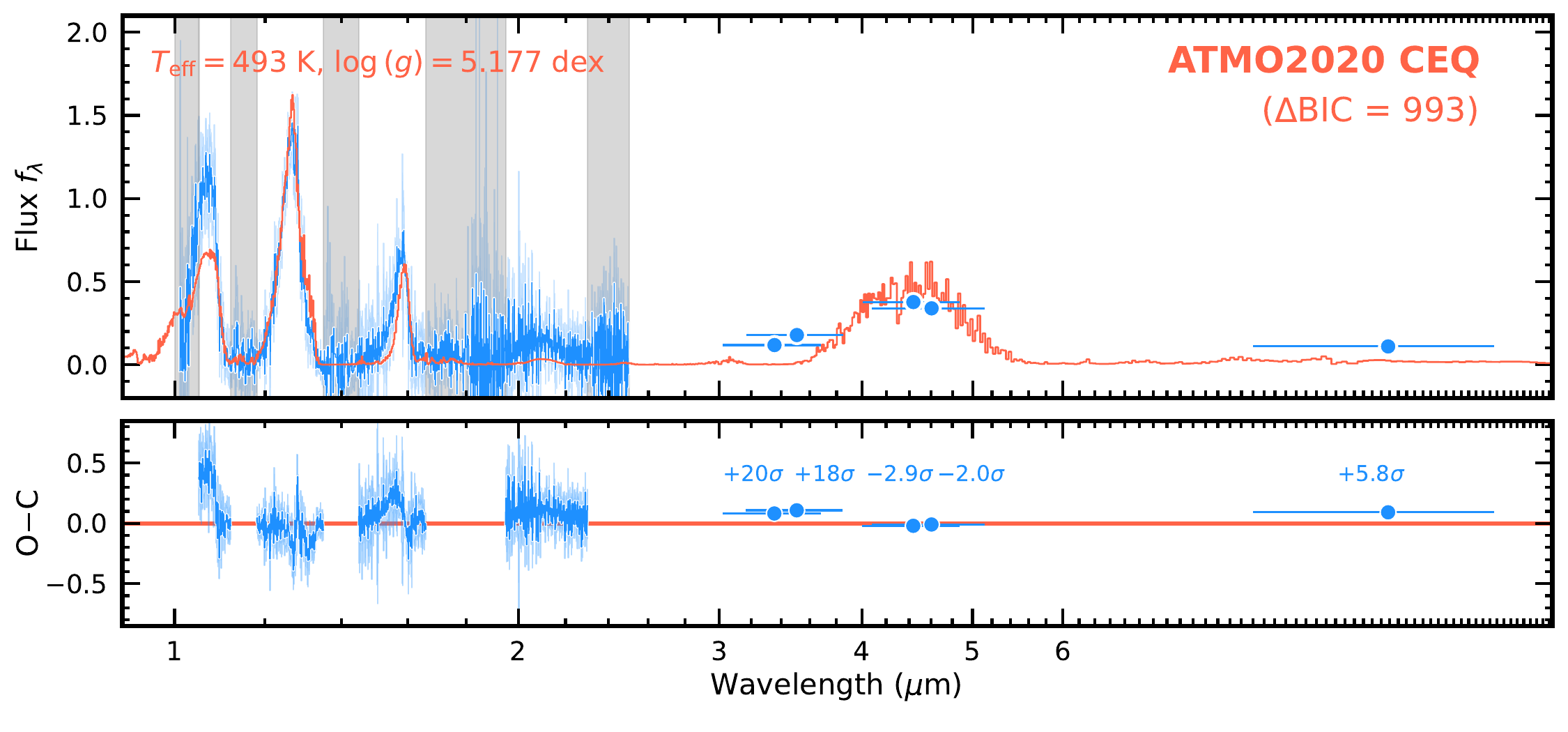}
\includegraphics[width=6.1in]{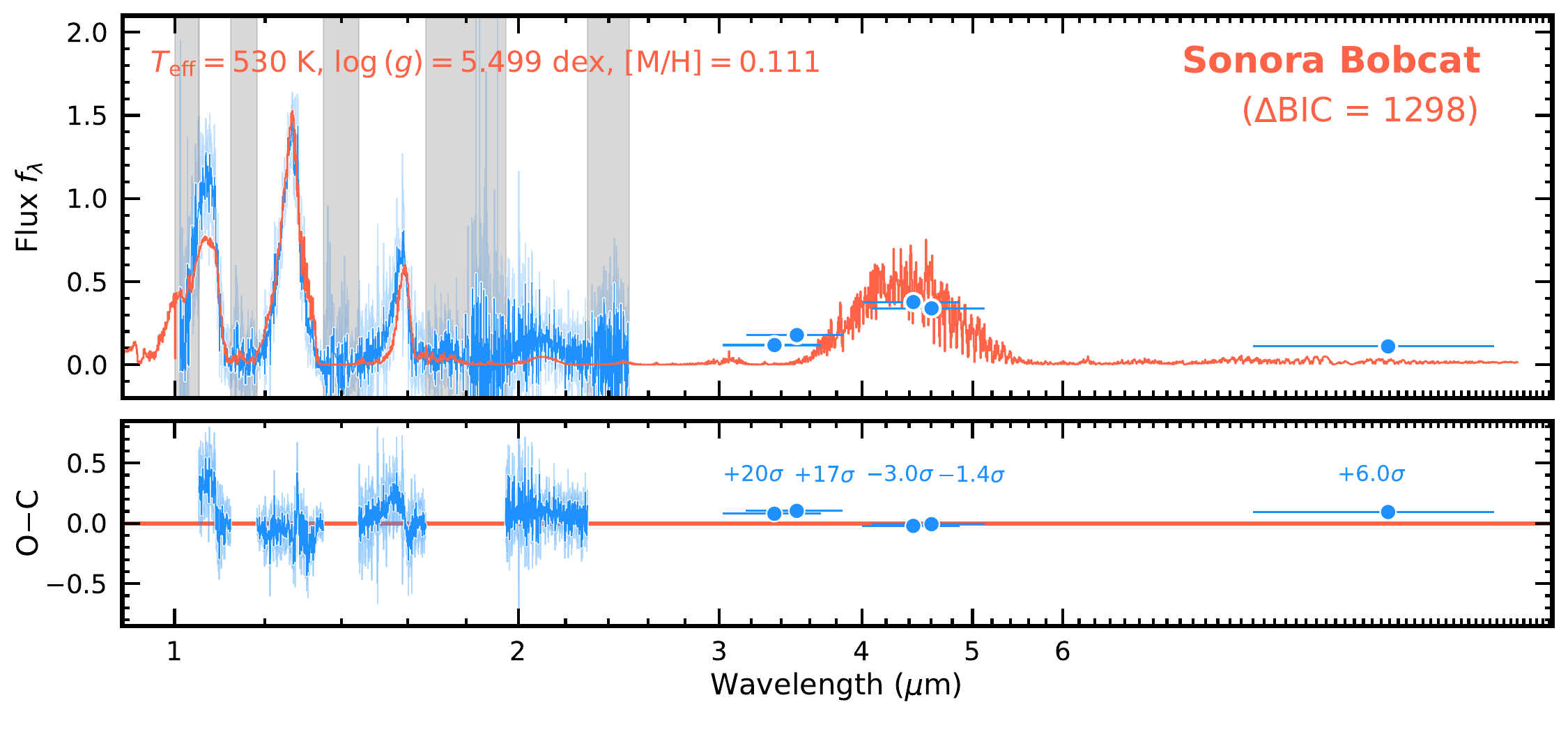}
\caption{Continued. }
\label{fig:w3}
\end{center}
\end{figure*}

\end{CJK*}

\clearpage
\bibliographystyle{aasjournal}
\bibliography{ms}{}

\end{document}